\documentclass[11pt]{article}
\pdfoutput=1
\usepackage{jheppub}
\usepackage{graphicx}
\usepackage{amssymb}
\usepackage{amsmath}
\usepackage{amsfonts}
\usepackage{latexsym}
\usepackage{hyperref}
\usepackage{physics}
\usepackage{bbm}
\usepackage{empheq}
\usepackage{tikz}
\usepackage{cancel}
\usepackage{hhline}
\usepackage{caption}
\usepackage{float}
\usetikzlibrary{arrows.meta}
\usetikzlibrary{decorations.markings}

\tikzset{->-/.style={decoration={
			markings,
			mark=at position #1 with {\arrow{>}}},postaction={decorate}}}

\newcommand\nn{\nonumber}
\newcommand\fft[2]{\frac{#1}{#2}}

\newcommand\tDelta{\widetilde{\Delta}}
\newcommand\tlambda{\widetilde{\lambda}}

\newcommand\tq{\tilde{q}}

\newcommand\mN{\mathcal{N}}
\def\ri{{\rm i}}
\newcommand\mR{\mathcal{R}}

\newcommand\mO{\mathcal{O}}

\newcommand\bDelta{\boldsymbol{\Delta}}
\newcommand\bmu{\boldsymbol{\mu}}
\newcommand\bnu{\boldsymbol{\nu}}
\newcommand\tir{\tilde{r}}
\newcommand\tpsi{\widetilde{\psi}}
\newcommand\mS{\mathcal{S}}
\newcommand\tS{\widetilde{S}}
\newcommand\mA{\mathcal{A}}
\newcommand\mfb{\mathfrak{b}}
\newcommand\mfc{\mathfrak{c}}
\newcommand\mD{\mathcal{D}}
\newcommand\mfm{\mathfrak{m}}

\makeatletter
\newcommand*{\rom}[1]{\expandafter\@slowromancap\romannumeral #1@}
\makeatother

\begin{document}
	
\title{Subleading analysis for $S^3$ partition functions of $\mN=2$ holographic SCFTs}
	
\author[a]{Seppe Geukens}
\author[b]{and Junho Hong}

\affiliation[a]{Theoretische natuurkunde, Vrije Universiteit Brussel (VUB),\\ Pleinlaan 2, B-1050 Brussels, Belgium}
\affiliation[b]{Department of Physics \& Center for Quantum Spacetime, Sogang University\,,\\ 35 Baekbeom-ro, Mapo-gu, Seoul 04107, Republic of Korea}

\emailAdd{seppe.geukens@vub.be}
\emailAdd{junhohong@sogang.ac.kr}

	
\abstract{We investigate the 3-sphere partition functions of various 3d $\mN=2$ holographic SCFTs arising from the $N$ stack of M2-branes in the 't~Hooft limit both analytically and numerically. We first employ a saddle point approximation to evaluate the free energy $F=-\log Z$ at the planar level, tracking the first subleading corrections in the large 't~Hooft coupling $\lambda$ expansion. Subsequently, we improve these results by determining the planar free energy to all orders in the large $\lambda$ expansion via numerical analysis. Remarkably, the resulting planar free energies turn out to take a universal form, supporting a prediction that these $S^3$ partition functions are all given in terms of an Airy function even beyond the special cases where the Airy formulae were derived analytically in the literature; in this context we also present new Airy conjectures in several examples. The subleading behaviors we captured encode a part of quantum corrections to the M-theory path integrals around dual asymptotically Euclidean AdS$_4$ backgrounds with the corresponding internal manifolds through holographic duality.}
	
\maketitle \flushbottom

\section{Introduction}\label{sec:intro}
\qquad The AdS/CFT correspondence \cite{Maldacena:1997re,Witten:1998qj} has provided a novel way to explore string or M-theory path integrals around asymptotically Euclidean AdS backgrounds in terms of dual CFT partition functions on the conformal boundaries, thereby shedding light on the quantum nature of gravity. A conventional perturbative analysis for CFT partition functions valid in the weak coupling regime could not be directly employed to comprehend the perturbative expansion of corresponding string/M-theory path integrals, however, since the AdS/CFT correspondence is featured as a strong-weak duality.

Supersymmetric localization \cite{Pestun:2007rz} has enabled our community to overcome this technical obstacle by providing finite-dimensional matrix models that capture various supersymmetric partition functions exactly, even in the strong coupling regime. In particular, various supersymmetric partition functions of holographic superconformal field theories (SCFT) have been extensively analyzed based on their matrix model representations as a means to unravel string/M-theory path integrals around dual supersymmetric asymptotically Euclidean AdS backgrounds.

Of particular interest in this line of research is the supersymmetric partition function on the round 3-sphere, whose corresponding free energy $F=-\log Z_{S^3}$ is regarded as a good measure for the degrees of freedom in a given theory \cite{Jafferis:2011zi,Casini:2011kv,Casini:2012ei}. Following the pioneering work of \cite{Kapustin:2009kz}, the $S^3$ partition function has been written in terms of a matrix model for general $\mN\geq2$ Chern-Simons-matter theories employing supersymmetric localization \cite{Kapustin:2010xq,Herzog:2010hf,Santamaria:2010dm,Jafferis:2010un,Martelli:2011qj,Cheon:2011vi,Jafferis:2011zi}. For the $S^3$ partition function of $\mN=2$ holographic SCFTs, in particular, the corresponding matrix models have been evaluated and compared with holographic dual quantities in a specific regime of parameters describing the theory. The first example was observed in a duality between M-theory on AdS$_4\times S^7/\mathbb{Z}_k$ and the U$(N)_k\times$U$(N)_{-k}$ ABJM theory arising from the $N$ stack of M2 branes probing the orbifold singularity of the cone over $S_7/\mathbb{Z}_k$ \cite{Aharony:2008ug}; in this example, the gravitational free energy of the M-theory captured by the classical on-shell action in the semi-classical limit was successfully matched with the ABJM $S^3$ free energy in the 't~Hooft limit where $\lambda=N/k$ plays the role of a 't~Hooft parameter \cite{Drukker:2010nc}. Such a leading order comparison between the $S^3$ free energy and a dual gravitational free energy has then been extended to more general $\mN\geq2$ holographic SCFTs dual to M-theory on AdS$_4\times X$ with $X$ labeling general 7d Sasaki-Einstein internal manifolds \cite{Herzog:2010hf,Santamaria:2010dm,Martelli:2011qj,Cheon:2011vi,Jafferis:2011zi,Freedman:2013oja}. See also \cite{Gaiotto:2009mv,Suyama:2010hr,Suyama:2013fua,Jafferis:2011zi,Marino:2011eh,Guarino:2015jca} for a similar comparison in holographic dualities involving massive IIA string theory. 

In some favorable cases, the $S^3$ partition function can be evaluated beyond the leading order in the 't~Hooft expansion and compared with a holographic dual quantity. Moreover, the perturbative expansion of the $S^3$ partition function of holographic SCFTs arising from the $N$ stack of M2 branes is beautifully captured by an Airy function for all known cases where the perturbative contribution has been evaluated explicitly. The examples include the ABJM theory possibly with real mass deformations \cite{Fuji:2011km,Marino:2011eh,Nosaka:2015iiw}, $\mN\geq4$ SCFTs with classical gauge groups \cite{Mezei:2013gqa}, and the ADHM theory (or $N_f$ matrix model) possibly with real mass and FI parameter deformations \cite{Grassi:2014vwa,Hatsuda:2021oxa,Chester:2023qwo}. Among these examples, the first Airy formula was derived for the undeformed ABJM $S^3$ partition function\footnote{The Airy formula for the ABJM partition function on a squashed 3-sphere with the squashing parameter $b=(\sqrt{3})^{\pm1}$ has also been obtained by employing a similar duality involving the topological strings on local $\mathbb{P}^2$, see \cite{Hatsuda:2016uqa} for details.\label{squashed:sqrt3}} by employing a chain of dualities linking the matrix model to the topological strings (B-model) living on the mirror dual of local $\mathbb{P}^1\times\mathbb{P}^1$ \cite{Fuji:2011km}. Subsequently, the free-Fermi gas formalism was established \cite{Marino:2011eh} and used extensively to derive similar Airy formulae in more general setting \cite{Nosaka:2015iiw,Mezei:2013gqa,Grassi:2014vwa,Hatsuda:2021oxa,Chester:2023qwo}. Reproducing these perturbatively exact $S^3$ partition functions beyond the semi-classical limit remains highly non-trivial on the gravity side. Still, there has been a remarkable achievement in the comparison of the first few subleading terms following the characteristic $N^{\fft32}$ leading order \cite{Klebanov:1996un,Drukker:2010nc} in the M-theory expansion ($N$ is taken large with $N/\lambda$ held fixed). The next-to-leading $N^\fft12$ orders in the $S^3$ free energies of the ABJM/ADHM theories, for example, were reproduced by employing the 4-derivative corrections to 4d $\mN=2$ minimal gauged supergravity \cite{Bobev:2020egg,Bobev:2021oku}. On the other hand, the next-next-to-leading $\log N$ order in the same M-theory expansion was derived by a 1-loop analysis in 11d supergravity focusing on the ABJM theory \cite{Bhattacharyya:2012ye}. As these successful subleading analyses on the gravity side imply, the beyond-the-leading order information in the $S^3$ partition function of holographic SCFTs is instrumental in exploring a dual string/M-theory path integral beyond the semi-classical limit; the leading order comparison is remarkable in providing non-trivial evidence for holographic duality but subleading behaviors are more relevant in the context of investigating quantum effects in quantum gravity path integrals beyond the semi-classical limit.

Given the prevalence of an Airy formula in the aforementioned holographic SCFTs, it is tempting to conjecture that the perturbative part of the $S^3$ partition function for general $\mN=2$ holographic SCFTs arising from $N$ coincident M2 branes is always captured by an Airy function. For example, the Airy formula for the $S^3$ partition function of the ABJM theory with generic real mass deformations was conjectured in this context and put through non-trivial consistency checks \cite{Bobev:2022jte,Bobev:2022eus,Hristov:2022lcw}, which extends the case focusing on a restricted set of real mass deformations studied via free Fermi-gas formalism \cite{Marino:2011eh,Nosaka:2015iiw}. Analytic proof of such an `Airy conjecture' remains a challenging problem though. In fact, even the first few subleading corrections in the $S^3$ partition function have not yet been analyzed in most $\mN=2$ SCFTs holographically dual to M-theory unless there is a proven Airy formula. A proper subleading analysis for the $S^3$ partition function of generic $\mN=2$ holographic SCFTs with M-theory duals is therefore crucial to test the Airy conjectures and thereby pave the way for the analysis of dual string/M-theory path integrals beyond the semi-classical limit.

In this paper we fill in this gap by tackling the subleading corrections to the $S^3$ partition functions in general $\mN=2$ holographic SCFTs arising from the $N$ stack of M2 branes, especially beyond the above listed examples with proven Airy formulae. In section \ref{sec:mm} we present various $\mN=2$ holographic SCFTs of our interest and the matrix models for their $S^3$ partition functions. In section \ref{sec:ana} we apply a saddle point approximation to these matrix models and consequently obtain the planar limit of the $S^3$ free energy, keeping track of the first subleading corrections in the large 't~Hooft coupling $\lambda$ expansion as in \cite{Mezei:2013gqa,Geukens2023}. In section \ref{sec:num} we improve the analytic result to all orders in the large $\lambda$ expansion by employing a numerical analysis. Remarkably, the resulting planar free energies turn out to take a universal form as a function of the 't~Hooft coupling in all examples of our interest, namely
\begin{equation}
	F_0(\lambda)=\mfb\big(\lambda-B\big)^\fft32+\fft{\mfc}{\lambda^2}+\mO(e^{-\#\sqrt{\lambda}})\,,\label{F0:universal:intro}
\end{equation}
where the relevant details including the theory-dependent constants $\mfb,\mfc,B$ can be found in the main text. In section \ref{sec:Airy} we compare the planar free energies with the known Airy conjectures and also conjecture new Airy formulae for several $S^3$ partition functions based on the universal form (\ref{F0:universal:intro}), serving as highly non-trivial evidence that the perturbative contribution to the $S^3$ partition function of interest is always given in terms of an Airy function. In section \ref{sec:discussion} we summarize the results and present open questions. In three Appendices, we present technical details involved in analytic calculation as well as the numerical data supporting our results.

\section{\texorpdfstring{$S^3$}{S3} partition functions of \texorpdfstring{$\mN=2$}{N=2} holographic SCFTs}\label{sec:mm}
In this section we introduce a class of $\mN=2$ holographic SCFTs of our interest and present the matrix models for their $S^3$ partition functions. 

To begin with, consider $\mN=2$ Chern-Simons-matter quiver gauge theories with the product unitary gauge group $G=\otimes_{r=1}^p\otimes\text{U}(N)_{k_r}$ where the subscript $k_r$ denotes the Chern-Simons (CS) level. The matter content of this CS-matter theory is described by various $\mN=2$ chiral multiplets collectively represented by $\Psi$ in the $\mR_\Psi$ representation with respect to the gauge group $G$. Such a theory may flow to the superconformal IR fixed point, and we study the $S^3$ partition function of the resulting $\mN=2$ IR SCFT via supersymmetric localization of the Lagrangian UV theory \cite{Jafferis:2010un}.

More specifically, supersymmetric localization provides a matrix model representation for the $S^3$ partition function of the $\mN=2$ CS-matter theory \cite{Kapustin:2009kz,Jafferis:2010un,Jafferis:2011zi,Martelli:2011qj,Hama:2010av},
\begin{equation}
\begin{split}
	Z_{S^3}&=\fft{1}{(N!)^p}\int\prod_{r=1}^p\Bigg[\prod_{i=1}^N\fft{d\mu_{r,i}}{2\pi}\,\exp^{\fft{\ri k_r}{4\pi}\sum_{i=1}^N\mu_{r,i}^2-\Delta_{r,m}\sum_{i=1}^N\mu_{r,i}}\prod_{i>j}^N4\sinh^2\fft{\mu_{r,i}-\mu_{r,j}}{2}\Bigg]\\
	&\kern5em~\times\prod_{\Psi}\prod_{\rho_\Psi}\exp[\ell\bigg(1-\Delta_{\Psi}+\ri\fft{\rho_\Psi(\mu)}{2\pi}\bigg)]\,,
\end{split}\label{Z:general}
\end{equation}
where the integral is over gauge zero modes $\mu_{r,i}$ that correspond to constant scalars in $\mN=2$ vector multiplets parametrizing the BPS manifold upon supersymmetric localization. In the matrix model (\ref{Z:general}), $\Delta_{m,r}$ denotes the bare monopole $R$-charge \cite{Jafferis:2011zi}, $\Delta_\Psi$ stands for the $R$-charge of a chiral multiplet $\Psi$, $\rho_\Psi$ runs over the weights of the representation $\mR_\Psi$, and the $\ell$-function used in the 1-loop contribution from chiral multiplets is introduced in Appendix \ref{app:fcts}. The $S^3$ partition function of the IR SCFT can be obtained by evaluating the matrix model (\ref{Z:general}) at the superconformal configuration of the $R$-charges that is determined by the $F$-maximization \cite{Jafferis:2010un,Jafferis:2011zi,Martelli:2011qj}.

\medskip

The matrix model (\ref{Z:general}) can be specialized further to describe the $S^3$ partition functions of various $\mN=2$ holographic SCFTs arising from the $N$ stack of M2 branes.\footnote{While analyzing the matrix model (\ref{Z:general}) for $\mN=2$ holographic SCFTs, we will not necessarily fix the $R$-charges to the exact superconformal values unless specified otherwise. Hence the resulting $S^3$ partition function will be given as a function of $R$-charges that break the conformal symmetry in general. We still call them partition functions of $\mN=2$ holographic SCFTs occasionally, however, for notational convenience. See \cite{Freedman:2013oja} for a holographic interpretation of such deformed partition functions.\label{foot:deform}} In the following subsections \ref{sec:mm:1-node} and \ref{sec:mm:2-node}, we summarize such $\mN=2$ holographic SCFTs that we explore throughout this paper.

\subsection{1-node}\label{sec:mm:1-node}
\begin{figure}[H]
	\centering
	\begin{tikzpicture}
		\draw[->-=0.52] (0.57,0.15) -- (2.43,0.15);
		\draw[->-=0.52] (2.43,-0.15) -- (0.57,-0.15);
		\draw[->-=0.45,->-=0.5,->-=0.55] (-0.2,0) arc (0:360:0.8) ;
		\node at (0,0) [circle,draw,scale=1.5,fill=white] {$N$};
		\node at (2.85,0) [rectangle,draw,scale=1.2,fill=white] {$N_f$};
		\node at (-2.3,0) {$\Psi_I$};
		\node at (1.55,0.5) {$\psi_q$};
		\node at (1.55,-0.6) {$\tpsi_q$};
	\end{tikzpicture}
	\caption{Quiver diagram for the ADHM/$V^{5,2}$ theories}\label{quiver:1-node}
\end{figure}
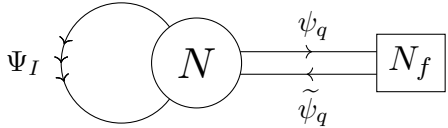

In this subsection we consider $\mN=2$ holographic SCFTs with a single unitary gauge group $G=\text{U}(N)$ and the vanishing CS level in the UV CS-matter description. The theory is also assumed to have three adjoint $\mN=2$ chiral multiplets and $N_f$ pairs of fundamental \& anti-fundamental $\mN=2$ chiral multiplets. The corresponding quiver diagram is presented in Fig.~\ref{quiver:1-node}. 
For these 1-node examples, the $S^3$ partition function (\ref{Z:general}) can be written more explicitly as
\begin{equation}
	\begin{split}
		Z_{S^3}^\text{1-node}&=\fft{1}{N!}\int\prod_{i=1}^N\fft{d\mu_{i}}{2\pi}\,\exp^{-\Delta_{m}\sum_{i=1}^N\mu_{i}}\times\prod_{i>j}^N4\sinh^2\fft{\mu_{i}-\mu_{j}}{2}\\
		&\quad\times\prod_{I=1}^3\prod_{i,j=1}^N\exp[\ell\bigg(1-\Delta_I+\ri\fft{\mu_i-\mu_j}{2\pi}\bigg)]\\
		&\quad\times\prod_{i=1}^N\exp[N_f\bigg\{\ell\bigg(1-\Delta+\ri\fft{\mu_i}{2\pi}\bigg)+\ell\bigg(1-\tDelta+\ri\fft{\mu_i}{2\pi}\bigg)\bigg\}]\,,
	\end{split}\label{Z:1-node}
\end{equation}
where we have assigned identical $R$-charges $\Delta$ \& $\tDelta$ to $N_f$ fundamental \& anti-fundamental chiral multiplets respectively.

\medskip

Imposing the superpotential marginality constraint on the $R$-charges appropriately, one can obtain the $S^3$ partition functions of various $\mN=2$ holographic SCFTs arising from M2 branes from the localization formula (\ref{Z:1-node}). In this paper we focus on the following two examples. 
\begin{itemize}
	\item ADHM theory (or $N_f$ matrix model) dual to M-theory on AdS$_4\times S^7/\mathbb{Z}_{N_f}$ with a fixed point \cite{Mezei:2013gqa,Grassi:2014vwa,Hosseini:2016ume,Bobev:2023lkx}. This theory has a superpotential
	\begin{equation}
		W=\Tr[\sum_{q=1}^{N_f}\tpsi_{q}\Psi_3\psi_q+\Psi_3[\Psi_1,\Psi_2]]\,,\label{ADHM:W}
	\end{equation}
	and the corresponding ``marginality'' --- following the terminology of \cite{Hosseini:2016ume} --- constrains the $R$-charges as
	\begin{equation}
		\sum_{I=1}^3\Delta_I=\Delta_3+\Delta+\tDelta=2\,.\label{ADHM:constraints}
	\end{equation}
	The superconformal configuration reads \cite{Jafferis:2011zi,Bobev:2023lkx}
	\begin{equation}
		\Delta_I=(\fft12,\fft12,1)\,,\qquad\Delta_m=0\,.\label{ADHM:constraints:sc}
	\end{equation}
	Throughout this paper, an overall numerical factor in the superpotential that does not affect the marginality constraints is neglected.
	
	\item SCFT$_3$ dual to M-theory on AdS$_4\times V^{5,2}/\mathbb{Z}_{N_f}$ dubbed as the $V^{5,2}$ theory \cite{Jafferis:2009th,Hosseini:2016ume,Bobev:2023lkx}. This theory has a superpotential
	\begin{equation}
		W=\Tr[\sum_{q=1}^{N_f}\psi_{\tq}(\Psi_1\Psi_2+\Psi_2\Psi_1-\Psi_3^2)\psi_q+\Psi_3[\Psi_1,\Psi_2]]\,,\label{V52:W}
	\end{equation}
	and the corresponding marginality constrains the $R$-charges as
	\begin{equation}
		\Delta_1+\Delta_2=\fft43\,,\qquad\Delta_3=\Delta+\tDelta=\fft23\,.\label{V52:constraints}
	\end{equation}
	The superconformal configuration corresponds to \cite{Jafferis:2011zi,Bobev:2023lkx}
	\begin{equation}
		\Delta_1=\Delta_2=\fft23\,,\qquad\Delta_m=0\,.\label{V52:constraints:sc}
	\end{equation}
	See \cite{Martelli:2009ga,Hosseini:2016ume} for an alternative UV description for the $V^{5,2}$ theory.
	
\end{itemize}
%

\subsection{2-node}\label{sec:mm:2-node}
In this subsection we consider $\mN=2$ holographic SCFTs with the gauge group $G=\text{U}(N)_k\times\text{U}(N)_{-k}$ and the opposite CS levels in the UV CS-matter description. The theories of our interest involve 2 pairs of $\mN=2$ bi-fundamental chiral multiplets and possibly more fundamental \& anti-fundamental pairs of $\mN=2$ chiral multiplets. For these 2-node examples, the matrix model for the $S^3$ partition function (\ref{Z:general}) reads
\begin{equation}
	\begin{split}
		Z_{S^3}^\text{2-node}&=\fft{1}{(N!)^2}\int\prod_{i=1}^N\fft{d\mu_i}{2\pi}\fft{d\nu_i}{2\pi}\,\exp^{\fft{\ri k}{4\pi}\sum_{i=1}^N(\mu_i^2-\nu_i^2)}\prod_{i>j}^N4\sinh^2\fft{\mu_{ij}}{2}4\sinh^2\fft{\nu_{ij}}{2}\\
		&\quad\times\prod_{i,j=1}^N\exp[\ell\bigg(1-\Delta_1+\ri\fft{\mu_i-\nu_j}{2\pi}\bigg)+\ell\bigg(1-\Delta_2+\ri\fft{\mu_i-\nu_j}{2\pi}\bigg)]\\
		&\quad\times\prod_{i,j=1}^N\exp[\ell\bigg(1-\Delta_3-\ri\fft{\mu_i-\nu_j}{2\pi}\bigg)+\ell\bigg(1-\Delta_4-\ri\fft{\mu_i-\nu_j}{2\pi}\bigg)]\\
		&\quad\times\prod_{i=1}^N\exp[\sum_{q=1}^{r_\mu}\ell\bigg(1-\Delta_{\mu_q}+\ri\fft{\mu_i}{2\pi}\bigg)+\sum_{q=1}^{\tir_\mu}\ell\bigg(1-\tDelta_{\mu_q}-\ri\fft{\mu_i}{2\pi}\bigg)]\\
		&\quad\times\prod_{i=1}^N\exp[\sum_{q=1}^{r_\nu}\ell\bigg(1-\Delta_{\nu_q}+\ri\fft{\nu_i}{2\pi}\bigg)+\sum_{q=1}^{\tir_\nu}\ell\bigg(1-\tDelta_{\nu_q}-\ri\fft{\nu_i}{2\pi}\bigg)]\,.
	\end{split}\label{Z:2-node}
\end{equation}

\medskip

As in the 1-node examples, we investigate the $S^3$ partition functions of various $\mN=2$ holographic SCFTs arising from the $N$ stack of M2 branes from the localization formula (\ref{Z:1-node}) by constraining the $R$-charges appropriately. The examples of our interest are listed below.
\begin{itemize}
	\item ABJM theory dual to M-theory on AdS$_4\times S^7/\mathbb{Z}_k$ without a fixed point \cite{Aharony:2008ug}. This theory is described without fundamental \& anti-chiral multiplets so $r_\mu=\tir_\mu=r_\nu=\tir_\nu=0$. The superpotential reads
	\begin{equation}
		W=\Tr[\Psi_1\Psi_3\Psi_2\Psi_4-\Psi_1\Psi_4\Psi_2\Psi_3]\,.\label{ABJM:W}
	\end{equation}
	The $R$-charges are therefore supposed to satisfy
	\begin{equation}
		\qquad\sum_{a=1}^4\Delta_a=2\,,\label{ABJM:constraints}
	\end{equation}
	where the superconformal configuration reads $\Delta_a=\fft12$ \cite{Jafferis:2011zi}. The ABJM quiver diagram is given by Fig.~\ref{quiver:ABJM}.
	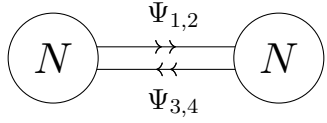
\begin{figure}
		\centering
		\begin{tikzpicture}
			\draw[->-=0.47,->-=0.57] (0.57,0.15) -- (2.43,0.15);
			\draw[->-=0.47,->-=0.57] (2.43,-0.15) -- (0.57,-0.15);
			\node at (0,0) [circle,draw,scale=1.5,fill=white] {$N$};
			\node at (3,0) [circle,draw,scale=1.5,fill=white] {$N$};
			\node at (1.6,0.55) {$\Psi_{1,2}$};
			\node at (1.6,-0.6) {$\Psi_{3,4}$};
		\end{tikzpicture}
		\caption{Quiver diagram for the ABJM theory}\label{quiver:ABJM}
	\end{figure}

	\item SCFT$_3$ dual to M-theory on AdS$_4\times N^{0,1,0}/\mathbb{Z}_{k}$ dubbed as the $N^{0,1,0}$ theory \cite{Gaiotto:2009tk,Hohenegger:2009as,Hikida:2009tp,Cheon:2011th,Hosseini:2016ume,Bobev:2023lkx}. This theory has $r=r_1+r_2$ pairs of fundamental \& anti-fundamental chiral multiplets where
	\begin{equation}
		r_1=r_\mu=\tir_\mu\,,\qquad r_2=r_\nu=\tir_\nu\,.\label{N010:r}
	\end{equation}
	One has to impose $r=k$ to describe the wanted holographic SCFT but we keep $r$ general for illustration \cite{Cheon:2011th,Bobev:2023lkx}, see \cite{Gaiotto:2009tk} for a proper interpretation of generic cases with $r\neq k$ involving D6-branes in IIA string theory backgrounds. The superpotential is given by
	\begin{equation}
		W=\Tr[\bigg(\Psi_1\Psi_4-\Psi_2\Psi_3-\sum_{q=1}^{r_1}\tpsi_{\mu_q}\psi_{\mu_q}\bigg)^2-\bigg(\Psi_1\Psi_4-\Psi_2\Psi_3-\sum_{q=1}^{r_2}\tpsi_{\nu_q}\psi_{\nu_q}\bigg)^2]\,,\label{N010:W}
	\end{equation}
	and the marginality constraints read
	\begin{equation}
		\Delta_1+\Delta_4=\Delta_2+\Delta_3=1\,,\qquad\Delta_{\mu_q}+\tDelta_{\mu_q}=\Delta_{\nu_q}+\tDelta_{\nu_q}=1\,.\label{N010:constraints}
	\end{equation}
	For this $N^{0,1,0}$ theory, we focus on the superconformal configuration \cite{Jafferis:2011zi,Bobev:2023lkx}
	\begin{equation}
		\Delta_a=\Delta_{\mu_q}=\tDelta_{\mu_q}=\Delta_{\nu_q}=\tDelta_{\nu_q}=\fft12\,,\label{N010:constraints:sc}
	\end{equation}
	and further assume
	\begin{equation}
		r_1=r_2=\fft{r}{2}\qquad(\alpha\equiv r/k)\label{N010:equal}
	\end{equation}
	for computational efficiency. See Appendix \ref{app:detail:N010} for analysis beyond the constraints (\ref{N010:constraints:sc}) and (\ref{N010:equal}) and related subtle issues. The $N^{0,1,0}$ quiver diagram is given by Fig.~\ref{quiver:N010}. 
	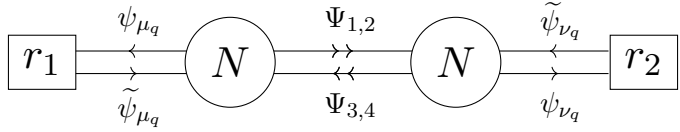
\begin{figure}[H]
		\centering
		\begin{tikzpicture}
			\draw[->-=0.47,->-=0.57] (0.57,0.15) -- (2.43,0.15);
			\draw[->-=0.47,->-=0.57] (2.43,-0.15) -- (0.57,-0.15);
			\draw[->-=0.5] (-0.57,0.15) -- (-2.13,0.15);
			\draw[->-=0.57] (-2.13,-0.15) -- (-0.57,-0.15);
			\draw[->-=0.5] (5.03,0.15) -- (3.47,0.15);
			\draw[->-=0.57] (3.47,-0.15) -- (5.03,-0.15);
			\node at (-2.5,0) [rectangle,draw,scale=1.5,fill=white] {$r_1$};
			\node at (0,0) [circle,draw,scale=1.5,fill=white] {$N$};
			\node at (3,0) [circle,draw,scale=1.5,fill=white] {$N$};
			\node at (5.5,0) [rectangle,draw,scale=1.5,fill=white] {$r_2$};
			\node at (1.6,0.55) {$\Psi_{1,2}$};
			\node at (1.6,-0.6) {$\Psi_{3,4}$};
			\node at (-1.2,0.55) {$\psi_{\mu_q}$};
			\node at (-1.2,-0.6) {$\tpsi_{\mu_q}$};
			\node at (4.4,0.55) {$\tpsi_{\nu_q}$};
			\node at (4.4,-0.6) {$\psi_{\nu_q}$};
		\end{tikzpicture}
		\caption{Quiver diagram for the $N^{0,1,0}$ theory}\label{quiver:N010}
	\end{figure}

	\item SCFT$_3$ dual to M-theory on AdS$_4\times Q^{1,1,1}/\mathbb{Z}_{N_f}$ dubbed as the $Q^{1,1,1}$ theory \cite{Benini:2009qs,Cremonesi:2010ae,Hosseini:2016ume,Bobev:2023lkx}. This theory has a vanishing CS level $k=0$ and $2N_f$ pairs of fundamental \& anti-fundamental chiral multiplets where
	\begin{equation}
		\tir_\mu=r_\nu=2N_f\,,\qquad r_\mu=\tir_\nu=0\,.\label{Q111:r}
	\end{equation}
	The superpotential reads
	\begin{equation}
		W=\Tr[\Psi_1\Psi_3\Psi_2\Psi_4-\Psi_1\Psi_4\Psi_2\Psi_3+\sum_{q=1}^{N_f}\tpsi_{\mu_q}\Psi_1\psi_{\nu_q}+\sum_{q=N_f+1}^{2N_f}\tpsi_{\mu_q}\Psi_2\psi_{\nu_q}]\,,\label{Q111:W}
	\end{equation}
	and consequently the $R$-charges are constrained as
	\begin{equation}
		\sum_{a=1}^4\Delta_a=2\,,\quad\Delta_1+\tDelta_{\mu_q}+\Delta_{\nu_{q}}=2~~(q\leq N_f)\,,\quad\Delta_2+\tDelta_{\mu_q}+\Delta_{\nu_q}=2~~(q>N_f)\,.\label{Q111:constraints}
	\end{equation}
	For this $Q^{1,1,1}$ theory, we focus on the superconformal configuration \cite{Jafferis:2011zi,Bobev:2023lkx}
	\begin{equation}
		\Delta_a=\fft12\,,\qquad\tDelta_{\mu_q}=\Delta_{\nu_q}=\fft34\,.\label{Q111:constraints:sc}
	\end{equation}
	The $Q^{1,1,1}$ quiver diagram is given by Fig.~\ref{quiver:Q111}. See \cite{Franco:2008um,Franco:2009sp} for an alternative UV description of the $Q^{1,1,1}$ theory.
	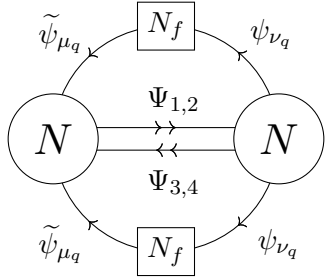
\begin{figure}[H]
		\centering
		\begin{tikzpicture}
			\draw[->-=0.47,->-=0.57] (0.57,0.15) -- (2.43,0.15);
			\draw[->-=0.47,->-=0.57] (2.43,-0.15) -- (0.57,-0.15);
			\draw[->-=0.28,->-=0.74] (3,0) arc (0:180:1.5);
			\draw[->-=0.28,->-=0.74] (3,0) arc (0:-180:1.5);
			\node at (0,0) [circle,draw,scale=1.5,fill=white] {$N$};
			\node at (3,0) [circle,draw,scale=1.5,fill=white] {$N$};
			\node at (1.5,1.5) [rectangle,draw,scale=1,fill=white] {$N_f$};
			\node at (1.5,-1.5) [rectangle,draw,scale=1,fill=white] {$N_f$};
			\node at (1.6,0.5) {$\Psi_{1,2}$};
			\node at (1.6,-0.6) {$\Psi_{3,4}$};
			\node at (2.9,1.4) {$\psi_{\nu_q}$};
			\node at (3,-1.4) {$\psi_{\nu_q}$};
			\node at (0.1,1.4) {$\tpsi_{\mu_q}$};
			\node at (0.1,-1.4) {$\tpsi_{\mu_q}$};
		\end{tikzpicture}
		\caption{Quiver diagram for the $Q^{1,1,1}$ theory}\label{quiver:Q111}
	\end{figure}
\end{itemize}

\medskip 

Throughout this paper we assume that $R$-charges take non-negative values (except the bare monopole $R$-charge for the 1-node case) to avoid otherwise complicated final expressions for the $S^3$ partition functions involving modded values of $R$-charges, in both 1-node and 2-node examples.

\section{Subleading structures of \texorpdfstring{$S^3$}{S3} partition functions; analytic approach}\label{sec:ana}
In this section we investigate the $S^3$ partition functions of various $\mN=2$ holographic SCFTs arising from the $N$ stack of M2 branes reviewed in the previous section \ref{sec:mm}, by applying the saddle point approximation to the corresponding matrix models in the following double scaling limit. We first take the large $N$ limit with a fixed 't~Hooft coupling $\lambda$ that will be specified later for various examples, where the holographic dual M-theory can be reduced to the type IIA string theory upon a dimensional reduction on the U(1) circle \cite{Nilsson:1984bj}. The 't~Hooft limit (or equivalently the IIA string theory limit) will then be followed by the large $\lambda$ limit. We call this 't~Hooft limit followed by the large $\lambda$ limit a double scaling limit. In this double scaling limit we focus on the planar contribution of order $N^2$ to the free energy
\begin{equation}
	F(N,\lambda)=-\log Z_{S^3}(N,\lambda)
\end{equation}
but investigate the planar contribution beyond the leading order in the large $\lambda$ limit. 

\medskip

To be more specific, we will investigate the genus-0 contribution $F_0(\lambda)$ in the genus expansion of the free energy
\begin{align}
    F(N,\lambda) = \sum_{g\geq 0} N^{2-2g}F_g(\lambda)\,, \label{F:genus}
\end{align}
which can be further expanded in the large $\lambda$ limit as
\begin{align}
     F_0(\lambda) = f_{1/2} \lambda^{-1/2} + f_{3/2} \lambda^{-3/2} + \cdots \, . \label{planar}
\end{align}
The leading $\lambda^{-1/2}$ behavior of the genus-0 free energy (\ref{planar}) has been confirmed in numerous $\mN=2$ holographic SCFTs arising from M2 branes via supersymmetric localization and the coefficients $f_{1/2}$ have also been evaluated explicitly and successfully compared with the holographic dual quantities, see \cite{Drukker:2010nc,Herzog:2010hf,Jafferis:2011zi} for example. We will therefore focus on the first subleading coefficients $f_{3/2}$, explaining how the saddle point approximation can be employed to yield $f_{3/2}$ as well as $f_{1/2}$. This is reminiscent of the analysis in \cite{Mezei:2013gqa,Geukens2023}, which will be extended to cover more general $\mN=2$ holographic SCFTs in the following subsections. 

\subsection{1-node}\label{sec:ana:1-node}
We first consider the 1-node examples in subsection \ref{sec:mm:1-node}. The 1-node partition function in  \eqref{Z:1-node} can be written compactly as  
\begin{equation}
   Z_{S^3}^\text{1-node}(N,\lambda) = \frac{1}{N!} \int \frac{d\bmu}{(2 \pi)^N} e^{-S^\text{1-node}[\bmu;N,\lambda]}\label{Z1}
\end{equation}
in terms of the effective action defined by
\begin{align}
    S^\text{1-node} [\bmu;N,\lambda] & = \Delta_m \sum^N_{i=1} \mu_i - 2\sum^N_{i > j} \log \left[2 \sinh \left(\frac{\mu_i - \mu_j}{2} \right) \right] - \sum^3_{I=1} \sum^N_{i,j=1} \ell \left(1 - \Delta_I +  \ri \frac{\mu_i - \mu_j}{2 \pi} \right) \nn \\
    &\quad- N_f \sum^N_{i=1} \left[\ell\left(1-\Delta + \ri \frac{\mu_i}{2 \pi}  \right) + \ell\left(1-\tDelta - \ri \frac{\mu_i}{2 \pi}  \right) \right] \,. \label{free1:def}
\end{align}
For notational convenience we omitted the dependence of the effective action on $R$-charges in the argument. Since the sum over $N$ eigenvalues effectively scales as $N$, we notice that the effective action (\ref{free1:def}) scales as $\sim N^2$ in the large $N$ limit with fixed
\begin{equation}
	\lambda\equiv\fft{N}{N_f}\quad(\text{'t~Hooft coupling})\qquad\&\qquad\chi\equiv\fft{\Delta_m}{N_f}\,,
\end{equation}
which defines the 't~Hooft limit in 1-node examples. Hence the matrix integral for the partition function (\ref{Z1}) can be analyzed using the saddle point approximation in the 't Hooft limit as
\begin{align}
    F^\text{1-node}(N,\lambda) = -\log Z_{S^3}^\text{1-node}(N,\lambda) = S^\text{1-node}[\bmu^\star;N,\lambda] + \text{(loop corrections)} \,,\label{free1:genus}
\end{align}
where $\bmu^\star$ stands for a saddle point configuration extremizing the effective action (\ref{free1:def}). 

\medskip

Based on the genus expansion (\ref{F:genus}) and the saddle point approximation (\ref{free1:genus}), one can determine the planar free energy by finding a saddle point $\bmu^\star$ and then evaluating the on-shell effective action as \cite{Herzog:2010hf,Jafferis:2011zi,Marino:2011nm,Mezei:2013gqa}
\begin{equation}
	N^2F^\text{1-node}_0(\lambda)= S^\text{1-node}[\bmu^\star;N,\lambda]\Big|_{N^2\text{-order}}\,,\label{planar1}
\end{equation}
where in the right hand side we are reading off the $N^2$ leading order in the 't~Hooft limit. In the following subsections we derive the planar free energy $F^\text{1-node}_0(\lambda)$ up to first subleading orders in the large $\lambda$ expansion based on the formula (\ref{planar1}). In due process we impose a single constraint
\begin{equation}
	\sum_{I=1}^3\Delta_I=2\label{1-node:constraints}
\end{equation}
shared by the ADHM/$V^{5,2}$ theories introduced in subsection \ref{sec:mm:1-node}, and then specialize to each example only in the last step of subsections \ref{sec:ana:1-node:lead} and \ref{sec:ana:1-node:sub} where we present their planar free energies.

\subsubsection{Effective action in the double scaling limit}\label{sec:ana:1-node:S}
Finding a saddle point configuration extremizing the effective action (\ref{free1:def}) is highly non-trivial, however, since it requires solving $N$ coupled non-linear equations. Hence a direct calculation of the planar free energy $F^\text{1-node}_0(\lambda)$ via the formula (\ref{planar1}) gets complicated. To simplify the problem, we will take advantage of the 't~Hooft limit first and then the large $\lambda$ expansion. 

\medskip
\noindent\textbf{Continuation in the 't~Hooft limit}
\medskip

To begin with, we rewrite the contributions from three adjoint $\mN=2$ chiral multiplets and the $\mN=2$ vector multiplet in (\ref{free1:def}) as
\begin{align}
	&S^\text{1-node} \left[\bmu;N,\lambda \right] \nn \\
	& = \Delta_m \sum_{i=1}^N \mu_i + \sum_{i> j}^N \sum^\infty_{n=1} \frac{2}{n} e^{-n (\mu_i-\mu_j)}\nn \\
	&\quad - \sum_{I=1}^3 \sum_{i>j}^N \sum^\infty_{n=1}\frac{2}{n} \left[ (1-\Delta_I) \cos 2 \pi n \Delta_I + \frac{\mu_i - \mu_j}{2 \pi} \sin 2 \pi n \Delta_I + \frac{1}{2\pi n} \sin 2 \pi n \Delta_I \right] e^{-n (\mu_i-\mu_j)}  \nn\\
	&\quad -N\sum_{I=1}^3\ell(1-\Delta_I) - N_f \sum^N_{i=1} \left[\ell\left(1-\Delta + \ri \frac{\mu_i}{2 \pi}  \right) + \ell\left(1-\tDelta- \ri \frac{\mu_i}{2 \pi}  \right) \right] \label{free1}
\end{align}
based on the series expansions (\ref{expl}) and (\ref{expu}), assuming the order of eigenvalues
\begin{equation}
	\Re[\mu_i]>\Re[\mu_j]\quad(i>j)
\end{equation}
without loss of generality. Then we take the 't~Hooft limit to explore the resulting expression (\ref{free1}) more explicitly. In the 't~Hooft limit, a discrete set of eigenvalues $\{\mu_i\}$ can be described by a continuous eigenvalue distribution $\mu:I\subseteq\mathbb{R}\to\mathbb{C}$ as
\begin{equation}
	\mu(q(i))=\mu_i\,,\label{mu:conti}
\end{equation}
where we have introduced a continuous increasing function $q:[1,N]\to I$ that maps discrete $N$ indices to a subset along the real line. Upon this continuation we define a normalized eigenvalue density $\rho:I\to\mathbb{R}$ as
\begin{equation}
	di=(N-1)\rho(q)dq\qquad\to\qquad \int_{q_l}^{q_r} dq\,\rho(q)=1\,,\label{rho1}
\end{equation}
where we have implicitly assumed that the function $q$ ranges over a certain interval $I=[q_l,q_r]$ that will be justified later. Consequently, the effective action (\ref{free1}) can be written in terms of continuous functions as
\begin{subequations}
\begin{align}
	S^\text{1-node} [\bmu;N,\lambda] & = N^2S^\text{1-node}_0[\rho,\mu;\lambda]+o(N^2)\,,\label{free1:conti:1}\\
	S^\text{1-node}_0[\rho,\mu;\lambda]& =\fft{\chi}{\lambda}\int_{q_l}^{q_r}dq\,\rho(q)\mu(q) +\int_{q_l}^{q_r}dq\,\rho(q)\int_{q_l}^qdq'\,\rho(q')\sum^\infty_{n=1} \frac{2}{n} e^{-n\left(\mu(q) - \mu(q')\right)} \nn\\
	&\quad - \sum^3_{I=1} \int_{q_l}^{q_r}dq\,\rho(q)\int_{q_l}^{q}dq'\,\rho(q')\sum^\infty_{n=1}\frac{2}{n} \bigg[ (1-\Delta_I) \cos 2 \pi n \Delta_I  \nn\\ 
	&\quad + \frac{\mu(q) - \mu(q')}{2 \pi} \sin 2 \pi n \Delta_I  + \frac{1}{2\pi n} \sin 2 \pi n \Delta_I \bigg]e^{-n(\mu(q)-\mu(q'))} \nn \\
	&\quad- \fft{1}{\lambda} \int_{q_l}^{q_r}dq\,\rho(q)\left[\ell\left(1-\Delta + \ri \frac{\mu(q)}{2 \pi}  \right) + \ell\left(1-\tDelta - \ri \frac{\mu(q)}{2 \pi}  \right) \right] \,, \label{free1:conti:2}
\end{align}\label{free1:conti}%
\end{subequations}
based on the Euler-Maclaurin formula. We refer the readers to \cite{Liu:2019tuk,Hong:2021bsb} for a detailed application of the Euler-Maclaurin formula to this type of continuation in the 't~Hooft limit but here we do not address complicated subleading corrections of order $o(N^2)$ since we are interested in the leading planar contribution only.

\medskip

Finding a saddle point $\bmu^\star$ extremizing the original effective action (\ref{free1}) now turns into a quest to find the continuous eigenvalue distribution $\mu(q)$ and the eigenvalue density $\rho(q)$ extremizing the planar effective action (\ref{free1:conti:2}) under the normalization constraint (\ref{rho1}). Accordingly we have replaced a discrete argument $\bmu$ of the effective action with continuous functions $\rho,\mu$ in the 't~Hooft limit.

\medskip
\noindent\textbf{Saddle point ansatz in the large 't~Hooft coupling limit}
\medskip

Finding $\mu(q)$ and $\rho(q)$ that extremize the planar effective action (\ref{free1:conti:2}) is still not straightforward, however, and therefore we make a saddle point ansatz for the continuous eigenvalue distribution $\mu(q)$ in the large $\lambda$ limit as\footnote{Note that we do not turn on subleading corrections of order $\mO(\lambda^{-1/2})$ in the saddle point ansatz (\ref{ansatz1}). This does not lose generality since one can always set the real part of the saddle point ansatz as $\lambda^{1/2}q$ by redefining the coordinate $q$, while the imaginary part does not require higher order corrections as commented below (\ref{free1:conti:split}).} 
\begin{equation} 
	\mu(q) = \lambda^{1/2} q +  \ri \pi (\tDelta - \Delta )\, . \label{ansatz1}
\end{equation}
The leading $\lambda^{1/2}$ order term in the ansatz (\ref{ansatz1}) is well known in the literature and captures the leading $\lambda^{-1/2}$ term in the planar free energy (\ref{planar}) precisely, see \cite{Herzog:2010hf,Martelli:2011qj,Jafferis:2011zi,Mezei:2013gqa} for example. On the other hand, the $\lambda^0$ behavior in the ansatz (\ref{ansatz1}) is new and turns out to be crucial to derive the first subleading $\lambda^{-3/2}$ behavior in the planar free energy (\ref{planar1}).\footnote{See \cite{Liu:2017vll,PandoZayas:2020iqr,PandoZayas:2019hdb,Bobev:2023lkx} for a similar shift in the eigenvalue distribution employed in the numerical analysis for 3d topologically twisted indices.} To see this explicitly, let us substitute the ansatz (\ref{ansatz1}) into the planar effective action (\ref{free1:conti:2}) and split the contributions into three parts as
\begin{subequations}
\begin{align}
	S^\text{1-node}_0[\rho;\lambda]&=S_0^\text{1-mono}[\rho;\lambda]+S_0^\text{1-adj}[\rho;\lambda]+S_0^\text{1-fun}[\rho;\lambda]\,,\\
	S_0^\text{1-mono}[\rho;\lambda]&=\lambda^{-1/2}\chi\int_{q_l}^{q_r}dq\,\rho(q)q+\lambda^{-1}\ri\pi(\tDelta-\Delta)\chi\,,\\
	S_0^\text{1-adj}[\rho;\lambda]&=\int_{q_l}^{q_r}dq\,\rho(q)\int_{q_l}^qdq'\,\rho(q')\sum_{n=1}^\infty\fft{2}{n}e^{-n\lambda^{1/2}(q-q')} \nn\\
	&\quad - \sum^3_{I=1} \int_{q_l}^{q_r}dq\,\rho(q)\int_{q_l}^{q}dq'\,\rho(q')\sum^\infty_{n=1}\frac{2}{n} \bigg[ (1-\Delta_I) \cos 2 \pi n \Delta_I  \nn\\ 
	&\quad + \frac{\lambda^{1/2}(q-q')}{2 \pi} \sin 2 \pi n \Delta_I  + \frac{1}{2\pi n} \sin 2 \pi n \Delta_I \bigg]e^{-n\lambda^{1/2}(q-q')}\,,\label{free1:conti:split:adj}\\
	S_0^\text{1-fun}[\rho;\lambda]&=-\lambda^{-1}\int_{q_l}^{q_r}dq\,\rho(q)\bigg[\ell\left(\fft{2-\Delta - \tDelta}{2} + \ri \frac{\lambda^{1/2}q}{2 \pi}  \right) \nn\\
	&\kern9em + \ell\left(\frac{2-\Delta - \tDelta}{2} - \ri \frac{\lambda^{1/2}q}{2 \pi}  \right) \bigg]\,,\label{free1:conti:split:fun}
\end{align}\label{free1:conti:split}%
\end{subequations}
where we have removed $\mu$ in the argument that is already specified by the ansatz (\ref{ansatz1}), and included the contribution from the vector multiplet into $S_0^\text{1-adj}$ for computational efficiency and notational convenience. Note that the $\lambda^0$ order constant term in the ansatz (\ref{ansatz1}) is chosen for the contributions from fundamental \& anti-fundamental $\mN=2$ chiral multiplets in (\ref{free1:conti:split:fun}) to be complex conjugate of each other beyond the $\lambda^{-1/2}$, which allows us to evaluate the planar free energy including the first $\lambda^{-3/2}$ subleading orders below. Since this complex conjugate property holds exactly regardless of the large $\lambda$ limit, the imaginary part in the ansatz (\ref{ansatz1}) does not involve higher order corrections.

\medskip

Thanks to the ansatz (\ref{ansatz1}) specifying the eigenvalue distribution, now it suffices to find the eigenvalue density $\rho(q)$ extremizing the planar effective action (\ref{free1:conti:split}) under the normalization constraint (\ref{rho1}). To solve this extremization problem, one should first expand the contributions from adjoint $\mN=2$ chiral multiplets (\ref{free1:conti:split:adj}) and fundamental \& anti-fundamental $\mN=2$ chiral multiplets (\ref{free1:conti:split:fun}) explicitly in the large $\lambda$ limit. We carry out this large $\lambda$ expansion in Appendix \ref{app:detail:isol1} in detail and present only the final results here: 
\begin{subequations}
\begin{align}
	S_0^\text{1-adj}[\rho;\lambda] &=  2\pi^2 \lambda^{-1/2}\Delta_1 \Delta_2 \Delta_3\int^{q_r}_{q_l} dq\,\rho(q)^2 +\lambda^{-1}\mS(\bDelta)\int_{q_l}^{q_r}dq\,\rho(q)\rho(q)' \nn\\
	&\quad - \pi^4 \lambda^{-3/2} \Delta_1 \Delta_2 \Delta_3 \left(\Delta_1^2+\Delta_2^2+\Delta_3^2-2 \right) \int^{q_r}_{q_l} dq\,\rho(q)\rho''(q)   \nn\\
	&\quad+ \mO(\lambda^{-2}) \,, \label{Free1adj}\\
	S_0^\text{1-fun}[\rho;\lambda]   &= \frac{1}{2}\lambda^{-1/2} \big( 2 - \Delta - \tDelta\big)\int^{q_r}_{q_l} dq\, \rho(q)|q|  \nn\\ 
	&\quad-\frac{1}{6}\lambda^{-3/2} \pi^2 \rho\left(0\right) (2-\Delta- \tDelta)(\Delta+\tDelta)(\Delta+\tDelta- 4) + \mO(\lambda^{-2}) \, . \label{Free1fun}
\end{align}\label{Free1:adj+fun}%
\end{subequations}
Note that we kept track of subleading corrections to the $\lambda^{-3/2}$ order according to our purpose, leaving $\mS(\bDelta)$ implicit since it won't affect the planar free energy at the end of the day as we will see in the next subsection. 

\medskip

Substituting (\ref{Free1:adj+fun}) back into (\ref{free1:conti:split}) we obtain the large $\lambda$ expansion of the planar effective action as
\begin{subequations}
\begin{align}
	S^\text{1-node}_0[\rho;\lambda]&=\lambda^{-1/2}S^\text{1-node}_{0,1/2}[\rho]+\lambda^{-1}S^\text{1-node}_{0,1}[\rho]+\lambda^{-3/2}S^\text{1-node}_{0,3/2}[\rho]+\mO(\lambda^{-2})\,,\\
	S^\text{1-node}_{0,1/2}[\rho]&= \chi  \int^{q_r}_{q_l} dq \, \rho(q) q + 2 \pi^2 \Delta_1 \Delta_2 \Delta_3\int^{q_r}_{q_l} dq \, \rho(q)^2 \nn \\
	& \quad +  \frac{2 - \Delta - \tDelta}{2}\int^{q_r}_{q_l} dq \, \rho(q) |q|\,, \label{free1:conti:largelambda:12}\\
	S^\text{1-node}_{0,1}[\rho]&=\ri\pi(\tDelta-\Delta)\chi+\mS(\bDelta)\int_{q_l}^{q_r}dq\,\rho(q)\rho(q)'\,, \label{free1:conti:largelambda:1}\\
	S^\text{1-node}_{0,3/2}[\rho]&=-  \pi^4  \Delta_1 \Delta_2 \Delta_3 \left(\Delta_1^2+\Delta_2^2+\Delta_3^2-2 \right) \int_{q_l}^{q_r}dq\,\rho(q)\rho''(q) \nn\\
	&\quad-\frac{1}{6}\pi^2 \rho\left(0\right) (2-\Delta- \tDelta)(\Delta+\tDelta)(\Delta+\tDelta- 4) \,.\label{free1:conti:largelambda:32}
\end{align}\label{free1:conti:largelambda}%
\end{subequations}
%

\subsubsection{Planar free energy from the effective action}\label{sec:ana:1-node:F}
The planar free energy of our interest can now be written in terms of the on-shell value of the planar effective action (\ref{free1:conti:largelambda}) based on the formula (\ref{planar1}) as
\begin{equation}
	F^\text{1-node}_0(\lambda)= S_0^\text{1-node}[\rho^\star;\lambda]\,,\label{planar1:genus0}
\end{equation}
where $\rho^\star(q)$ denotes the eigenvalue density extremizing the planar effective action (\ref{free1:conti:largelambda}) under the normalization constraint (\ref{rho1}). Hence the first step toward the evaluation of the planar free energy is to analyze the saddle point configuration $\rho^\star(q)$.

\medskip

From the large $\lambda$ expansion of the planar effective action (\ref{free1:conti:largelambda}), we expect that the saddle point configuration be expanded in the large $\lambda$ limit as
\begin{equation}
	\rho^\star(q)=\rho_0(q)+\lambda^{-1/2}\rho_{1/2}(q)+\lambda^{-1}\rho_1(q)+\mO(\lambda^{-3/2})\,.\label{rho1:exp}
\end{equation}
The expansion coefficients in (\ref{rho1:exp}) can be determined by solving the saddle point equation
\begin{equation}
	\fft{\delta\tS^\text{1-node}_0[\rho;\lambda]}{\delta\rho(q)}\bigg|_{\rho=\rho^\star}=0\label{saddle1}
\end{equation}
order by order, where we have defined $\tS^\text{1-node}_0[\rho;\lambda]$ by introducing $L$ as a Lagrange multiplier for the normalization condition in (\ref{rho1}) as
\begin{equation}
	\tS^\text{1-node}_0[\rho;\lambda]\equiv S^\text{1-node}_0[\rho;\lambda]-\lambda^{-1/2}L\bigg(\int_{q_l}^{q_r}dq\,\rho(q)-1\bigg)\,.
\end{equation}
To be more explicit, the first three terms in the large $\lambda$ expansion of the saddle point equation (\ref{saddle1}) read
\begin{subequations}
\begin{align}
	&\lambda^{-1/2}&:\quad 0&=S_{0,1/2}^\text{1-node}\,{}'[\rho_0;\lambda]-L\,,\label{saddle1:exp:12}\\[0.5em]
	&\lambda^{-1}&:\quad 0&=S_{0,1/2}^\text{1-node}\,{}''[\rho_0;\lambda]\rho_{1/2}(q)+S_{0,1}^\text{1-node}\,{}'[\rho_0;\lambda]\,,\label{saddle1:exp:1}\\[0.5em]
	&\lambda^{-3/2}&:\quad 0&=\fft12S_{0,1/2}^\text{1-node}\,{}'''[\rho_0;\lambda]\rho_{1/2}(q)^2+S_{0,1/2}^\text{1-node}\,{}''[\rho_0;\lambda]\rho_1(q)\nn\\
	&&&\quad+S_{0,1}^\text{1-node}\,{}''[\rho_0;\lambda]\rho_{1/2}(q)+S_{0,3/2}^\text{1-node}\,{}'[\rho_0;\lambda]\,,\label{saddle1:exp:32}
\end{align}\label{saddle1:exp}%
\end{subequations}
where the prime denotes a functional derivative with respect to $\rho(q)$ in this context. The $\lambda^{-1/2}$ order equation (\ref{saddle1:exp:12}) determines the leading order saddle point configuration $\rho_0(q)$. The $\lambda^{-1}$ order equation (\ref{saddle1:exp:1}) is then solved trivially as
\begin{equation}
	S_{0,1}^\text{1-node}\,{}'[\rho;\lambda]=0~~\text{for any}~~\rho(q)\qquad\to\qquad \rho_{1/2}(q)=0\,,\label{rho1:exp:12}
\end{equation}
since the $\rho$-dependent term in the planar effective action at the $\lambda^{-1}$ order (\ref{free1:conti:largelambda:1}) is captured by a total derivative as
\begin{equation}
	\int_{q_l}^{q_r}dq\,\fft{d}{dq}\Big[\rho(q)^2\Big]\,.
\end{equation}
Hence the first non-trivial correction to the leading order eigenvalue density $\rho_0(q)$ becomes $\rho_1(q)$, and it is determined by solving the $\lambda^{-3/2}$ order equation (\ref{saddle1:exp:32}). 

\medskip

Employing the above perturbative analysis for the large $\lambda$ expansion of the saddle point configuration $\rho^\star(q)$, the planar free energy (\ref{planar1:genus0}) can be expanded as
\begin{equation}
\begin{split}
	F_0^\text{1-node}(\lambda)&=S^\text{1-node}_0[\rho^\star;\lambda]=\tS^\text{1-node}_0[\rho^\star;\lambda]\\[0.5em]
	&=\lambda^{-1/2}S^\text{1-node}_{0,1/2}[\rho_0]+\lambda^{-1}\ri\pi(\tDelta-\Delta)\chi+\lambda^{-3/2}S^\text{1-node}_{0,3/2}[\rho_0]+\mO(\lambda^{-2})\,.
\end{split}\label{planar1:S}
\end{equation}
To arrive at the 2nd line of (\ref{planar1:S}), we have used the leading order saddle point equation (\ref{saddle1:exp:12}) together with the observation (\ref{rho1:exp:12}). Note that the first non-trivial subleading correction to the saddle point configuration, $\rho_1(q)$, does not affect the planar free energy to the $\lambda^{-3/2}$ order as observed in \cite{Mezei:2013gqa,Geukens2023}. Another remark is that the real part of the $\lambda^{-1}$ order planar free energy vanishes on-shell as seen in the aforementioned references, resulting in the typical form of the expansion presented in \eqref{planar} provided we ignore the imaginary part of the free energy that is only determined modulo $2\pi\mathbb{Z}$ after all. This imaginary term corresponds to a constant phase in the $S^3$ partition function that can be determined easily by shifting the dummy integration variables in the matrix model (\ref{Z:1-node}), see the second last comment following (\ref{F:ADHM:num}) for details.

\medskip

To evaluate the large $\lambda$ expansion of the planar free energy (\ref{planar1:S}) explicitly, we will now determine the leading order saddle point configuration $\rho_0(q)$ and then evaluate the planar free energy up to $\lambda^{-3/2}$ order in the following subsections.

\subsubsection{Planar free energy at leading order} \label{sec:ana:1-node:lead}
Now we present the planar free energies for two different examples of our interest, namely ADHM and $V^{5,2}$ theories. In this subsection we first review the leading order saddle point configuration and the planar free energy in the large $\lambda$ limit \cite{Martelli:2011qj,Cheon:2011vi,Jafferis:2011zi}, and then investigate the first subleading order corrections in the next subsection.  

\medskip

The leading order saddle point configuration $\rho_0(q)$ solving (\ref{saddle1:exp:12}) reads \cite{Martelli:2011qj,Cheon:2011vi,Jafferis:2011zi}
\begin{subequations} \label{1-sol}
    \begin{equation}
        \begin{aligned} 
            \rho_0(q) = \begin{cases} 
            	0  \quad & q \leq q_l \\ 
            \frac{L + \tDelta_3q}{4 \pi^2 \tDelta_1 \tDelta_2 \left(2- \tDelta_1 - \tDelta_2 \right)} \quad & q_l<q\leq 0\\
            \frac{L - \tDelta_4q}{4 \pi^2 \tDelta_1 \tDelta_2 \left(2- \tDelta_1 - \tDelta_2 \right)} \quad & 0<q\leq q_r \\
            0  \quad & q_r < q
            \end{cases}
        \end{aligned}
    \end{equation}
    in terms of the Lagrange multiplier
    \begin{align}
    	L = 2 \pi \sqrt{\frac{2 (2- \tDelta_1 - \tDelta_2)}{\tDelta_3 + \tDelta_4}\tDelta_1 \tDelta_2 \tDelta_3 \tDelta_4} \,,
    \end{align}
    where the endpoints of the interval are given by
    \begin{align}
        q_l = -\frac{L}{\tDelta_3} \, , \qquad q_r = \frac{L}{\tDelta_4}\, .
    \end{align}
\end{subequations}
In the above expressions we have employed $\tDelta$ parameters defined as
\begin{align}
	\widetilde{\bDelta} = \left(\Delta_1, \Delta_2, \frac{2-\Delta-\tDelta}{2}- \chi,\frac{2-\Delta-\tDelta}{2}+ \chi \right) \,. \label{eq:not}
\end{align}
Note that the existence of a well-behaving saddle point (\ref{1-sol}) justifies the assumption made in (\ref{rho1}) that the eigenvalue density $\rho(q)$ is supported along a single interval $I=[q_l,q_r]$. 

\medskip

Substituting the saddle point solution (\ref{1-sol}) back into the leading order planar effective action (\ref{free1:conti:largelambda:12}), we obtain
\begin{equation}
	S^\text{1-node}_{0,1/2}[\rho_0]=\frac{2L}{3} \,. \label{S:112}
\end{equation}
The $\lambda^{-1/2}$ leading contribution to the planar free energy is then given from the formula (\ref{planar1:S}) for the two examples discussed in subsection \ref{sec:mm:1-node} as follows \cite{Martelli:2011qj,Cheon:2011vi,Jafferis:2011zi}. 
\begin{itemize}
    \item Planar free energy of the ADHM theory:
    \begin{align}
           F^\text{ADHM}_0[\lambda]= \frac{4\pi}{3} \sqrt{2\tDelta_1 \tDelta_2 \tDelta_3 \tDelta_4} \, \lambda^{-1/2} + \mO(\lambda^{-1}) \,,\label{F:ADHM:lead}
    \end{align}
    with the constraint $\sum_{a=1}^4\tDelta_a=2$. 
    
    \item Planar free energy of the $V^{5,2}$ theory:
    \begin{align}
        F^{V^{5,2}}_0[\lambda] = \frac{4\pi}{3} \sqrt{\tDelta_1 \tDelta_2 \tDelta_3 \tDelta_4} \, \lambda^{-1/2} + \mO(\lambda^{-1}) \, ,\label{F:V52:lead}
    \end{align}
    with the constraint $\tDelta_1+\tDelta_2=\tDelta_3+\tDelta_4 =\fft43$.
\end{itemize}

\medskip

\noindent The $F$-maximization \cite{Jafferis:2010un,Jafferis:2011zi} then tells us that the configuration of $R$-charges maximizing the leading order planar free energy under the given constraints determines the superconformal fixed points for each case (see \eqref{ADHM:constraints:sc} and \eqref{V52:constraints:sc}). We will not restrict ourselves to the superconformal configuration as declared in footnote \ref{foot:deform}, however, and investigate the first subleading corrections to the planar free energies (\ref{F:ADHM:lead}) and (\ref{F:V52:lead}) as functions of $R$-charges in the next subsection.

\subsubsection{Planar free energy beyond the leading order} \label{sec:ana:1-node:sub}
According to the large $\lambda$ expansion of the planar free energy (\ref{planar1:S}), the first subleading corrections to the planar free energy can be obtained by substituting the leading order saddle point configuration (\ref{1-sol}) into the $\lambda^{-3/2}$ order planar effective action (\ref{free1:conti:largelambda:32}). This calculation is subtle in that one of the integrands in the expression (\ref{free1:conti:largelambda:32}) involves the 2nd derivative of a piecewise function $\rho_0(q)$ whose 1st derivative is already not continuous at $q\in\{q_l,0,q_r\}$ on the domain. This means that $\rho_0''(q)$ has to be evaluated carefully around those points where the Dirac-Delta function arises, and the result is given as 
\begin{equation}
	\rho_0''(q) = \begin{cases} 
       	0 \quad & q < q_l \\
       \frac{ \tDelta_3}{4 \pi^2 \tDelta_1 \tDelta_2 \left(2- \tDelta_1 - \tDelta_2 \right)}\delta(q-q_l)  \quad & q = q_l \\ 
        0 \quad &q_l <q<0\\
        \frac{ - \left(\tDelta_4 + \tDelta_3\right)}{4 \pi^2 \tDelta_1 \tDelta_2 \left(2- \tDelta_1 - \tDelta_2 \right)}\delta(q)   \quad & q=0 \\
        0 \quad &0<q<q_r \\
       \frac{ - \tDelta_4}{4 \pi^2 \tDelta_1 \tDelta_2 \left(2- \tDelta_1 - \tDelta_2 \right)}\delta(q-q_r)  \quad &q = q_r \\
        0 \quad &q_r < q 
         \end{cases} \, .\label{rho1:2nd}
\end{equation}
See Appendix \ref{app:detail:delta} for a detailed analysis regarding this matter. 

\medskip

Evaluating (\ref{free1:conti:largelambda:32}) on the leading saddle point configuration (\ref{1-sol}) with the help of (\ref{rho1:2nd}), we obtain 
\begin{equation}
\begin{split}
	S^\text{1-node}_{0,3/2}[\rho_0]&=-\frac{4\pi}{3} \sqrt{\fft{2(2-\tDelta_1-\tDelta_2)}{\tDelta_3+\tDelta_4}\tDelta_1 \tDelta_2 \tDelta_3 \tDelta_4}\\
	&\quad\times\fft{1-2(2-\tDelta_1-\tDelta_2)-(\tDelta_1+\tDelta_2)^2+(\tDelta_3+\tDelta_4)^2+\tDelta_1\tDelta_2}{24\tDelta_1\tDelta_2}\,.
\end{split}
\end{equation}
The $\lambda^{-3/2}$ subleading contribution to the planar free energy is then given from the formula (\ref{planar1:S}) for the two examples discussed in subsection \ref{sec:mm:1-node} as follows, where we have combined the subleading results with the leading behaviors (\ref{F:ADHM:lead}) and (\ref{F:V52:lead}) for a complete presentation.
\begin{itemize}
    \item Planar free energy of the ADHM theory:
    \begin{align}
         F_0^\text{ADHM}[\lambda] &=  \frac{4\pi}{3} \sqrt{2\tDelta_1 \tDelta_2 \tDelta_3 \tDelta_4} \bigg(\lambda^{-1/2} - \frac{1-2(\tDelta_1+\tDelta_2)+\tDelta_1\tDelta_2}{16\tDelta_1\tDelta_2}\lambda^{-3/2}\bigg) \nn\\[0.5em]
         &\quad +\lambda^{-1}\ri\pi(\tDelta-\Delta)\chi + \mO(\lambda^{-2}) \, ,\label{F:ADHM:sub}
    \end{align}
    with the constraint $\sum_{a=1}^4\tDelta_a =2$. 
    
    \item Planar free energy of the $V^{5,2}$ theory:
    \begin{align}
         F_0^{V^{5,2}}[\lambda] &= \frac{4\pi}{3} \sqrt{\tDelta_1 \tDelta_2 \tDelta_3 \tDelta_4}\bigg(\lambda^{-1/2} - \frac{1-(\tDelta_1+\tDelta_2)+\tDelta_1\tDelta_2}{8\tDelta_1\tDelta_2}\lambda^{-3/2}\bigg)  + \mO(\lambda^{-2}) \nn\\[0.5em]
         &\quad +\lambda^{-1}\ri\pi(\tDelta-\Delta)\chi + \mO(\lambda^{-2}) \, ,\label{F:V52:sub}
    \end{align}
    with the constraint  $\tDelta_1+\tDelta_2=\tDelta_3+\tDelta_4 =\fft43$.
\end{itemize}

\medskip

\noindent These results (\ref{F:ADHM:sub}) and (\ref{F:V52:sub}) improve the previous work of \cite{Martelli:2011qj,Cheon:2011vi,Jafferis:2011zi} restricted to the $\lambda^{-1/2}$ leading behaviors of the $S^3$ planar free energies by incorporating subleading corrections up to $\lambda^{-3/2}$ orders. We will improve them further to all orders in the large $\lambda$ expansion by employing an appropriate numerical analysis in section \ref{sec:num:1-node} and discuss related physics in sections \ref{sec:Airy} and \ref{sec:discussion}.

\subsection{2-node}\label{sec:ana:2-node}
Next we consider the 2-node examples in subsection \ref{sec:mm:2-node}. The 2-node partition function in \eqref{Z:2-node} can be written as
\begin{equation}
   Z_{S^3}^\text{2-node}(N,\lambda) = \frac{1}{(N!)^2} \int \frac{d\bmu}{(2 \pi)^N} \frac{d\bnu}{(2 \pi)^N} e^{-S^\text{2-node}\left[\bmu, \bnu;N,\lambda\right]}\label{Z2}
\end{equation}
in terms of the effective action defined by 
\begin{align}
        S^\text{2-node}\left[\bmu, \bnu;N,\lambda \right] & = -\frac{\ri k}{4\pi} \sum_{i=1}^N\left(\mu_i^2 - \nu^2_i\right) -2\sum_{i >j}^N \log \left[4 \sinh \left(\frac{\mu_i - \mu_j}{2}\right)\sinh \left(\frac{\nu_i - \nu_j}{2} \right)\right] \nn \\ &\quad-\sum_{a=1}^4\sum_{i,j=1}^N \ell\left(1-\Delta_a+\ri\sigma_a\frac{\mu_i - \nu_j}{2\pi}\right)\nn \\
        &\quad-\sum_{i=1}^N \left[\sum^{r_\mu}_{q=1}\ell\left(1- \Delta_{\mu_q} + \ri\frac{\mu_i}{2\pi} \right)+\sum^{\Tilde{r}_\mu}_{q=1}\ell\left(1- \tDelta_{\mu_q} - \ri\frac{\mu_i}{2\pi}  \right) \right. \nn \\ 
        & \kern3em~ \quad \left. + \sum^{r_\nu}_{q=1}\ell\left(1- \Delta_{\nu_q} + \ri\frac{\nu_i}{2\pi} \right) +\sum^{\Tilde{r}_\nu}_{q=1} \ell\left(1- \tDelta_{\nu_q} - \ri\frac{\nu_i}{2\pi} \right) \right] \, ,\label{free2}
\end{align}
where we have introduced $\sigma_a\equiv(1,1,-1,-1)$ for a compact expression. For notational convenience we omitted the dependence of the effective action on $R$-charges in the argument as in the 1-node case. We notice that the effective action (\ref{free2}) scales as $\sim N^2$ in the large $N$ limit with fixed
\begin{equation}
	\lambda\equiv\begin{cases}
		\displaystyle \fft{N}{k} & (\text{ABJM},\,N^{0,1,0}~\text{theories}) \\[1em]
		\displaystyle \fft{N}{N_f} & (Q^{1,1,1}~\text{theory})
	\end{cases}\,,~~ \lambda_\mu\equiv\fft{N}{r_\mu}\,,~~\lambda_\nu\equiv\fft{N}{r_\nu}\,~~\tlambda_\mu\equiv\fft{N}{\tilde{r}_\mu}\,,~~\tlambda_\nu\equiv\fft{N}{\tilde{r}_\nu}\,,\label{2-node:fixed}
\end{equation}
which defines the 't~Hooft limit in 2-node examples. Hence the matrix integral for the partition function (\ref{Z2}) can be analyzed using the saddle point approximation in the 't Hooft limit as in the 1-node case. In particular, the planar free energy can be obtained by finding a saddle point $(\bmu^\star,\bnu^\star)$ that extremizes the effective action (\ref{free2}) and then reading off the $N^2$ leading behavior of the on-shell effective action as  \cite{Herzog:2010hf,Martelli:2011qj,Cheon:2011vi,Jafferis:2011zi,Marino:2011nm}
\begin{equation}
	N^2F^\text{2-node}_0(\lambda)= S^\text{2-node}[\bmu^\star,\bnu^\star;N,\lambda]\Big|_{N^2\text{-order}}\label{planar2}
\end{equation}
based on the genus expansion (\ref{F:genus}) and the saddle point approximation. In the following subsections we derive the planar free energy $F^\text{2-node}_0(\lambda)$ up to first subleading corrections in the large $\lambda$ expansion based on the formula (\ref{planar2}). In doing so, we impose the constraints
\begin{equation}
	\sum_{a=1}^4\Delta_a=2\,,\quad\tilde{\lambda}_\mu=\lambda_\nu\,,\quad \lambda_\mu=\tilde{\lambda}_\nu\,,\quad \Delta=\Delta_{\mu_q}=\Delta_{\nu_q}=\tDelta_{\mu_q}=\tDelta_{\nu_q}\,, \label{2-node:constraints}
\end{equation}
shared by the ABJM/$N^{0,1,0}$/$Q^{1,1,1}$ theories introduced in subsection \ref{sec:mm:2-node} (recall that for the latter two examples we focus on the superconformal configuration, see Appendix \ref{app:detail:N010} for comments on the less constrained $N^{0,1,0}$ theory), and then specialize to each example only in the last step of subsections \ref{sec:ana:2-node:lead} and \ref{sec:ana:2-node:sub} where we present their planar free energies.

\subsubsection{Effective action in the double scaling limit}\label{sec:ana:2-node:S}
As in the 1-node case, it is highly complicated to evaluate the on-shell effective action in (\ref{planar2}) exactly and therefore we take advantage of the double scaling limit. First we introduce continuous eigenvalue distributions $\mu(q)$ and $\nu(q)$ over an interval $q\in[q_a,q_d]\subseteq\mathbb{R}$ in the 't~Hooft limit, while introducing the eigenvalue density normalized as
\begin{equation}
	\int_{q_a}^{q_d} dq\,\rho(q)=1\,.\label{rho2}
\end{equation}
Then in the large $\lambda$ limit we make a saddle point ansatz \cite{Herzog:2010hf}
\begin{equation}
	\mu(q) = \lambda^{1/2} q + \ri p(q) \, , \qquad \nu(q) = \lambda^{1/2} q - \ri p(q) \,. \label{ansatz2}
\end{equation}
Substituting the ansatz (\ref{ansatz2}) into the effective action (\ref{free2}) and focusing on the planar term
\begin{equation}
	S^\text{2-node}\left[\bmu, \bnu;N,\lambda \right]=N^2S_0^\text{2-node}\left[\rho,p;\lambda \right]+o(N^2)\,,
\end{equation}
we obtain
\begin{subequations}
	\begin{align}
		S^\text{2-node}_0[\rho,p;\lambda]&=S_0^\text{2-bif}[\rho,p;\lambda]+S_0^\text{2-fun}[\rho,p;\lambda]\,,\\
		S_0^\text{2-bif}[\rho,p;\lambda]&=\int_{q_a}^{q_d} dq\, \rho(q) \Bigg[ \int_{q_a}^{q} dq'\,\rho(q')  \sum^\infty_{n=1} \frac{2}{n}e^{-n\lambda^{1/2}(q-q')}\bigg\{2 \cos n \left(p(q) -p(q') \right)  \nonumber\\
		&\quad-\sum_{a=1}^4\left(1- \Delta_{a} -\sigma_a\frac{p(q) + p(q')}{2 \pi}\right) \cos  n \left(2 \pi\Delta_a +\sigma_a(p(q) + p(q'))\right)  \label{free2:conti:split:bif}\\
		&\quad- \frac{\frac{1}{n} + \lambda^{1/2}(q-q')}{2 \pi} \sum_{a=1}^4\sin  n\left(2 \pi\Delta_{a} +\sigma_a(p(q) + p(q'))\right)\bigg\} + \frac{\lambda^{-1/2}}{\pi}  q p(q)  \Bigg]\,,\nn\\
		S_0^\text{2-fun}[\rho,p;\lambda]&=- \int_{q_a}^{q_d} dq\,\rho(q)\bigg[\lambda^{-1}_\mu \ell\bigg(1- \Delta - \frac{p(q)}{2\pi}+ \ri\lambda^{1/2}\frac{ q}{2\pi} \bigg)  \nonumber \\ 
		& \quad +\lambda^{-1}_\nu \ell\bigg(1- \Delta + \frac{p(q)}{2 \pi} - \ri \lambda^{1/2}\frac{q}{2\pi}  \bigg) + \lambda^{-1}_\nu \ell\bigg(1- \Delta + \frac{p(q)}{2 \pi} + \ri \lambda^{1/2}\frac{q}{2\pi} \bigg) \nn\\
		&\quad+ \lambda^{-1}_\mu  \ell\bigg(1- \Delta - \frac{p(q)}{2 \pi} -  \ri \lambda^{1/2}\frac{q}{2\pi} \bigg) \bigg]\,, \label{free2:conti:split:fun}
	\end{align}\label{free2:conti:split}%
\end{subequations}
where we have used the Euler-Maclaurin formula as in the 1-node case. Note that we included the contribution from the classical CS action and $\mN=2$ vector multiplets into $S_0^\text{2-bif}$ for computational efficiency and notational convenience. 

\medskip

Expanding the contributions from bi-fundamental $\mN=2$ chiral multiplets (\ref{free2:conti:split:bif}) and fundamental \& anti-fundamental $\mN=2$ chiral multiplets (\ref{free2:conti:split:fun}) explicitly in the large $\lambda$ limit with $\lambda_\mu,\lambda_\nu$ scaling together with $\lambda$, we obtain
\begin{subequations}
	\begin{align}
		S_0^\text{2-bif}[\rho,p;\lambda] &=  \lambda^{-1/2}\int_{q_a}^{q_d}  dq\, \rho(q) \bigg[-4 \rho(q) p(q)^2 - 4\pi (\Delta_{1} \Delta_{2} - \Delta_{3} \Delta_{4}) \rho(q) p(q)  \nn\\ 
		&\quad  +2 \pi^2 \big(\Delta_{1} \Delta_{2}(\Delta_{3} +\Delta_{4}) +\Delta_{3} \Delta_{4}(\Delta_{1} +\Delta_{2})\big) \rho(q)+ \frac{1}{\pi}  p(q) q  \bigg] \nn\\
		&\quad+\lambda^{-3/2}S^\text{2-bif}_{0,3/2}[\rho,p]+\mO(\lambda^{-2})  \,, \label{Free2bif}\\[1em]
		S_0^\text{2-fun}[\rho,p;\lambda] &= \lambda^{1/2} \int_{q_a}^{q_d} dq\,\rho(q) |q| \left[ \lambda^{-1}_\mu\left(1-\Delta - \frac{p(q)}{2 \pi} \right) +  \lambda^{-1}_\nu\left(1-\Delta + \frac{p(q)}{2 \pi} \right) \right] \nn\\ 
		&\quad - 4 \pi^2\lambda^{-1/2} \rho(0)  \Bigg[ \lambda^{-1}_\mu\left(1- \Delta - \frac{p(0)}{2 \pi}\right)  \text{B}_2\left( \Delta + \frac{p(0)}{2 \pi} \right) \nn\\
		&\quad +  \frac{2}{3}\lambda^{-1}_\mu  \text{B}_3\left( \Delta + \frac{p(0)}{2 \pi} \right) 
		+\lambda^{-1}_\nu\left(1- \Delta + \frac{p(0)}{2 \pi}\right) \text{B}_2\left( \Delta - \frac{p(0)}{2 \pi} \right) \nn\\
		&\quad + \frac23\lambda^{-1}_\nu \text{B}_3 \left(\Delta - \frac{p(0)}{2 \pi} \right)\Bigg] + \mO(\lambda^{-2})\,. \label{Free2fun}
	\end{align}\label{Free2:bif+fun}%
\end{subequations}
See Appendix \ref{app:detail:isol2} for detailed derivations. The $\lambda^{-3/2}$ order contribution $S^\text{2-bif}_{0,3/2}[\rho,p]$ from bi-fundamental representations is quite cumbersome so we refer the readers for its concrete expression to (\ref{free2bis}). 

\medskip

Substituting (\ref{Free2:bif+fun}) back into (\ref{free2:conti:split}) we obtain the large $\lambda$ expansion of the planar effective action as 
\begin{subequations}
	\begin{align}
		S^\text{2-node}_0[\rho,p;\lambda]&=\lambda^{-1/2}S^\text{2-node}_{0,1/2}[\rho,p]+\lambda^{-3/2}S^\text{2-node}_{0,3/2}[\rho,p]+\mO(\lambda^{-2})\,,\\
		S^\text{2-node}_{0,1/2}[\rho,p]&= \int_{q_a}^{q_d}  dq\, \rho(q)  \bigg[-4 \rho(q) p(q)^2 - 4\pi (\Delta_{1} \Delta_{2} - \Delta_{3} \Delta_{4}) \rho(q) p(q)  \nn\\ 
		&\quad  +2 \pi^2 \big(\Delta_{1} \Delta_{2}(\Delta_{3} +\Delta_{4}) +\Delta_{3} \Delta_{4}(\Delta_{1} +\Delta_{2})\big) \rho(q)+ \frac{1}{\pi}  p(q) q  \bigg] \nn \\
		&\quad+\int_{q_a}^{q_d} dq\,\rho(q) |q| \left[  \fft{\lambda}{\lambda_\mu}\left(1-\Delta - \frac{p(q)}{2 \pi} \right) + \fft{\lambda}{\lambda_\nu} \left(1-\Delta + \frac{p(q)}{2 \pi} \right) \right]\,, \label{free2:conti:largelambda:12}\\
		S^\text{2-node}_{0,3/2}[\rho,p]&=- 4 \pi^2\rho(0)  \Bigg[ \fft{\lambda}{\lambda_\mu}\left(1- \Delta - \frac{p(0)}{2 \pi}\right)  \text{B}_2\left( \Delta + \frac{p(0)}{2 \pi} \right) +  \frac{2}{3}\fft{\lambda}{\lambda_\mu}\text{B}_3\left( \Delta + \frac{p(0)}{2 \pi} \right)  \nn\\
		&\quad 
		+\fft{\lambda}{\lambda_\nu}\left(1- \Delta + \frac{p(0)}{2 \pi}\right) \text{B}_2\left( \Delta - \frac{p(0)}{2 \pi} \right) + \frac{3}{2}\fft{\lambda}{\lambda_\nu}\text{B}_3 \left(\Delta - \frac{p(0)}{2 \pi} \right)\Bigg] \nn\\
		&\quad+S^\text{2-bif}_{0,3/2}[\rho,p]\,.\label{free2:conti:largelambda:32}
	\end{align}\label{free2:conti:largelambda}%
\end{subequations}
%

\subsubsection{Planar free energy from the effective action}\label{sec:ana:2-node:F}
The planar free energy of our interest can now be written in terms of the on-shell value of the planar effective action (\ref{free2:conti:largelambda}) based on the formula (\ref{planar2}) as
\begin{equation}
	F^\text{2-node}_0(\lambda)= S_0^\text{2-node}[\rho^\star,p^\star;\lambda]\,,\label{planar2:genus0}
\end{equation}
where $\rho^\star(q)$ and $p^\star(q)$ denote the eigenvalue density and the imaginary part of the eigenvalue distribution extremizing the planar effective action (\ref{free2:conti:largelambda}) under the normalization constraint (\ref{rho2}) respectively. Hence the first step to evaluate the planar free energy is to analyze the saddle point configuration $(\rho^\star(q),\rho^\star(q))$.

\medskip

Motivated by the large $\lambda$ expansion of the planar effective action (\ref{free2:conti:largelambda}), we expand the saddle point configuration $(\rho^\star(q),p^\star(q))$ in the large $\lambda$ limit as
\begin{equation}
\begin{split}
	\rho^\star(q)&=\rho_0(q)+\lambda^{-1}\rho_1(q)+\mO(\lambda^{-3/2})\,,\\
	p^\star(q)&=p_0(q)+\lambda^{-1}p_1(q)+\mO(\lambda^{-3/2})\,.
\end{split}\label{rho2p2:exp}
\end{equation}
The expansion coefficients in (\ref{rho2p2:exp}) can be determined by solving the saddle point equation
\begin{equation}
\begin{split}
	\fft{\delta}{\delta \rho(q)}\bigg[S^\text{2-node}_0[\rho,p;\lambda]-\lambda^{-1/2}L\bigg(\int_{q_a}^{q_d}dq\,\rho(q)-1\bigg)\bigg]_{(\rho,p)=(\rho^\star,p^\star)} &=0\,,\\
	\fft{\delta}{\delta p(q)}\bigg[S^\text{2-node}_0[\rho,p;\lambda]-\lambda^{-1/2}L\bigg(\int_{q_a}^{q_d}dq\,\rho(q)-1\bigg)\bigg]_{(\rho,p)=(\rho^\star,p^\star)} &=0\,,
\end{split}\label{saddle2}
\end{equation}
order by order, where we have introduced $L$ as a Lagrange multiplier for the normalization condition in (\ref{rho2}). 

\medskip

As we have already discussed in subsection \ref{sec:ana:1-node:F} for the 1-node case, however, it suffices to solve the saddle point equation (\ref{saddle2}) at the $\lambda^{-1/2}$ leading order, namely
\begin{equation}
\begin{split}
	0&=\fft{\delta S_{0,1/2}^\text{2-node}[\rho,p;\lambda]}{\delta\rho(q)}\bigg|_{(\rho,p)=(\rho_0,p_0)}-L\,,\\
	0&=\fft{\delta S_{0,1/2}^\text{2-node}[\rho,p;\lambda]}{\delta p(q)}\bigg|_{(\rho,p)=(\rho_0,p_0)}\,,
\end{split}\label{saddle2:exp:12}
\end{equation}
to determine the planar free energy (\ref{planar2}) up to $\lambda^{-3/2}$ order. To be specific, the planar free energy (\ref{planar2:genus0}) can be obtained by evaluating the planar effective action (\ref{free2:conti:largelambda}) on the leading order saddle point configuration $\rho_0(q)$ and $p_0(q)$ solving (\ref{saddle2:exp:12}) as
\begin{equation}
	F_0^\text{2-node}(\lambda)=\lambda^{-1/2}S^\text{2-node}_{0,1/2}[\rho_0,p_0]+\lambda^{-3/2}S^\text{2-node}_{0,3/2}[\rho_0,p_0]+\mO(\lambda^{-2})\,.\label{planar2:S}
\end{equation}
The effective action at order $\lambda^{-1}$ again vanishes on-shell as in the 1-node case, see Appendix \ref{app:detail:isol2:bif} for details. 

\medskip

To evaluate the large $\lambda$ expansion of the planar free energy (\ref{planar2:S}) explicitly, we will now determine the leading order saddle point configuration $\rho_0(q)$ \& $p_0(q)$ and then evaluate the planar free energy up to $\lambda^{-3/2}$ order in the following subsections.

\subsubsection{Planar free energy at leading order}\label{sec:ana:2-node:lead}
Now we present the planar free energies for three different examples of our interest, namely ABJM, $N^{0,1,0}$, $Q^{1,1,1}$ theories. In this subsection we first review the leading order saddle point configuration and the planar free energy in the large $\lambda$ limit \cite{Herzog:2010hf,Martelli:2011qj,Cheon:2011vi,Jafferis:2011zi,Marino:2011eh}, and then investigate the first subleading order corrections in the next subsection.  

\medskip

For the ABJM \& $N^{0,1,0}$ theories, the leading order saddle point configuration $\rho_0(q)$ \& $p_0(q)$ solving (\ref{saddle2:exp:12}) read \cite{Jafferis:2011zi} 
\begin{subequations} \label{2-sol:ABJMN010}
	\begin{align}
		\rho_0(q) &= \begin{cases}
			\displaystyle \frac{2 \pi L +2 \pi \Delta_2 q + \pi \alpha q}{ 8 \pi^3 (\Delta_2 +\Delta_4) (\Delta_2 + \Delta_3) (\Delta_1 - \Delta_2) } & (q_a<q<q_b) \\[1em]
			\displaystyle \frac{\pi q (\Delta_{1} \Delta_{2} - \Delta_{3} \Delta_{4}) + 2 \pi L - \pi \alpha |q| }{4 \pi^3 (\Delta_1 + \Delta_3)(\Delta_1 + \Delta_4)(\Delta_2 + \Delta_3)(\Delta_2 + \Delta_4)} & (q_b<q<q_c) \\[1em]
			\displaystyle \frac{2 \pi L -2 \pi \Delta_4 q - \pi \alpha q}{ 8 \pi^3 (\Delta_1 +\Delta_4) (\Delta_2 + \Delta_4) (\Delta_3 - \Delta_4) } & (q_c<q<q_d) \\
		\end{cases}\,,\\
		p_0(q) &= \begin{cases}
			\displaystyle -\pi\Delta_2 & (q_a<q<q_b) \\
			\displaystyle \fft{q}{8\pi \rho_0(q)}-\fft{\pi}{2}(\Delta_1\Delta_2-\Delta_3\Delta_4) & (q_b<q<q_c) \\
			\displaystyle \pi\Delta_4 & (q_c<q<q_d) \\
		\end{cases}\,,
	\end{align}
	in terms of $\alpha\equiv r/k$ and the Lagrange multiplier
	\begin{align}
		L = 2 \pi \sqrt{\frac{2 (\Delta_1 +\Delta_3)(\Delta_2 + \Delta_4)\left(\Delta_1 + \frac{\alpha}{2} \right)\left(\Delta_2 + \frac{\alpha}{2} \right)\left(\Delta_3 + \frac{\alpha}{2} \right)\left(\Delta_4 + \frac{\alpha}{2} \right)}{(\Delta_1 +\Delta_3)(\Delta_2 + \Delta_4) + \frac{\alpha}{2}\left((2+\alpha)^2 + 2 \left( \Delta_3 - \Delta_3^2 + \Delta_4 - \Delta_4^2\right) \right)}} \,,
	\end{align}
	where the interval $[q_a,q_d]$ is split into three parts by the breaking/endpoints
	\begin{align}
		q_a= \frac{-2 L}{2 \Delta_2 + \alpha}\,,\qquad q_b=\frac{-2 L}{2 \Delta_1 + \alpha}\,,\qquad q_c=\frac{2 L}{2 \Delta_3 + \alpha}\,,\qquad q_d=\frac{2 L}{2 \Delta_4 + \alpha}\,.
	\end{align}
\end{subequations}
In the above expressions we have assumed $\Delta_1\geq\Delta_2$ and $\Delta_3\geq\Delta_4$ without loss of generality as in Appendix \ref{app:detail:isol2:bif}; note that the effective action (\ref{free2}) is invariant under interchanging $\Delta_1 \leftrightarrow \Delta_2$ and $\Delta_3 \leftrightarrow \Delta_4$.

\medskip

Substituting the saddle point solution (\ref{2-sol:ABJMN010}) back into the leading order planar effective action (\ref{free2:conti:largelambda:12}) now yields\footnote{For the ABJM/$N^{0,1,0}$ theories described in subsection \ref{sec:mm:2-node}, it suffices to derive the expression (\ref{S:212}) under the extra constraints $\Delta=\frac{1}{2}$ and $\lambda_\mu=\lambda_\nu$ (or $r_1=r_2$) on top of the starting constraints \eqref{2-node:constraints}. As shown in \ref{app:detail:N010}, however, the same expression is valid for the $N^{0,1,0}$ theory with general $r_{1,2}$ values and weaker constraints on $R$-charges \eqref{N010:constraints}.} 
\begin{equation}
	S^\text{2-node}_{0,1/2}[\rho_0,p_0]=\frac{2L}{3}\,.\label{S:212}
\end{equation}
Remarkably, this is the same expression as for the 1-node case \eqref{S:112}, albeit with a different Lagrange multiplier. The $\lambda^{-1/2}$ leading contribution to the planar free energy is then obtained through the formula (\ref{planar2:S}). The results are given as follows \cite{Jafferis:2011zi,Marino:2011eh}.
\begin{itemize}
	\item Planar free energy of the ABJM theory:
	\begin{equation}
		F_0^\text{ABJM}[\lambda] = \frac{4\pi}{3} \sqrt{2\Delta_1 \Delta_2 \Delta_3 \Delta_4} \lambda^{-1/2} + \mO(\lambda^{-3/2}) \, , \label{F:ABJM:lead}
	\end{equation}
	with the constraint $\sum_{a=1}^4\Delta_a =2$. 
	
	\item Planar free energy of the $N^{0,1,0}$ theory at the superconformal configuration: 
	\begin{equation}
		F_0^{N^{0,1,0}}[\lambda] = \frac{2\pi(1+\alpha)}{3\sqrt{2+\alpha}} \lambda^{-1/2} + \mO(\lambda^{-3/2}) \,. \label{F:N010:lead}
	\end{equation}
\end{itemize}

\noindent Note that we have imposed the constraints presented in subsection \ref{sec:mm:2-node} specialized to each theory in (\ref{F:ABJM:lead}) and (\ref{F:N010:lead}) respectively, on top of the shared constraints (\ref{2-node:constraints}). The same $\lambda^{-1/2}$ leading order result for the $N^{0,1,0}$ theory can be obtained for more general cases, however, see Appendix \ref{app:detail:N010} for details. 

\medskip

For the $Q^{1,1,1}$ theory, the leading order saddle point configuration $\rho_0(q)$ \& $p_0(q)$ solving (\ref{saddle2:exp:12}) read \cite{Cheon:2011vi,Jafferis:2011zi} 
\begin{subequations} \label{2-sol:Q111}
	\begin{align}
		\rho_0(q) &= \frac{1}{2 q_0} + \frac{1}{4 \pi^2} \left(\frac{1}{2}q_0 - |q| \right)\,, \\
		p_0(q) &= \frac{|q|}{8 \pi \rho_0(q)}  \,,
	\end{align}
	where the endpoints of the interval are given by
	\begin{align}
		q_d=-q_a=q_0=\fft{2\pi}{\sqrt{3}}\,.
	\end{align}
\end{subequations}

\medskip

Substituting the saddle point solution (\ref{2-sol:Q111}) back into the leading order planar effective action (\ref{free2:conti:largelambda:12}) now yields the $\lambda^{-1/2}$ leading contribution to the planar free energy through the formula (\ref{planar2:S}). The result is given as follows \cite{Jafferis:2011zi}.
\begin{itemize}
	\item Planar free energy of the superconformal $Q^{1,1,1}$ theory:
	\begin{equation}
		F_0^{Q^{1,1,1}}[\lambda] = \frac{4 \pi}{3 \sqrt{3}} \lambda^{-1/2} + \mO(\lambda^{-3/2}) \, . \label{F:Q111:lead}
	\end{equation}
\end{itemize}
Note that we have imposed the constraint (\ref{Q111:constraints:sc}) specialized to the $Q^{1,1,1}$ theory at the superconformal configuration to obtain (\ref{F:Q111:lead}), on top of the shared constraints (\ref{2-node:constraints}).

\subsubsection{Planar free energy beyond the leading order} \label{sec:ana:2-node:sub}
As in the 1-node case, the large $\lambda$ expansion of the planar free energy (\ref{planar2:S}) tells us that the $\lambda^{-3/2}$ order contribution to the planar free energy can be obtained by substituting the leading order saddle point configuration, (\ref{2-sol:ABJMN010}) or (\ref{2-sol:Q111}), into the planar effective action at the $\lambda^{-3/2}$ order, (\ref{free2:conti:largelambda:32}). In due process we again encounter the Dirac-Delta functions arising in the 2nd derivatives $\rho_0''(q)$ and $p_0''(q)$ due to their piecewise nature, which can be determined following the analysis in Appendix \ref{app:detail:delta}. 

\medskip

For the ABJM theory, we find
\begin{subequations}
\begin{align}
	\rho_0''(q) &= \begin{cases}
		0 & (q_a< q < q_b)\\[0.3em]
		\frac{-\Delta_1}{4 \pi^2 (\Delta_1+\Delta_3) (\Delta_1 - \Delta_2)}\delta(q-q_b) & (q = q_b) \\[0.3em]
		0 & (q_b< q < q_c)\\[0.3em]
		\frac{-\Delta_3}{4 \pi^2 (\Delta_1+\Delta_3) (\Delta_3 - \Delta_4)}\delta(q-q_c) & (q = q_c) \\[0.3em]
		0 & (q_c< q < q_d)
	\end{cases}\,,\\
	p_0''(q) &= \begin{cases}
		0 & (q_a< q < q_b) \\[0.3em]
		\frac{\pi (\Delta_2 + \Delta_4)\Delta_1^2}{(\Delta_1 + \Delta_3) L}\delta(q-q_b) & (q = q_b) \\[0.3em]
		\fft{1}{8\pi}\big(\fft{q}{\rho_0(q)}\big)'' & (q_b< q < q_c)\\[0.3em]
		-\frac{\pi (\Delta_2 + \Delta_4)\Delta_3^2}{(\Delta_1 + \Delta_3) L}\delta(q-q_c) & (q = q_c) \\[0.3em]
		0 & (q_c< q < q_d)
	\end{cases}\,.
\end{align}\label{rho2:2nd:ABJM}%
\end{subequations}
Evaluating the $\lambda^{-3/2}$ order planar effective action (\ref{free2:conti:largelambda:32}) on the leading saddle point configuration (\ref{2-sol:ABJMN010}) with the help of (\ref{rho2:2nd:ABJM}), and then substituting the result in conjunction with the $\lambda^{-1/2}$ leading behavior (\ref{F:ABJM:lead}) into (\ref{planar2:S}) yields the planar free energy. The result is given as follows.
\begin{itemize}
	\item Planar free energy of the ABJM theory:
	\begin{equation}
		F_0^\text{ABJM}[\lambda] = \frac{4\pi}{3} \sqrt{2\Delta_1 \Delta_2 \Delta_3 \Delta_4} \left( \lambda^{-1/2} - \frac{1}{16} \lambda^{-3/2}\right) + \mO( \lambda^{-2}) \, .\label{F:ABJM:sub}
	\end{equation}
\end{itemize}
It is remarkable that the planar effective action at the $\lambda^{-3/2}$ order (\ref{free2:conti:largelambda:32}), which involves the lengthy expression (\ref{free2bis}) with non-trivial dependence on $R$-charges, turns out to yield such a simple contribution in (\ref{F:ABJM:sub}) at the leading saddle point configuration (\ref{2-sol:ABJMN010}) for the ABJM theory. We will comment on this point further in subsection \ref{sec:num:2-node} and section \ref{sec:Airy}.

\medskip

For the $N^{0,1,0}$ theory at the superconformal configuration (\ref{N010:constraints:sc}), where the leading saddle point configuration (\ref{2-sol:ABJMN010}) is simplified significantly as
\begin{equation}
\begin{split}
	\rho_0(q) &= \frac{\sqrt{2+\alpha}}{4 \pi} + \frac{\alpha}{4 \pi^2} \left(\frac{\pi}{\sqrt{2+\alpha}} - |q| \right)\,,\\
	p_0(q) &= \frac{q}{8 \pi \rho(q)} \,, \\
	q_c=q_d=-q_a=-q_b&=\fft{\pi}{\sqrt{2+\alpha}}\,,
\end{split}
\end{equation}
we find
\begin{equation}
\begin{split}
	\rho_0''(q) &= \begin{cases}
		0 \quad & (q_a<q<0) \\
		-\frac{\alpha}{2\pi^2}\delta(q) \quad & (q=0) \\
		0 \quad & (0<q<q_d)
	\end{cases} \,, \\
	p_0''(q) &= \fft{1}{8\pi}\bigg(\fft{q}{\rho_0(q)}\bigg)''\,. 
\end{split}\label{rho2:2nd:N010}
\end{equation}
Note that the imaginary part of the eigenvalue distribution, $p_0(q)$, is written in terms of the eigenvalue density $\rho_0(q)$ involving the absolute value of $q$ so in principle its 2nd derivative could have a Dirac-Delta function at the breaking point $q=0$. But it turns out that the 1st derivative $p_0'(q)$ has identical left/right limits at $q=0$ and therefore the Dirac-Delta function does not arise for $p_0''(q)$ at $q=0$. The planar free energy of the $N^{0,1,0}$ theory can then be obtained by following the same procedure described for the ABJM theory. The result is given as follows.
\begin{itemize}
	\item Planar free energy of the superconformal $N^{0,1,0}$ theory:
	\begin{equation}
		F_0^{N^{0,1,0}}[\lambda]  = \frac{2 (1+\alpha) \pi}{3 \sqrt{2+\alpha}} \left( \lambda^{-1/2} + \frac{3 \alpha - 2}{32} \lambda^{-3/2}\right) + \mO( \lambda^{-2}) \, .\label{F:N010:sub}
	\end{equation}
\end{itemize}

\medskip

For the $Q^{1,1,1}$ theory at the superconformal configuration (\ref{Q111:constraints:sc}), we find
\begin{equation}
	\begin{split}
		\rho_0''(q) &= \begin{cases}
			0 \quad & (-q_0<q<0) \\
			-\frac{1}{2\pi^2}\delta(q) \quad & (q=0) \\
			0 \quad & (0<q<q_0)
		\end{cases} \,, \\
		p_0''(q) &= \begin{cases}
			-\fft{1}{8\pi}\big(\fft{q}{\rho_0(q)}\big)'' \quad & (-q_0<q<0) \\[0.3em]
			\fft{\sqrt{3}}{4}\delta(q) \quad & (q=0) \\[0.3em]
			\fft{1}{8\pi}\big(\fft{q}{\rho_0(q)}\big)'' \quad & (0<q<q_0)
		\end{cases} \,. 
	\end{split}\label{rho2:2nd:Q111}
\end{equation}
The planar free energy of the $Q^{1,1,1}$ theory can then be obtained by following the same procedure described for the ABJM theory. The result is given as follows.
\begin{itemize}
	\item Planar free energy of the superconformal $Q^{1,1,1}$ theory:
	\begin{equation}
		F_0^{Q^{1,1,1}}[\lambda] = \frac{4 \pi}{3 \sqrt{3}} \left(\lambda^{-1/2} +\frac{1}{8}\lambda^{-3/2} \right) + \mO( \lambda^{-2}) \, .\label{F:Q111:sub}
	\end{equation}
\end{itemize}

\medskip

\noindent These results (\ref{F:ABJM:sub}), (\ref{F:N010:sub}), and (\ref{F:Q111:sub}) improve the previous work of \cite{Herzog:2010hf,Martelli:2011qj,Cheon:2011vi,Jafferis:2011zi} restricted to the $\lambda^{-1/2}$ leading behaviors of the $S^3$ planar free energies by incorporating subleading corrections up to $\lambda^{-3/2}$ orders. They will be improved to all orders in the large $\lambda$ expansion via numerical analysis in section \ref{sec:num:2-node} and its implication on physics will be explored in sections \ref{sec:Airy} and \ref{sec:discussion}.

\section{Subleading structures of \texorpdfstring{$S^3$}{S3} partition functions; numerical approach}\label{sec:num}
In this section we improve on the planar $S^3$ free energy of $\mN=2$ holographic SCFTs computed in the previous section \ref{sec:ana} based on the saddle point approximation to the corresponding matrix models, to all orders in the large $\lambda$ expansion by employing a numerical technique delineated for 1-node and 2-node examples in subsections \ref{sec:num:1-node} and \ref{sec:num:2-node} respectively. As a result, we will arrive at the universal form of the planar free energies presented in (\ref{F0:universal:intro}). For clarity, in this section we have restored $\bDelta$ explicitly in the argument to represent the $R$-charge dependence.

\subsection{1-node}\label{sec:num:1-node}
For the 1-node theories, the numerical analysis proceeds as follows.
\begin{enumerate}
	\item Derive the saddle point equation 
	\begin{equation}
		\begin{split}
			0&=\Delta_m-\sum_{j=1\,(\neq i)}^N\coth\fft{\mu_i-\mu_j}{2}\\
			&\quad+\fft12\sum_{I=1}^3\sum_{j=1}^N\Bigg[\bigg(1-\Delta_I+\ri\fft{\mu_i-\mu_j}{2\pi}\bigg)\coth(\fft{\mu_i-\mu_j}{2}-\ri\pi(1-\Delta_I))\\
			&\kern7em+\bigg(1-\Delta_I-\ri\fft{\mu_i-\mu_j}{2\pi}\bigg)\coth(\fft{\mu_i-\mu_j}{2}+\ri\pi(1-\Delta_I))\Bigg]\\
			&\quad+\fft{N_f}{2}\Bigg[\bigg(1-\Delta+\ri\fft{\mu_i}{2\pi}\bigg)\coth(\fft{\mu_i}{2}-\ri\pi(1-\Delta))\\
			&\kern5em+\bigg(1-\tDelta-\ri\fft{\mu_i}{2\pi}\bigg)\coth(\fft{\mu_i}{2}+\ri\pi(1-\tDelta))\Bigg]
		\end{split}\label{1-node:saddle}%
	\end{equation}
	by taking the 1st derivative of the effective action (\ref{free1:def}) with respect to the gauge holonomies. Find a numerical solution $\{\bmu^\star\}$ to the saddle point equation (\ref{1-node:saddle}) for given $(N,\lambda,\bDelta)$ configurations using the \texttt{FindRoot} in \textit{Mathematica} at \texttt{WorkingPrecision} 200, employing the leading order solution (\ref{1-sol}) as initial conditions.
	
	\item Evaluate the on-shell effective action
	\begin{equation}
		S^\text{1-node} [\bmu^\star;N,\lambda,\bDelta]\label{1-node:S-onshell}
	\end{equation}
	by substituting the numerical solution $\{\bmu^\star\}$ into the effective action (\ref{free1:def}). In this step we find that the imaginary part of the on-shell effective action is given by
	\begin{equation}
		\Im S^\text{1-node} [\bmu^\star;N,\lambda,\bDelta]=\lambda^{-1}\pi(\tDelta-\Delta)\chi N^2\label{1-node:S-onshell:Im}
	\end{equation}
	for all numerical solutions.
	
	\item \texttt{LinearModelFit} the real part of the on-shell effective action (\ref{1-node:S-onshell}) with respect to large values of $N$ ($100\sim350$ in step of $10$) for given $(\lambda,\bDelta)$ configurations as
	\begin{equation}
	\begin{split}
		\Re S^\text{1-node} [\bmu^\star;N,\lambda,\bDelta]&=S_0^\text{1-node,(lmf)} (\lambda,\bDelta)N^2+S_\text{1/2-log}^\text{1-node,(lmf)} (\lambda,\bDelta)N\log N\\
		&\quad+S_{1/2}^\text{1-node,(lmf)} (\lambda,\bDelta)N+S_\text{1-log}^\text{1-node,(lmf)} (\lambda,\bDelta)\log N\\
		&\quad+\sum_{\ell=0}^{18}S_{\ell/2+1}^\text{1-node,(lmf)} (\lambda,\bDelta)N^{-\ell}\,.
	\end{split}\label{1-node:lmf}
	\end{equation}
	Note that we have included extra terms absent in a typical genus expansion (\ref{F:genus}) in the \texttt{LinearModelFit} (\ref{1-node:lmf}), which are supposed to be cancelled by loop corrections \cite{Liu:2019tuk,Hong:2021bsb}. Here we do not need to explore such cancellation since we are interested in the planar contribution fully captured by the classical on-shell action without loop corrections. According to the formula (\ref{planar1}), reading off the coefficient of the $N^2$ leading term in the \texttt{LinearModelFit} (\ref{1-node:lmf}) yields the numerical value of the planar free energy as
	\begin{equation}
		F_0^\text{1-node}(\lambda,\bDelta)=S_0^\text{1-node,(lmf)} (\lambda,\bDelta)+\ri\lambda^{-1}\pi(\tDelta-\Delta)\chi\,,\label{1-node:F0}
	\end{equation}
	where we have also included the imaginary contribution from (\ref{1-node:S-onshell:Im}).
	
	\item Repeating the above 1-3 steps for various $(\lambda,\bDelta)$ configurations, deduce the analytic expression of the planar free energy $F_0^\text{1-node}(\lambda,\bDelta)$ as a function of the 't~Hooft coupling $\lambda$ and the $R$-charges $\bDelta$.  
\end{enumerate}
The final results and details involved in the above-described numerical analysis will be presented below for the ADHM \& $V^{5,2}$ theories in order. For a clear presentation, we provide the final results first and then explain how the numerical data support them through the aforementioned procedure.

\subsubsection{ADHM theory}\label{sec:num:1-node:ADHM}
Applying the numerical analysis described above to the ADHM theory, we obtain the ADHM planar free energy
\begin{empheq}[box=\fbox]{equation}
\begin{split}
	F_0^\text{ADHM}(\lambda,\bDelta)&=\fft{4\pi\sqrt{2\tDelta_1\tDelta_2\tDelta_3\tDelta_4}}{3}\fft{\Big(\lambda-\fft{1-2(\tDelta_1+\tDelta_2)+\tDelta_1\tDelta_2}{24\tDelta_1\tDelta_2}\Big)^\fft32}{\lambda^2}+\fft{\mathfrak{c}^\text{ADHM}(\bDelta)}{\lambda^2}\\
	&\quad+\lambda^{-1}\ri\pi(\tDelta-\Delta)\chi+\mO(e^{-\#\sqrt{\lambda}})\,,\\
\end{split}\label{F:ADHM:num}
\end{empheq}
which takes the universal form (\ref{F0:universal:intro}) as advertised. The analytic expression of $\mathfrak{c}^\text{ADHM}(\bDelta)$ has been deduced under the extra constraint $\Delta_3=1$ as
\begin{equation}
	\mathfrak{c}^\text{ADHM}(\bDelta)\Big|_{\Delta_3=1}=\fft{\mA(2\tDelta_1)+\mA(2\tDelta_2)}{4}-\fft{\zeta(3)}{8\pi^2}\big(\tDelta_3^2+\tDelta_4^2\big)\,,\label{ADHM:mfc:special}
\end{equation}
where the special function $\mA(k)$ is defined as 
\begin{equation}
	\begin{split}
		\mathcal{A}(k) =&\; \frac{2\zeta(3)}{\pi^2 k}\Bigl(1 - \frac{k^3}{16}\Bigr) + \frac{k^2}{\pi^2}\int_0^{\infty}dx\,\frac{x}{e^{kx} - 1}\log\bigl(1 - e^{-2x}\bigr)\\[1mm]
		=&\; -\frac{\zeta(3)}{8\pi^2}\,k^2 + \frac12\log(2) + 2\,\zeta'(-1) + \frac16\log\Bigl(\frac{\pi}{2k}\Bigr) \\
		&\; + \sum_{n\geq2}\Bigl(\frac{2\pi}{k}\Bigr)^{2n-2}\,\frac{(-1)^{n}4^{n-1}|B_{2n}B_{2n-2}|}{n(2n-2)(2n-2)!} \, ,
	\end{split}
\end{equation}
see \cite{Hatsuda:2014vsa,Marino:2016new,Bobev:2022eus} for example. Several comments on the ADHM planar free energy (\ref{F:ADHM:num}) are as follows.
\begin{itemize}
	\item The ADHM planar free energy (\ref{F:ADHM:num}) is valid to all orders in the large $\lambda$ expansion up to exponentially suppressed non-perturbative corrections, and consistent with the result from an analytic approach (\ref{F:ADHM:sub}) as expected.
	
	\item As anticipated from the imaginary shift in the eigenvalue distribution ansatz (\ref{ansatz1}), it is straightforward to prove
	\begin{equation}
		Z_{S^3}^\text{1-node}(\Delta,\tDelta)=e^{\ri\pi N\Delta_m(\Delta-\tDelta)}Z_{S^3}^\text{1-node}(\fft{\Delta+\tDelta}{2},\fft{\Delta+\tDelta}{2})\label{1-node:simple}
	\end{equation}
	by renaming the dummy integration variables as $\mu_i\to\mu_i+\ri\pi(\tDelta-\Delta)$ in the matrix model (\ref{Z:1-node}), where we have temporarily presented the dependence of the partition function on $(\Delta,\tDelta)$ explicitly to avoid confusion. This means that the difference of $R$-charges for fundamental/anti-fundamental chiral multiplets, $\Delta-\tDelta$, affects the phase of the partition function only, which is fully captured already at the planar level as observed numerically in the second line of (\ref{F:ADHM:num}); in other words, there is no imaginary contribution to the free energy beyond the planar level. Since (\ref{1-node:simple}) applies to the generic 1-node matrix model (\ref{Z:1-node}) we will observe the same phase difference in the $V^{5,2}$ theory below.
	
	\item The ADHM planar free energy (\ref{F:ADHM:num}) is consistent with the results from the free Fermi-gas formalism of \cite{Hatsuda:2014vsa} and \cite{Hatsuda:2021oxa,Chester:2023qwo}, where the former focuses on the superconformal configuration (\ref{ADHM:constraints:sc}) and the latter generalize the former under the weaker constraints
	\begin{subequations}
		\begin{alignat}{3}
			&\text{\cite{Hatsuda:2021oxa}}:&\quad\Delta_3&=1\,,&\qquad \Delta&=\tDelta=\fft12\,,\label{ADHM:special:Hatsuda}\\
			&\text{\cite{Chester:2023qwo}}:&\quad\Delta_3&=1\,,&\qquad \chi&=0\,.\label{ADHM:special:Chester}
		\end{alignat}\label{ADHM:special}%
	\end{subequations}
	To be more explicit, we focus on the case with the constraints (\ref{ADHM:special:Hatsuda}) where the ADHM free energy is given for a finite $N_f=N/\lambda$ as \cite{Hatsuda:2021oxa}
	\begin{subequations}
		\begin{align}
			F^\text{ADHM}(N,\lambda,\bDelta)\Big|_\text{(\ref{ADHM:special:Hatsuda})}&=-\log\text{Ai}\Big[C^{-1/3}(N-B)\Big]-\fft13\log C \nn\\
			&\quad+\mA+\mO(e^{-\sqrt{N}})\,,\\
			C&=\fft{1}{8\pi^2 N_f\tDelta_1\tDelta_2\tDelta_3\tDelta_4}\,,\\
			B&=\fft{1}{24}\bigg(1-\fft{1}{\tDelta_1\tDelta_2}\bigg)\bigg(N_f-\fft{1}{\tDelta_3\tDelta_4N_f}\bigg)\,,
		\end{align}\label{F:ADHM:Airy}%
	\end{subequations}
	where we have identified their parameters with ours as\footnote{The parameters of \cite{Chester:2023qwo} can be identified to ours as
		\begin{equation}
			N_f=k^\text{\cite{Chester:2023qwo}}\,,\qquad \tDelta_1=\fft12+\ri M^\text{\cite{Chester:2023qwo}}\,,\qquad\tDelta_2=\fft12-\ri M^\text{\cite{Chester:2023qwo}}\,.\label{ADHM:Chester}
		\end{equation}
	}
	\begin{equation}
		l^\text{\cite{Hatsuda:2021oxa}}=N_f\,,\quad \zeta^\text{\cite{Hatsuda:2021oxa}}=\ri\Delta_m\,,\quad \epsilon_1^\text{\cite{Hatsuda:2021oxa}}=2\pi\,,\quad\epsilon_2^\text{\cite{Hatsuda:2021oxa}}=-2\pi\tDelta_1\,,\quad \epsilon_3^\text{\cite{Hatsuda:2021oxa}}=-2\pi\tDelta_2\,.\label{ADHM:Hatsuda}
	\end{equation}
	The $N$-independent constant $\mA$ is known explicitly only under both constraints in (\ref{ADHM:special}) as \cite{Chester:2023qwo}
	\begin{equation}
		\mA\big|_\text{(\ref{ADHM:special})}=\fft{\mA(2\tDelta_1)+\mA(2\tDelta_2)}{4}N_f^2+\fft12\mA(N_f)\,.\label{ADHM:mA:special}
	\end{equation}
	One can check that (\ref{F:ADHM:Airy}) is consistent with the planar free energy (\ref{F:ADHM:num}) we obtained from the numerical analysis in the IIA string theory limit under the constraints (\ref{ADHM:special:Hatsuda}) --- the comparison of $\mA$ term is available only under both constraints in (\ref{ADHM:special}) --- using the asymptotic expansion of the Airy function
	\begin{equation}
		\begin{split}
			\text{Ai}[z] &= \frac{\text{exp}\bigl[-\frac23\,z^{3/2}\bigr]}{2\sqrt{\pi}\,z^{1/4}}\,\sum_{n = 0}^{+\infty}\,\left(-\tfrac32\right)^n\,u_n\,z^{-3n/2}\,,\\
			u_n &= \frac{(6n-5)(6n-3)(6n-1)}{216(2n-1)n}\,u_{n-1}\qquad(n\geq1\,,~u_0=1)\,.
		\end{split}\label{Airy}
	\end{equation}
	It is worth emphasizing, however, that our planar free energy expression (\ref{F:ADHM:num}) covers more general cases with $\Delta_3\neq1$, which is beyond the scope of the free Fermi-gas approach. The analytic expression of $\mathfrak{c}^\text{ADHM}(\bDelta)$ given in (\ref{ADHM:mfc:special}) is also new even though it is restricted to the constraint $\Delta_3=1$; in particular, (\ref{ADHM:mfc:special}) motivates the generalization of the $N$-independent constant in (\ref{ADHM:mA:special}) for a non-vanishing $\chi$ as
	\begin{equation}
		\mA\big|_\text{(\ref{ADHM:special:Hatsuda})}=\fft{\mA(2\tDelta_1)+\mA(2\tDelta_2)}{4}N_f^2+\fft{\mA(2\tDelta_3N_f)+\mA(2\tDelta_4N_f)}{4}\,.\label{ADHM:mA:general}
	\end{equation}
	We will comment further on this point in section \ref{sec:Airy}.
	
	\item Numerical precision is not good enough to analyze the leading exponentially suppressed terms of order $\mO(e^{-\#\sqrt{\lambda}})$, where $\#$ represents the undetermined order 1 coefficient. These non-perturbative corrections are supposed to capture the world-sheet (WS) instanton behaviors in a dual IIA string theory side, see \cite{Hatsuda:2014vsa} for example. We leave a complete numerical analysis on such instanton corrections for future research.
\end{itemize} 

\medskip

Numerical data that support the analytic expression for the ADHM planar free energy (\ref{F:ADHM:num}) is provided in Appendix \ref{app:num:ADHM}. Here we focus on a particular set of $(\lambda,\bDelta)$ configuration
\begin{equation}
	\lambda\in\{30,32,34,36,38,40\}\,,\quad \Delta_I=(\fft23,\fft59,\fft79)\,,\quad\chi=\fft{1}{12}\,,\quad(\Delta,\tDelta)=(\fft{13}{18},\fft12)\,,\label{ADHM:example}
\end{equation}
and show that the corresponding numerical analysis described at the beginning of this subsection \ref{sec:num:1-node} is indeed consistent with the expression (\ref{F:ADHM:num}). 

\medskip

\begin{figure}[t]
	\centering
	\includegraphics[width=0.5\textwidth]{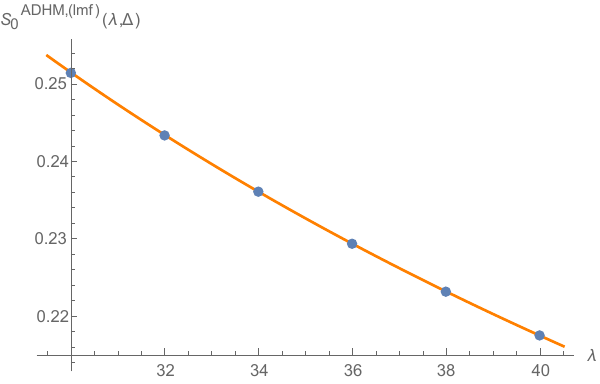}
	\caption{Blue dots represent the numerical values of $S_0^\text{ADHM,(lmf)}(\lambda,\bDelta)$ and the orange line corresponds to the fitting curve in (\ref{ADHM:lmf:lambda}) for the configuration (\ref{ADHM:example}).}
	\label{fig:ADHM}
\end{figure}

To begin with, we find that the coefficients of logarithmic contributions in the \texttt{LinearModelFit} with respect to $N$, (\ref{1-node:lmf}), are given by
\begin{equation}
	S_\text{1/2-log}^\text{1-node,(lmf)} (\lambda,\bDelta)\simeq-1\,,\qquad S_\text{1-log}^\text{1-node,(lmf)} (\lambda,\bDelta)\simeq-\fft{1}{12}\,,\label{ADHM:log}
\end{equation}
which are independent of given $(\lambda,\bDelta)$ configurations. Subtracting these universal logarithmic terms from the numerical values of the on-shell effective action and removing the logarithmic behaviors from the fitting functions, we improve on the \texttt{LinearModelFit} (\ref{1-node:lmf}) as
\begin{align}
	\Re S^\text{1-node} [\bmu^\star;N,\lambda,\bDelta]+N\log N+\fft{1}{12}\log N&=S_0^\text{1-node,(lmf)} (\lambda,\bDelta)N^2+S_{1/2}^\text{1-node,(lmf)} (\lambda,\bDelta)N\nn\\
	&\quad+\sum_{\ell=0}^{20}S_{\ell/2+1}^\text{1-node,(lmf)} (\lambda,\bDelta)N^{-\ell}\,.\label{1-node:lmf:improve}
\end{align}%
Indeed, we confirmed that the precision of the numerical fitting coefficients in the \texttt{LinearModelFit} (\ref{1-node:lmf:improve}) is better than that of (\ref{1-node:lmf}). We then \texttt{LinearModelFit} the real part of the improved $N^2$ leading order coefficients with respect to $\lambda\in\{30,32,34,36,38,40\}$, employing two fitting functions as
\begin{equation}
	S_0^\text{ADHM,(lmf)}(\lambda,\bDelta)=\mfb^\text{ADHM,(lmf)}(\bDelta)\fft{\Big(\lambda-\fft{1-2(\tDelta_1+\tDelta_2)+\tDelta_1\tDelta_2}{24\tDelta_1\tDelta_2}\Big)^\fft32}{\lambda^2}+\mfc^\text{ADHM,(lmf)}(\bDelta)\fft{1}{\lambda^2}\,.\label{ADHM:lmf:lambda}
\end{equation}
The first fitting coefficient $\mfb^\text{ADHM,(lmf)}(\bDelta)$ turns out to be consistent with the analytic result (\ref{F:ADHM:sub}) as 
\begin{equation}
	\mfb^\text{ADHM,(lmf)}(\bDelta)\simeq 	\mfb^\text{ADHM}(\bDelta)\equiv \fft{4\pi\sqrt{2\tDelta_1\tDelta_2\tDelta_3\tDelta_4}}{3}\,.\label{ADHM:mfb}
\end{equation}
The error ratio for the estimate (\ref{ADHM:mfb}) defined by
\begin{equation}
	R^{\mfb}(\bDelta)\equiv\fft{\mfb^\text{1-node,(lmf)}(\bDelta)-\mfb^\text{1-node}(\bDelta)}{\mfb^\text{1-node}(\bDelta)}\label{1-node:R}
\end{equation}
and the standard errors in the \texttt{LinearModelFit} (\ref{ADHM:lmf:lambda}) are presented in Table~\ref{tab:ADHM-1}.
\begin{table}[t]
	\centering
	\footnotesize
	\renewcommand{\arraystretch}{1.2}
	\begin{tabular}{ |c|c||c|c| } 
		\hline
		$R^{\text{ADHM},\mfb}(\bDelta)$ & $\sigma^{\text{ADHM},\mfb}(\bDelta)$ & $\mfc^\text{ADHM,(lmf)}(\bDelta)$ & $\sigma^{\text{ADHM},\mfc}(\bDelta)$ \\
		\hline\hline
		$-3.899\times10^{-16}$ & $4.645\times10^{-17}$ & $-0.064488340394333171725$ & $9.260\times10^{-15}$ \\
		\hline
	\end{tabular}\caption{Fitting data for the ADHM planar free energy for the configuration (\ref{ADHM:example}). Small values of error ratios and standard errors imply that the \texttt{LinearModelFit} (\ref{ADHM:lmf:lambda}) is consistent with a true analytic behavior.}\label{tab:ADHM-1}
\end{table}
See also Fig.~\ref{fig:ADHM} for the \texttt{LinearModelFit} (\ref{ADHM:lmf:lambda}) compared with numerical data points.
Both Table~\ref{tab:ADHM-1} and Fig.~\ref{fig:ADHM} strongly support that the \texttt{LinearModelFit} (\ref{ADHM:lmf:lambda}) captures the analytic behavior of the real part of the planar effective action precisely. Based on this observation, substituting (\ref{ADHM:lmf:lambda}) with (\ref{ADHM:mfb}) back into (\ref{1-node:F0}), we arrive at the analytic expression for the planar free energy (\ref{F:ADHM:num}). A similar numerical analysis has been implemented for different $(\lambda,\bDelta)$ configurations in Appendix \ref{app:num:ADHM}, which finally confirms the ADHM planar free energy (\ref{F:ADHM:num}) for general $(\lambda,\bDelta)$ configurations. 

\medskip

Note that in this subsection we have focused on the configuration (\ref{ADHM:example}) where we do not know the analytic form of $\mfc^\text{ADHM}(\bDelta)$ yet; hence we provided the numerical estimation $\mfc^\text{ADHM,(lmf)}(\bDelta)$ in Table~\ref{tab:ADHM-1} instead of the unknown error ratio $R^{\mfc}(\bDelta)$. In Appendix \ref{app:num:ADHM} we also consider the cases satisfying the extra constraints $\Delta_3=1$ where the analytic form of $\mfc^\text{ADHM}(\bDelta)$ is successfully confirmed as (\ref{ADHM:mfc:special}).

\subsubsection{\texorpdfstring{$V^{5,2}$}{V52} theory}\label{sec:num:1-node:V52}
The numerical analysis described in the beginning of this subsection \ref{sec:num:1-node} yields the $V^{5,2}$ planar free energy consistent with the universal form (\ref{F0:universal:intro}), namely
\begin{empheq}[box=\fbox]{equation}
	\begin{split}
		F_0^{V^{5,2}}(\lambda,\bDelta)&=\fft{4\pi\sqrt{\tDelta_1\tDelta_2\tDelta_3\tDelta_4}}{3}\fft{\Big(\lambda-\fft{1-(\tDelta_1+\tDelta_2)+\tDelta_1\tDelta_2}{12\tDelta_1\tDelta_2}\Big)^\fft32}{\lambda^2}+\fft{\mathfrak{c}^{V^{5,2}}(\bDelta)}{\lambda^2}\\
		&\quad+\lambda^{-1}\ri\pi(\tDelta-\Delta)\chi+\mO(e^{-\#\sqrt{\lambda}})\,,\\
	\end{split}\label{F:V52:num}
\end{empheq}
where the analytic expression of $\mathfrak{c}^{V^{5,2}}(\bDelta)$ has not yet been known. As in the ADHM theory case, the planar free energy (\ref{F:V52:num}) is valid to all orders in the large $\lambda$ expansion up to exponentially suppressed non-perturbative corrections and consistent with the result from an analytic approach (\ref{F:V52:sub}). We leave a more detailed numerical analysis on non-perturbative corrections for future research. 

\medskip

What is different from the ADHM theory case is that the analytic expression (\ref{F:V52:num}) has not yet been known even for the superconformal configuration (\ref{V52:constraints:sc}) since the free Fermi-gas approach \cite{Marino:2011eh,Hatsuda:2014vsa} does not apply to the $V^{5,2}$ theory. It would be very interesting to derive the $V^{5,2}$ planar free energy (\ref{F:V52:num}) analytically and we will comment on this open question further in section \ref{sec:Airy} and \ref{sec:discussion}.

\medskip

Numerical data that support the analytic expression for the $V^{5,2}$ planar free energy (\ref{F:V52:num}) is provided in Appendix \ref{app:num:V52}. Here we focus on a particular set of $(\lambda,\bDelta)$ configuration with the superconformal values (\ref{V52:constraints:sc}), namely
\begin{equation}
	\lambda\in\{30,32,34,36,38,40\}\,,\quad \Delta_I=(\fft23,\fft23,\fft23)\,,\quad\chi=0\,,\quad(\Delta,\tDelta)=(\fft13,\fft13)\,,\label{V52:example}
\end{equation}
and show that the corresponding numerical analysis leads to the expression (\ref{F:V52:num}). 

\medskip

\begin{figure}[t]
	\centering
	\includegraphics[width=0.5\textwidth]{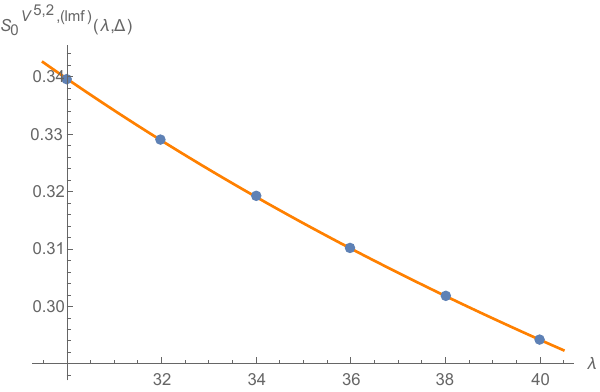}
	\caption{Blue dots represent the numerical values of $S_0^{V^{5,2},\text{(lmf)}}(\lambda,\bDelta)$ and the orange line corresponds to the fitting curve in (\ref{V52:lmf:lambda}) for the configuration (\ref{V52:example}).}
	\label{fig:V52}
\end{figure}

As we did in the previous subsection \ref{sec:num:1-node:ADHM} for the ADHM theory, we first improve the \texttt{LinearModelFit} (\ref{1-node:lmf}) as (\ref{1-node:lmf:improve}) based on the same observation (\ref{ADHM:log}) for universal logarithmic coefficients. We then \texttt{LinearModelFit} the numerical values of $S_0^{V^{5,2}} (\lambda,\bDelta)$ in the improved fitting (\ref{1-node:lmf:improve}) with respect to $\lambda$, employing two fitting functions as
\begin{equation}
	S_0^{V^{5,2},\text{(lmf)}}(\lambda,\bDelta)=\mfb^{V^{5,2},\text{(lmf)}}(\bDelta)\fft{\Big(\lambda-\fft{1-(\tDelta_1+\tDelta_2)+\tDelta_1\tDelta_2}{12\tDelta_1\tDelta_2}\Big)^\fft32}{\lambda^2}+\mfc^{V^{5,2},\text{(lmf)}}(\bDelta)\fft{1}{\lambda^2}\,.\label{V52:lmf:lambda}
\end{equation}
The resulting numerical estimation $\mfb^{V^{5,2},\text{(lmf)}}(\bDelta)$ turns out to be consistent with the analytic result (\ref{F:V52:sub}) as 
\begin{equation}
	\mfb^{V^{5,2},\text{(lmf)}}(\bDelta)\simeq 	\mfb^{V^{5,2}}(\bDelta)\equiv \fft{4\pi\sqrt{\tDelta_1\tDelta_2\tDelta_3\tDelta_4}}{3}\,.\label{V52:mfb}
\end{equation}
The error ratio for the estimate (\ref{V52:mfb}) defined as (\ref{1-node:R}) and the standard errors in the \texttt{LinearModelFit} (\ref{V52:lmf:lambda}) are presented in Table~\ref{tab:V52-1}.
\begin{table}[t]
	\centering
	\footnotesize
	\renewcommand{\arraystretch}{1.2}
	\begin{tabular}{ |c|c||c|c| } 
		\hline
		$R^{V^{5,2},\mfb}(\bDelta)$ & $\sigma^{V^{5,2},\mfb}(\bDelta)$ & $\mfc^{V^{5,2},\text{(lmf)}}(\bDelta)$ & $\sigma^{V^{5,2},\mfc}(\bDelta)$ \\
		\hline\hline
		$2.062\times10^{-17}$ & $1.219\times10^{-17}$ & $0.065067276994384483579$ & $2.416\times10^{-15}$ \\
		\hline
	\end{tabular}\caption{Fitting data for the $V^{5,2}$ planar free energy for the configuration (\ref{V52:example}). Small values of error ratios and standard errors imply that the \texttt{LinearModelFit} (\ref{V52:lmf:lambda}) is consistent with a true analytic behavior.}\label{tab:V52-1}
\end{table}
See also Fig.~\ref{fig:V52} for the \texttt{LinearModelFit} (\ref{V52:lmf:lambda}) compared with numerical data points.
Both Table~\ref{tab:V52-1} and Fig.~\ref{fig:V52} strongly support that the \texttt{LinearModelFit} (\ref{V52:lmf:lambda}) captures the analytic behavior of the real part of the planar effective action precisely. Based on this observation, substituting (\ref{V52:lmf:lambda}) with (\ref{V52:mfb}) back into (\ref{1-node:F0}), we arrive at the analytic expression for the planar free energy (\ref{F:V52:num}). Various other $(\lambda,\bDelta)$ configurations have been explored in a similar manner in Appendix \ref{app:num:V52} and all the results support the $V^{5,2}$ planar free energy (\ref{F:V52:num}).

\subsection{2-node}\label{sec:num:2-node}
For the 2-node theories, the numerical analysis proceeds as follows.
\begin{enumerate}
	\item Derive the saddle point equations 
	\begin{subequations}
		\begin{align}
			0&=\fft{N}{\lambda}\fft{\ri}{2\pi}\mu_i+\sum_{j=1\,(\neq i)}^N\coth\fft{\mu_i-\mu_j}{2}\nn\\
			&\quad+\sum_{a=1}^4\sum_{j=1}^N\bigg(\fft{\Delta_a-1}{2}-\ri\sigma_a\fft{\mu_i-\nu_j}{4\pi}\bigg)\coth(\fft{\mu_i-\nu_j}{2}-\ri\sigma_a\pi(1-\Delta_a))\nn\\
			&\quad+\fft{N}{\lambda_\mu}\bigg(\fft{\Delta-1}{2}-\ri\fft{\mu_i}{4\pi}\bigg)\coth\bigg(\fft{\mu_i}{2}-\ri\pi(1-\Delta)\bigg)\nn\\
			&\quad+\fft{N}{\lambda_\nu}\bigg(\fft{\Delta-1}{2}+\ri\fft{\mu_i}{4\pi}\bigg)\coth\bigg(\fft{\mu_i}{2}+\ri\pi(1-\Delta)\bigg)\,,\\
			0&=-\fft{N}{\lambda}\fft{\ri}{2\pi}\nu_i+\sum_{j=1\,(\neq i)}^N\coth\fft{\nu_i-\nu_j}{2}\nn\\
			&\quad+\sum_{a=1}^4\sum_{j=1}^N\bigg(\fft{\Delta_a-1}{2}+\ri\sigma_a\fft{\nu_i-\mu_j}{4\pi}\bigg)\coth(\fft{\nu_i-\mu_j}{2}+\ri\sigma_a\pi(1-\Delta_a))\nn\\
			&\quad+\fft{N}{\lambda_\nu}\bigg(\fft{\Delta-1}{2}-\ri\fft{\nu_i}{4\pi}\bigg)\coth\bigg(\fft{\nu_i}{2}-\ri\pi(1-\Delta)\bigg)\nn\\
			&\quad+\fft{N}{\lambda_\mu}\bigg(\fft{\Delta-1}{2}+\ri\fft{\nu_i}{4\pi}\bigg)\coth\bigg(\fft{\nu_i}{2}+\ri\pi(1-\Delta)\bigg)\,,
		\end{align}\label{2-node:saddle}%
	\end{subequations}
	by taking the 1st derivative of the effective action (\ref{free2}) with respect to the gauge holonomies under the constraints (\ref{2-node:constraints}). Find a numerical solution $\{\bmu^\star,\bnu^\star\}$ to the saddle point equation (\ref{2-node:saddle}) for given $(N,\lambda,\bDelta)$ configurations using the \texttt{FindRoot} in \textit{Mathematica} at \texttt{WorkingPrecision} 200, employing the leading order solution (\ref{2-sol:ABJMN010}) or (\ref{2-sol:Q111}) as initial conditions.
	
	\item Evaluate the on-shell effective action
	\begin{equation}
		S^\text{2-node} [\bmu^\star,\bnu^\star;N,\lambda,\bDelta]\label{2-node:S-onshell}
	\end{equation}
	by substituting the numerical solution $\{\bmu^\star,\bnu^\star\}$ into the effective action (\ref{free2}). In this step we confirmed that the imaginary part of the on-shell effective action vanishes for all the examples of our interest listed in subsection \ref{sec:mm:2-node}; recall that we focus on the superconformal configuration for the $N^{0,1,0}$/$Q^{1,1,1}$ theories.
	
	\item \texttt{LinearModelFit} the on-shell effective action (\ref{2-node:S-onshell}) with respect to large values of $N$ ($100\sim350$ in step of $10$) for given $(\lambda,\bDelta)$ configurations as
	\begin{equation}
		\begin{split}
			S^\text{2-node} [\bmu^\star,\bnu^\star;N,\lambda,\bDelta]&=S_0^\text{2-node,(lmf)} (\lambda,\bDelta)N^2+S_\text{1/2-log}^\text{2-node,(lmf)} (\lambda,\bDelta)N\log N\\
			&\quad+S_{1/2}^\text{2-node,(lmf)} (\lambda,\bDelta)N+S_\text{1-log}^\text{2-node,(lmf)} (\lambda,\bDelta)\log N\\
			&\quad+\sum_{\ell=0}^{18}S_{\ell/2+1}^\text{2-node,(lmf)} (\lambda,\bDelta)N^{-\ell}\,,
		\end{split}\label{2-node:lmf}
	\end{equation}
	as we did in the 1-node examples. According to the formula (\ref{planar2}), reading off the coefficient of the $N^2$ leading term in the \texttt{LinearModelFit} (\ref{2-node:lmf}) yields the numerical value of the planar free energy as
	\begin{equation}
		F_0^\text{2-node}(\lambda,\bDelta)=S_0^\text{2-node,(lmf)} (\lambda,\bDelta)\,.\label{2-node:F0}
	\end{equation}

	\item Repeating the above 1-3 steps for various $(\lambda,\bDelta)$ configurations, deduce the analytic expression of the planar free energy $F_0^\text{2-node}(\lambda,\bDelta)$ as a function of the 't~Hooft coupling $\lambda$ and the $R$-charges $\bDelta$.  
\end{enumerate}
The final results and details involved in the above-described numerical analysis will be presented below for the ABJM, $N^{0,1,0}$, and $Q^{1,1,1}$ theories in order. As in the 1-node cases, we provide the final results first and then explain how the numerical data support them through the above procedure.

\subsubsection{ABJM theory}\label{sec:num:2-node:ABJM}
Applying the numerical analysis described above to the ABJM theory, we obtain the ABJM planar free energy
\begin{empheq}[box=\fbox]{equation}
	\begin{split}
		F_0^\text{ABJM}(\lambda,\bDelta)&=\fft{4\pi\sqrt{2\Delta_1\Delta_2\Delta_3\Delta_4}}{3}\fft{(\lambda-\fft{1}{24})^\fft32}{\lambda^2}+\fft{\mathfrak{c}^\text{ABJM}(\bDelta)}{\lambda^2}+\mO(e^{-\#\sqrt{\lambda}})\,,\\
	\end{split}\label{F:ABJM:num}
\end{empheq}
where the analytic expression of $\mathfrak{c}^\text{ABJM}(\bDelta)$ has been deduced as
\begin{align}
	\mathfrak{c}^\text{ABJM}(\bDelta)&=\fft{\zeta(3)}{8\pi^2}\Bigg[4-\sum_{a=1}^4\Delta_a^2-4\sum_{a=1}^4\fft{\Delta_1\Delta_2\Delta_3\Delta_4}{\Delta_a}\label{ABJM:mfc}\\
	&\kern3em~~+\fft{(\Delta_1+\Delta_3-\Delta_2-\Delta_4)^2(\Delta_1+\Delta_4-\Delta_2-\Delta_3)^2}{4(\Delta_1+\Delta_3)(\Delta_1+\Delta_4)(\Delta_2+\Delta_3)(\Delta_2+\Delta_4)}\sum_{a=1}^4\fft{\Delta_1\Delta_2\Delta_3\Delta_4}{\Delta_a}\Bigg]\,.\nn
\end{align}%
The result (\ref{F:ABJM:num}) again takes the universal form (\ref{F0:universal:intro}) as in the 1-node cases. Several comments on the ABJM planar free energy (\ref{F:ABJM:num}) are as follows. 
\begin{itemize}
	\item The ABJM planar free energy (\ref{F:ABJM:num}) is valid to all orders in the large $\lambda$ expansion up to exponentially suppressed non-perturbative corrections, and consistent with the result from a saddle point approximation (\ref{F:ABJM:sub}) as expected.
	
	\item The ABJM planar free energy (\ref{F:ABJM:num}) is consistent with the results from the free Fermi-gas approach of \cite{Marino:2011eh} and \cite{Nosaka:2015iiw}, where the former focuses on the superconformal configuration $\Delta_a=\fft12$ (see \cite{Fuji:2011km} for an alternative derivation) and the latter generalizes the former under the weaker constraint
	\begin{equation}
		\Delta_1+\Delta_3=1\qquad\text{or}\qquad\Delta_1+\Delta_4=1\,.\label{ABJM:special}
	\end{equation}
	To be more explicit, the ABJM free energy was obtained under the constraints (\ref{ABJM:special}) via the free Fermi-gas approach for a finite $k=N/\lambda$ as \cite{Nosaka:2015iiw}
	\begin{subequations}
		\begin{align}
			F^\text{ABJM}(N,\lambda,\bDelta)\Big|_\text{(\ref{ABJM:special})}&=-\log\text{Ai}\Big[C^{-1/3}(N-B)\Big]-\fft13\log C+\fft14\sum_{a=1}^4\mA(2\Delta_a k)\nn\\
			&\quad+\mO(e^{-\sqrt{N}})\,,\label{F:ABJM:Airy-special:mA}\\[0.5em]
			C&=\fft{1}{8\pi^2 k\Delta_1\Delta_2\Delta_3\Delta_4}\,,\\
			B&=\fft{k}{24}+\fft{1}{12k}\bigg(\fft{1-\fft14\sum_{a=1}^4\Delta_a^2}{\Delta_1\Delta_2\Delta_3\Delta_4}-\sum_{a=1}^4\fft{1}{\Delta_a}\bigg)\,.
		\end{align}\label{F:ABJM:Airy-special}%
	\end{subequations}
	One can check that (\ref{F:ABJM:Airy-special}) is consistent with the planar free energy (\ref{F:ABJM:num}) we obtained via the numerical analysis in the IIA string theory limit under the constraints (\ref{ABJM:special}), using the asymptotic expansion of the Airy function (\ref{Airy}).	But our planar free energy expression (\ref{F:ABJM:num}) has been obtained without imposing the constraints (\ref{ABJM:special}), which motivates an appropriate generalization of the Airy formula (\ref{F:ABJM:Airy-special}) to generic $R$-charge configurations as in \cite{Bobev:2022jte,Bobev:2022eus,Hristov:2022lcw}. We will comment further on this point in section \ref{sec:Airy}.
	
	\item To deduce the analytic expression (\ref{ABJM:mfc}) based on numerical data, we have employed the known result \cite{Nosaka:2015iiw}
	\begin{equation}
		\mathfrak{c}^\text{ABJM}(\bDelta)\Big|_\text{(\ref{ABJM:special})}=\fft{\zeta(3)}{8\pi^2}\sum_{a=1}^4\Delta_a^2\label{ABJM:mfc:special}
	\end{equation}
	together with the symmetry that the free energy is invariant under
	\begin{equation}
		\Delta_1\leftrightarrow\Delta_2\,,\qquad \Delta_3\leftrightarrow\Delta_4\,,\qquad (\Delta_1,\Delta_2)\leftrightarrow (\Delta_3,\Delta_4)\,,\label{ABJM:sym}
	\end{equation}
	originating from the structure of the matrix model (\ref{Z:2-node}). Note that the 2nd line of (\ref{ABJM:mfc}) vanishes under the special constraint (\ref{ABJM:special}), which is required to employ the free Fermi-gas formalism, and respects the symmetry (\ref{ABJM:sym}). But it does not respect the full permutation symmetry among $\Delta_a$, which is reminiscent of the observation made in \cite{Chester:2021gdw} for the exact partition function with small $(N,k)$ values.
	
	\item As in the ADHM theory case, numerical precision is not good enough to analyze exponentially suppressed terms. These non-perturbative corrections are supposed to capture the WS instanton behaviors in a dual IIA string theory side, see \cite{Cagnazzo:2009zh,Drukker:2010nc,Hatsuda:2012dt,Nosaka:2015iiw,Gautason:2023igo,Beccaria:2023ujc,Nosaka:2024gle} for example. We leave a complete numerical analysis on such instanton corrections for future research.
\end{itemize} 

\medskip

Numerical data that support the analytic expression for the ABJM planar free energy (\ref{F:ABJM:num}) is provided in Appendix \ref{app:num:ABJM}. Here we focus on a particular set of $(\lambda,\bDelta)$ configuration 
\begin{equation}
	\lambda\in\{30,32,34,36,38,40\}\,,\qquad \Delta_a=(\fft{3}{5},\fft{3}{10},\fft{9}{10},\fft{2}{10})\,,\label{ABJM:example}
\end{equation}
and show that the corresponding numerical analysis described in the beginning of this subsection \ref{sec:num:2-node} is indeed consistent with the expression (\ref{F:ABJM:num}). Note that (\ref{ABJM:example}) does not satisfy the special constraint (\ref{ABJM:special}) employed in \cite{Nosaka:2015iiw} so there is no known Airy formula like (\ref{F:ABJM:Airy-special}) for the free energy with the configuration (\ref{ABJM:example}); only the $\lambda^{-1/2}$ leading order (\ref{F:ABJM:lead}) has been known so far in the literature \cite{Jafferis:2011zi}, which is improved to the first subleading correction of order $\lambda^{-3/2}$ in (\ref{F:ABJM:sub}) analytically and extended to all orders in the large $\lambda$ expansion as (\ref{F:ABJM:num}) via numerical analysis. 

\medskip

\begin{figure}[t]
	\centering
	\includegraphics[width=0.5\textwidth]{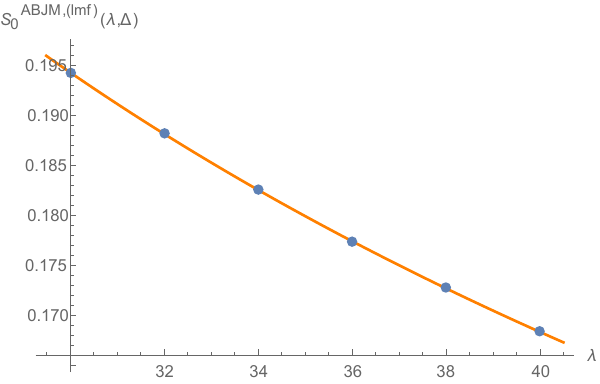}
	\caption{Blue dots represent the numerical values of $S_0^\text{ABJM,(lmf)}(\lambda,\bDelta)$ and the orange line corresponds to the fitting curve in (\ref{ABJM:lmf:lambda}) for the configuration (\ref{ABJM:example}).}
	\label{fig:ABJM}
\end{figure}

The first step is to observe universal logarithmic coefficients in the \texttt{LinearModelFit} with respect to $N$ (\ref{2-node:lmf}) as in the 1-node examples, namely
\begin{equation}
	S_\text{1/2-log}^\text{2-node,(lmf)} (\lambda,\bDelta)\simeq-2\,,\qquad S_\text{1-log}^\text{2-node,(lmf)} (\lambda,\bDelta)\simeq-\fft{1}{6}\,.\label{ABJM:log}
\end{equation}
Subtracting these logarithmic terms from the numerical values of the on-shell effective action improves the \texttt{LinearModelFit} (\ref{2-node:lmf}) as
\begin{align}
	S^\text{2-node} [\bmu^\star;N,\lambda,\bDelta]+2N\log N+\fft16\log N&=S_0^\text{2-node,(lmf)} (\lambda,\bDelta)N^2+S_{1/2}^\text{2-node,(lmf)} (\lambda,\bDelta)N\nn\\
	&\quad+\sum_{\ell=0}^{20}S_{\ell/2+1}^\text{2-node,(lmf)} (\lambda,\bDelta)N^{-\ell}\,.\label{2-node:lmf:improve}
\end{align}%
We then \texttt{LinearModelFit} the improved numerical estimations for the $N^2$ leading order coefficients with respect to $\lambda$, employing two fitting functions as
\begin{equation}
	S_0^\text{ABJM,(lmf)}(\lambda,\bDelta)=\mfb^\text{ABJM,(lmf)}(\bDelta)\fft{(\lambda-\fft{1}{24})^\fft32}{\lambda^2}+\mfc^\text{ABJM,(lmf)}(\bDelta)\fft{1}{\lambda^2}\,.\label{ABJM:lmf:lambda}
\end{equation}
The numerical fitting coefficients turn out to be consistent with the analytic expression (\ref{F:ABJM:num}) as 
\begin{equation}
\begin{split}
	\mfb^\text{ABJM,(lmf)}(\bDelta)&\simeq 	\mfb^\text{ABJM}(\bDelta)\equiv \fft{4\pi\sqrt{2\Delta_1\Delta_2\Delta_3\Delta_4}}{3}\,,\\
	\mfc^\text{ABJM,(lmf)}(\bDelta)&\simeq\mfc^\text{ABJM}(\bDelta)\,,
\end{split}\label{ABJM:mf}
\end{equation}
where the latter coefficient $\mfc^\text{ABJM}(\bDelta)$ is defined in (\ref{ABJM:mfc}). The error ratios for the estimates (\ref{ABJM:mf}) defined by
\begin{equation}
	R^{\mfb}(\bDelta)\equiv\fft{\mfb^\text{2-node,(lmf)}(\bDelta)-\mfb^\text{2-node}(\bDelta)}{\mfb^\text{2-node}(\bDelta)}\,,\quad R^{\mfc}(\bDelta)\equiv\fft{\mfc^\text{2-node,(lmf)}(\bDelta)-\mfc^\text{2-node}(\bDelta)}{\mfc^\text{2-node}(\bDelta)}\,,\label{2-node:R}
\end{equation}
and the standard errors in the \texttt{LinearModelFit} (\ref{ABJM:lmf:lambda}) are presented in Table~\ref{tab:ABJM-1}.
\begin{table}[t]
	\centering
	\footnotesize
	\renewcommand{\arraystretch}{1.2}
	\begin{tabular}{ |c|c||c|c| } 
		\hline
		$R^{\text{ABJM},\mfb}(\bDelta)$ & $\sigma^{\text{ABJM},\mfb}(\bDelta)$ & $R^{\text{ABJM},\mfc}(\bDelta)$ & $\sigma^{\text{ABJM},\mfc}(\bDelta)$ \\
		\hline\hline
		$1.550\times10^{-11}$ & $3.139\times10^{-12}$ & $-2.056\times10^{-7}$ & $6.213\times10^{-10}$ \\
		\hline
	\end{tabular}\caption{Fitting data for the ABJM planar free energy for the configuration (\ref{ABJM:example}). Small values of error ratios and standard errors imply that the \texttt{LinearModelFit} (\ref{ABJM:lmf:lambda}) is consistent with a true analytic behavior.}\label{tab:ABJM-1}
\end{table}
See also Fig.~\ref{fig:ABJM} for the \texttt{LinearModelFit} (\ref{ABJM:lmf:lambda}) compared with numerical data points.
Both Table~\ref{tab:ABJM-1} and Fig.~\ref{fig:ABJM} strongly support that the \texttt{LinearModelFit} (\ref{ABJM:lmf:lambda}) captures the analytic behavior of the planar effective action precisely. Based on this observation, substituting (\ref{ABJM:lmf:lambda}) with (\ref{ABJM:mf}) back into (\ref{2-node:F0}), we obtain the analytic expression for the planar free energy (\ref{F:ABJM:num}). A similar numerical analysis for various other $(\lambda,\bDelta)$ configurations also confirms the ABJM planar free energy (\ref{F:ABJM:num}), see Appendix \ref{app:num:ABJM} for details.

\subsubsection{\texorpdfstring{$N^{0,1,0}$}{N010} theory}\label{sec:num:2-node:N010}
The numerical analysis described in the beginning of this subsection \ref{sec:num:2-node} yields the $N^{0,1,0}$ planar free energy ($\alpha\equiv r/k$)
\begin{empheq}[box=\fbox]{equation}
	\begin{split}
		F_0^{N^{0,1,0}}(\lambda,\alpha)=\fft{2(1+\alpha)\pi}{3\sqrt{2+\alpha}}\fft{(\lambda-\fft{2-3\alpha}{48})^\fft32}{\lambda^2}+\fft{\mathfrak{c}^{N^{0,1,0}}(\alpha)}{\lambda^2}+\mO(e^{-\#\sqrt{\lambda}})\\
	\end{split}\label{F:N010:num}
\end{empheq}
for the superconformal configuration (\ref{N010:constraints:sc}), where the analytic expression of $\mathfrak{c}^{N^{0,1,0}}(\alpha)$ has not yet been known. The result (\ref{F:N010:num}) is not new and can be read off from the known Airy formula for the $S^3$ partition function \cite{Marino:2011eh,Bobev:2023lkx} in the IIA string theory limit. Hence in this example our numerical analysis serves only as a consistency check, and therefore we relegate all the details involved in the numerical analysis to Appendix \ref{app:num:N010}. It would be very interesting to investigate the $N^{0,1,0}$ planar free energy beyond the reach of the free Fermi-gas formalism that yields the Airy formula, for instance by removing the constraint $r_1=r_2$ introduced in (\ref{N010:equal}), and see if the planar free energy still takes the universal form (\ref{F0:universal:intro}). See Appendix \ref{app:detail:N010} for more detailed discussion on this point.

\subsubsection{\texorpdfstring{$Q^{1,1,1}$}{Q111} theory}\label{sec:num:2-node:Q111}
Applying the numerical analysis summarized in the beginning of this subsection \ref{sec:num:2-node} to the $Q^{1,1,1}$ theory at the superconformal configuration (\ref{Q111:constraints:sc}), we obtain the $Q^{1,1,1}$ planar free energy
\begin{empheq}[box=\fbox]{equation}
	F_0^{Q^{1,1,1}}(\lambda)=\fft{4\pi}{3\sqrt{3}}\fft{(\lambda+\fft{1}{12})^\fft32}{\lambda^2}+\fft{\mathfrak{c}^{Q^{1,1,1}}}{\lambda^2}+\mO(e^{-\#\sqrt{\lambda}})\,,\label{F:Q111:num}
\end{empheq}
where the analytic expression of $\mathfrak{c}^{Q^{1,1,1}}$ has not yet been deduced. As in the previous examples, the planar free energy (\ref{F:Q111:num}) is valid to all orders in the large $\lambda$ expansion up to exponentially suppressed non-perturbative corrections, and consistent with the result from an analytic approach (\ref{F:Q111:sub}). We leave a more detailed numerical analysis on non-perturbative corrections for future research. 

\medskip

Similar to the $V^{5,2}$ theory case, the analytic expression (\ref{F:Q111:num}) has not yet been known even for the superconformal configuration (\ref{Q111:constraints:sc}) since the free Fermi-gas formalism is not applicable to the matrix model for the $S^3$ partition function of the $Q^{1,1,1}$ theory. Hence the analytic derivation of a simple expression (\ref{F:Q111:num}), again consistent with the universal form (\ref{F0:universal:intro}), remains an interesting open problem and we will comment on this issue in section \ref{sec:Airy}.

\medskip

\begin{figure}[t]
	\centering
	\includegraphics[width=0.5\textwidth]{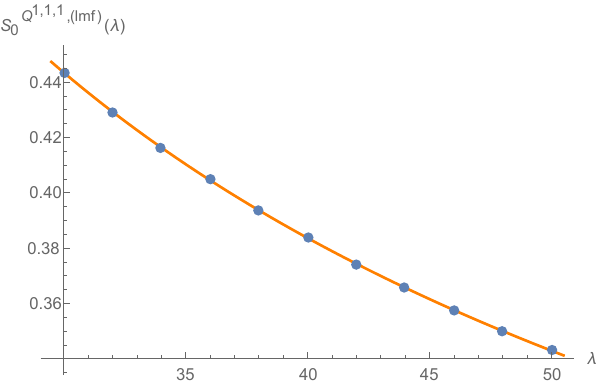}
	\caption{Blue dots represent the numerical values of $S_0^{Q^{1,1,1},\text{(lmf)}}(\lambda)$ and the orange line corresponds to the fitting curve in (\ref{Q111:lmf:lambda}).}
	\label{fig:Q111}
\end{figure}

To derive the $Q^{1,1,1}$ planar free energy (\ref{F:Q111:num}), we implement the numerical analysis for $\lambda=30\sim50$ (in step of $2$). As in the ABJM theory case explored in subsection \ref{sec:num:2-node:ABJM}, we first improve the  \texttt{LinearModelFit} (\ref{2-node:lmf}) as (\ref{2-node:lmf:improve}) by subtracting the universal logarithmic coefficients (\ref{ABJM:log}). We then \texttt{LinearModelFit} the improved numerical values of $S_0^{Q^{1,1,1}} (\lambda)$ in (\ref{2-node:lmf:improve}) with respect to $\lambda$, employing two fitting functions as
\begin{equation}
	S_0^{Q^{1,1,1},\text{(lmf)}}(\lambda)=\mfb^{Q^{1,1,1},\text{(lmf)}}\fft{(\lambda+\fft{1}{12})^\fft32}{\lambda^2}+\mfc^{Q^{1,1,1},\text{(lmf)}}\fft{1}{\lambda^2}\,.\label{Q111:lmf:lambda}
\end{equation}
The fitting coefficient $\mfb^{Q^{1,1,1},\text{(lmf)}}$ is confirmed to be given as 
\begin{equation}
	\mfb^{Q^{1,1,1},\text{(lmf)}}\simeq 	\mfb^{Q^{1,1,1}}\equiv \fft{4\pi}{3\sqrt{3}}\,,\label{Q111:mfb}
\end{equation}
which is consistent with the analytic results (\ref{F:Q111:sub}). The error ratio for the estimate (\ref{Q111:mfb}) defined as (\ref{2-node:R}) and the standard errors in the \texttt{LinearModelFit} (\ref{Q111:lmf:lambda}) are presented in Table~\ref{tab:Q111-1}.
\begin{table}[t]
	\centering
	\footnotesize
	\renewcommand{\arraystretch}{1.2}
	\begin{tabular}{ |c|c||c|c| } 
		\hline
		$R^{Q^{1,1,1},\mfb}$ & $\sigma^{Q^{1,1,1},\mfb}$ & $\mfc^{Q^{1,1,1},\text{(lmf)}}$ & $\sigma^{Q^{1,1,1},\mfc}$ \\
		\hline\hline
		$2.263\times10^{-11}$ & $1.291\times10^{-11}$ & $0.060896898392804042312$ & $2.917\times10^{-9}$ \\
		\hline
	\end{tabular}\caption{Fitting data for the $Q^{1,1,1}$ planar free energy for $\lambda=30\sim50$ (in step of $2$). Small values of error ratios and standard errors imply that the \texttt{LinearModelFit} (\ref{Q111:lmf:lambda}) is consistent with a true analytic behavior.}\label{tab:Q111-1}
\end{table}
See also Fig.~\ref{fig:Q111} for the \texttt{LinearModelFit} (\ref{Q111:lmf:lambda}) compared with numerical data points.
Both Table~\ref{tab:Q111-1} and Fig.~\ref{fig:Q111} confirm that the \texttt{LinearModelFit} (\ref{Q111:lmf:lambda}) captures the analytic behavior of the planar effective action precisely. Based on this observation, substituting (\ref{Q111:lmf:lambda}) with (\ref{Q111:mfb}) back into (\ref{2-node:F0}), we arrive at the analytic expression for the planar free energy (\ref{F:Q111:num}).

\section{Airy conjectures}\label{sec:Airy}
In this section we discuss Airy formulae for the $S^3$ partition functions of $\mN=2$ holographic SCFTs introduced in section \ref{sec:mm}. In subsection \ref{sec:Airy:old} we first compare the planar free energies obtained in the previous sections \ref{sec:ana} and \ref{sec:num} with the known Airy conjectures for the $S^3$ partition functions of the ABJM/ADHM theories. In subsection \ref{sec:Airy:new} we move on to other examples, and conjecture new Airy formulae for the $S^3$ partition functions based on their planar free energies. In due process, the universal form of the planar free energy (\ref{F0:universal:intro}), which we repeat here as
\begin{equation}
	F^\text{SCFT}_0(\lambda)=\mfb^\text{SCFT}\big(\lambda-B^\text{SCFT}\big)^\fft32+\fft{\mfc^\text{SCFT}}{\lambda^2}+\mO(e^{-\#\sqrt{\lambda}})\,,\label{F0:universal}
\end{equation}
will play the key role. In (\ref{F0:universal}) the theory-dependent constants $\mfb^\text{SCFT}$, $\mfc^\text{SCFT}$, and $B^\text{SCFT}$ can be read off from the results in the previous section \ref{sec:num} for various examples of our interest.

\subsection{Comparison with the known Airy conjectures}\label{sec:Airy:old}
To begin with, recall that the Airy formula for the ABJM $S^3$ partition function (\ref{F:ABJM:Airy-special}), derived via the free Fermi-gas formalism under the extra constraint (\ref{ABJM:special}) in \cite{Nosaka:2015iiw}, has recently been extended to generic $R$-charges under the original superpotential marginality constraint (\ref{ABJM:constraints}) as \cite{Bobev:2022jte,Bobev:2022eus,Hristov:2022lcw}
\begin{subequations}
	\begin{align}
		F^\text{ABJM}(N,\lambda,\bDelta)&=-\log\text{Ai}\Big[C^{-1/3}(N-B)\Big]-\fft13\log C+\mA^\text{ABJM}(k,\bDelta)+\mO(e^{-\sqrt{N}})\,,\\[0.5em]
		C&=\fft{1}{8\pi^2 k\Delta_1\Delta_2\Delta_3\Delta_4}\,,\\
		B&=\fft{k}{24}+\fft{1}{12k}\bigg(\fft{1-\fft14\sum_{a=1}^4\Delta_a^2}{\Delta_1\Delta_2\Delta_3\Delta_4}-\sum_{a=1}^4\fft{1}{\Delta_a}\bigg)\,,\label{F:ABJM:Airy:B}
	\end{align}\label{F:ABJM:Airy}%
\end{subequations}
where the analytic expression for the $N$-independent contribution $\mA^\text{ABJM}(k,\bDelta)$ is left unknown. The conjecture (\ref{F:ABJM:Airy}) has not yet been analytically proven, but passed various non-trivial consistency checks \cite{Bobev:2022eus}. Our planar free energy result (\ref{F:ABJM:num}) provides a new independent non-trivial test for the Airy conjecture (\ref{F:ABJM:Airy}) by confirming that the $\fft{k}{24}$ piece in (\ref{F:ABJM:Airy:B}) is valid for generic $R$-charges even without the constraint (\ref{ABJM:special}), see section \ref{sec:discussion} for comments on this shift from the holographic viewpoint. Furthermore, the planar free energy (\ref{F:ABJM:num}) tells us that the large $k$ expansion of the unknown function $\mA^\text{ABJM}(k,\bDelta)$ is supposed to be given as
\begin{equation}
	\mA^\text{ABJM}(k,\bDelta)=\mfc^\text{ABJM}(\bDelta)k^2+o(k^2)\,,\label{ABJM:mA}
\end{equation}
where the analytic expression for the coefficient $\mfc^\text{ABJM}(\bDelta)$ is determined explicitly in (\ref{ABJM:mfc}). 

\medskip

Another Airy conjecture we would like to compare with our planar free energy results is for the $S^3$ partition function of the ADHM theory with an $\mN=4$ symmetry breaking mass term \cite{Minahan:2021pfv,Bobev:2023lkx}. Turning on the $\mN=4$ symmetry breaking mass parameter $\mfm$ corresponds to choosing a specific configuration of $R$-charges in the 1-node matrix model (\ref{Z:1-node}), namely \cite{Bobev:2023lkx}
\begin{equation}
	\Delta_1=\Delta_2=\Delta=\tDelta=\fft{1-\ri\mfm}{2}\,,\qquad\Delta_3=1+\ri\mfm\,,\qquad\Delta_m=0\,,\label{ADHM:N=4}
\end{equation}
and the corresponding Airy conjecture for the $S^3$ partition function is given as \cite{Bobev:2023lkx}\footnote{Here we turned off the squashing parameter as $b=1$ and replaced the $\mN=4$ symmetry breaking mass parameter symbol as $\mu^\text{there}=\mfm^\text{here}$.}
\begin{subequations}
\begin{align}
	F^\text{ADHM}(N,\lambda,\bDelta)\Big|_\text{(\ref{ADHM:N=4})}&=-\log\text{Ai}[C^{-\fft13}(N-B)]-\fft13\log C+\mA+\mO(e^{-\sqrt{N}})\,,\\
	C&=\fft{2}{\pi^2N_f}\fft{1}{(1+\ri\mfm)^2(1-\ri\mfm)^2}\,,\\
	B&\overset{!}{=}N_f\bigg(\fft{1}{24}-\fft{1+3\mfm^2}{6(1+\ri\mfm)^2(1-\ri\mfm)^2}\bigg)+\fft{1}{N_f}\fft{3+\mfm^2}{6(1+\ri\mfm)^2(1-\ri\mfm)^2}\,,\label{F:ADHM:Airy:N=4:B}
\end{align}\label{F:ADHM:Airy:N=4}%
\end{subequations}
where the $N$-independent constant $\mA$ has not yet been conjectured. Note that (\ref{F:ADHM:Airy:N=4}) is beyond the reach of the Airy formula (\ref{F:ADHM:Airy}) derived via the free Fermi-gas formalism in \cite{Chester:2023qwo} since the configuration (\ref{ADHM:N=4}) is distinguished from (\ref{ADHM:special}). Unlike the ABJM Airy conjecture discussed above, in this case our planar free energy result (\ref{F:ADHM:num}) is \emph{not} consistent with the Airy formula (\ref{F:ADHM:Airy:N=4}) at the planar level in the 't~Hooft limit under the parameter identifications (\ref{eq:not}) and (\ref{ADHM:N=4}); the key difference is between the shift of the 't~Hooft coupling in (\ref{F:ADHM:num}) and the first term in the $B$ parameter (\ref{F:ADHM:Airy:N=4:B}). This implies that, even though the conjecture (\ref{F:ADHM:Airy:N=4}) passed various consistency checks \cite{Bobev:2023lkx}, it does not capture the $S^3$ partition function precisely. It is non-trivial to derive the correct Airy formula analytically and thereby fix (\ref{F:ADHM:Airy:N=4:B}) in this case, however, since (\ref{ADHM:N=4}) does not satisfy the simplifying constraint $\Delta_3=1$ in (\ref{ADHM:special}) required for the application of the free Fermi-gas approach. Hence the first step toward addressing this issue is to explore if one can formulate an Airy conjecture for the squashed $S^3_b$ partition function of the ADHM with the $\mN=4$ symmetry breaking mass term as in \cite{Bobev:2023lkx}, but in a way consistent with our planar free energy result (\ref{F:ADHM:num}). A useful guidance for this research direction will be briefly discussed in the next subsection.

\subsection{New Airy conjectures}\label{sec:Airy:new}
For other holographic SCFTs considered in this paper whose $S^3$ partition functions are beyond the scope of the free Fermi-gas formalism --- namely the ADHM theory with the most general $R$-charges, the $V^{5,2}$ theory, the $N^{0,1,0}$ theory with generic $r_{1,2}$ values, and the $Q^{1,1,1}$ theory --- there is no known Airy conjecture for the $S^3$ partition function. Motivated by the above observation that our planar free energy calculation provides non-trivial evidence for the Airy conjecture for the ABJM $S^3$ partition function with generic $R$-charges, however, we can now formulate an Airy conjecture for such cases employing the universal planar free energy form (\ref{F0:universal}). Below we focus on the ADHM theory with generic $R$-charges and the $V^{5,2}$/$Q^{1,1,1}$ theories in this regard and leave the Airy conjectures for the $N^{0,1,0}$ theory with generic $r_{1,2}$ values for future work.

\medskip

For the ADHM theory with generic $R$ charges, we can deduce the Airy conjecture based on the planar free energy result (\ref{F:ADHM:num}) and the IR duality between the ADHM theory with $N_f=1$ and the ABJM theory with $k=1$ \cite{Aharony:2008ug,Kapustin:2010xq}. The resulting conjecture reads
\begin{subequations}
	\begin{align}
		F^\text{ADHM}(N,\lambda,\bDelta)&=-\log\text{Ai}[C^{-\fft13}(N-B)]-\fft13\log C+\mA\nn\\
		&\quad+\ri\pi N\Delta_m(\tDelta-\Delta)+\mO(e^{-\sqrt{N}})\,,\\
		C&=\fft{1}{8\pi^2N_f\tDelta_1\tDelta_2\tDelta_3\tDelta_4}\,,\\
		B&=\fft{1-2(\tDelta_1+\tDelta_2)+\tDelta_1\tDelta_2}{24\tDelta_1\tDelta_2}N_f\nn\\
		&\quad-\fft{\tDelta_1^2+\tDelta_2^2+5\tDelta_1\tDelta_2-2(1+\tDelta_1\tDelta_2)(\tDelta_1+\tDelta_2)}{24\tDelta_1\tDelta_2\tDelta_3\tDelta_4}\fft{1}{N_f}\,,\label{F:ADHM:Airy:general:B}
	\end{align}\label{F:ADHM:Airy:general}%
\end{subequations}
where the $N$-independent constant $\mA$ is supposed to reproduce (\ref{ADHM:mfc:special}) and (\ref{ADHM:mA:special}) under the corresponding special cases but left unknown in general. The imaginary contribution is read off from the property (\ref{1-node:simple}) immediately. The $C$ coefficient and the first $N_f$-linear term in the $B$ coefficient are determined unambiguously relying on the planar free energy result (\ref{F:ADHM:num}) and the asymptotic expansion of the Airy function (\ref{Airy}). The IR equivalence between the ADHM theory with $N_f=1$ and the ABJM theory with $k=1$ is then used to determine the second term in the $B$-coefficient; the ADHM $B$ coefficient (\ref{F:ADHM:Airy:general:B}) is designed to match the ABJM $B$ coefficient (\ref{F:ABJM:Airy:B}) under $N_f=k=1$ and the identification $\tDelta_a\leftrightarrow\Delta_a$.\footnote{In this step, it suffices to employ the map $\tDelta_a\leftrightarrow\Delta_a$ as in \cite{Bobev:2023lkx} since the ABJM $B$ coefficient (\ref{F:ABJM:Airy:B}) is fully symmetric under the permutation of $\Delta_a$ parameters. The $N$-independent constant $\mA$ does not respect the full permutation symmetry, however, as implied in the expression (\ref{ABJM:mA}) with (\ref{ABJM:mfc}). Hence one should be more careful when identifying $\tDelta_a$ and $\Delta_a$ parameters between the two theories in principle. For instance, the comparison of the constraints required for explicit expressions of the $\mA$ constants in (\ref{ADHM:mA:general}) and (\ref{F:ABJM:Airy-special:mA}), namely $\tDelta_1+\tDelta_2=1$ and $\Delta_1+\Delta_3=1$ respectively, implies that the parameters are supposed to be mapped as $(\tDelta_1,\tDelta_3,\tDelta_2,\tDelta_4)$ $\leftrightarrow$ $(\Delta_1,\Delta_2,\Delta_3,\Delta_4)$. This is precisely consistent with the parameter identification for the IR duality obtained by matching the matrix models of the ABJM/ADHM $S^3$ partition functions \cite{Chester:2020jay}.} 

One can check that the general ADHM Airy conjecture (\ref{F:ADHM:Airy:general}) reproduces the proven Airy formula (\ref{F:ADHM:Airy}) under the constraint (\ref{ADHM:special:Hatsuda}). However, it is not consistent with the Airy conjecture (\ref{F:ADHM:Airy:N=4}) for the $\mN=4$ symmetry breaking mass term \cite{Bobev:2023lkx} as we already commented on in the previous subsection. To be more specific, the inconsistency is captured by 
\begin{equation}
\begin{split}
	B\Big|_{(\ref{F:ADHM:Airy:general:B})\,\text{w/}\,(\ref{ADHM:N=4})}-B\Big|_{(\ref{F:ADHM:Airy:N=4:B})}=-\fft{\ri\mfm^3}{8(1+\mfm^2)^2}\bigg(N_f-\fft{1}{N_f}\bigg)\,,
\end{split}
\end{equation}
and it would be very interesting to modify the Airy conjecture for the $S^3_b$ partition function in \cite{Bobev:2023lkx} incorporating this difference. We leave this problem for future work.

\medskip

For the $V^{5,2}$/$Q^{1,1,1}$ theories, we first use the recent observation on the relation between the topologically twisted index and the superconformal index of 3d $\mN=2$ SCFTs \cite{Bobev:2022wem,Bobev:2024} in conjunction with the structure of the 4-derivative corrections to dual $\mN=2$ minimal gauged supergravity \cite{Bobev:2020egg,Bobev:2021oku}, which determines the 3d large $N$ partition functions to the first subleading order in the M-theory regime \cite{Bobev:2020egg,Bobev:2021oku,Bobev:2022wem,Bobev:2024}. For the $V^{5,2}$/$Q^{1,1,1}$ theories of our interest, the resulting $S^3$ free energies are given by \cite{Bobev:2024}
\begin{align}
	F^{V^{5,2}}(\lambda)&=\fft{16\pi\sqrt{N_f}}{27}\bigg[N^\fft32-\fft{N_f^2+15}{32N_f}N^\fft12\bigg]+\mO(\log N)\,,\label{F:V52:largeN}\\
	F^{Q^{1,1,1}}(\lambda)&=\fft{4\pi\sqrt{N_f}}{3\sqrt{3}}\bigg[N^\fft32+\fft{N_f^2-3}{8N_f}N^\fft12\bigg]+\mO(\log N)\,,\label{F:Q111:largeN}
\end{align}
where $N_f$ is treated as a finite parameter in the large $N$ limit. In the IIA string theory limit where $N$ is taken large with a fixed 't~Hooft coupling $\lambda\equiv N/N_f$, these are consistent with the planar free energy results we obtained in (\ref{F:V52:num}) and (\ref{F:Q111:num}) respectively. Now combining  (\ref{F:V52:num}) with (\ref{F:V52:largeN}), and similarly (\ref{F:Q111:num}) with (\ref{F:Q111:largeN}), we conjecture the Airy formulae for their $S^3$ partition functions as (recall $F=-\log Z$)
\begin{align}
	Z^{V^{5,2}}(N,\lambda)&=\bigg(\fft{81}{64\pi^2N_f}\bigg)^{-\fft13}e^{\mA^{V^{5,2}}(N_f)}\text{Ai}\bigg[\bigg(\fft{81}{64\pi^2N_f}\bigg)^{-\fft13}\bigg(N-\fft{N_f^2+15}{48N_f}\bigg)\bigg](1+\mO(e^{-\sqrt{N}}))\,,\label{V52:Airy}\\
	Z^{Q^{1,1,1}}(N,\lambda)&=\bigg(\fft{3}{4\pi^2N_f}\bigg)^{-\fft13}e^{\mA^{Q^{1,1,1}}(N_f)}\text{Ai}\bigg[\bigg(\fft{3}{4\pi^2N_f}\bigg)^{-\fft13}\bigg(N+\fft{N_f^2-3}{12N_f}\bigg)\bigg](1+\mO(e^{-\sqrt{N}}))\,,\label{Q111:Airy}
\end{align}
where the $\mA$-functions have the following large $N_f$ behaviors:
\begin{equation}
	\mA^{V^{5,2}}(N_f)=\mfc^{V^{5,2}}(\bDelta)\Big|_\text{(\ref{V52:constraints:sc})}N_f^2+o(N_f^2)\,,\qquad \mA^{Q^{1,1,1}}(N_f)=\mfc^{Q^{1,1,1}}N_f^2+o(N_f^2)\,.\label{mA:V52Q111}
\end{equation}
In fact, one may derive the above Airy conjectures relying solely on (\ref{F:V52:largeN}) and (\ref{F:Q111:largeN}), mimicking the known structures in (\ref{F:ADHM:Airy}) and (\ref{F:ABJM:Airy}). This was first done in the work in progress \cite{Hong:2024} exploring the squashed 3-sphere partition functions. The important role of our planar free energy calculations (\ref{F:V52:num}) and (\ref{F:Q111:num}), however, is to provide non-trivial evidence for the Airy conjectures to all orders in the large $\lambda$ expansion at the planar level, which is not captured by the previous large $N$ partition functions (\ref{F:V52:largeN}) and (\ref{F:Q111:largeN}). It would be very interesting to confirm the Airy conjectures (\ref{V52:Airy}) and (\ref{Q111:Airy}) either beyond the planar level in the IIA string theory limit or beyond the $N^\fft12$ order in the M-theory limit where $N$ is taken large with a fixed $N_f$. In section \ref{sec:discussion} we will comment on this research direction further.

\section{Discussion}\label{sec:discussion}
We studied the $S^3$ partition functions of various 3d $\mN=2$ holographic SCFTs arising from the $N$ stack of M2 branes at the tip of the cone over 7d Sasaki-Einstein manifolds, focusing on the planar limit where $N$ is taken large with a fixed 't~Hooft coupling $\lambda$. Based on the matrix model representation of such $S^3$ partition functions given from supersymmetric localization, we first employed the saddle point approximation to evaluate the planar $S^3$ free energy analytically to the first subleading corrections in the large 't~Hooft coupling limit. These analytic results are then improved to all orders in the large $\lambda$ expansion via numerical analysis. It is remarkable that the resulting planar free energies always take the universal form (\ref{F0:universal}), which we compared with the known Airy conjectures for the ABJM/ADHM theories and furthermore employed to propose new Airy conjectures for the $V^{5,2}$/$Q^{1,1,1}$ theories as well as the ADHM theory with generic $R$-charges. Below we present various open questions. 

\medskip

One of the most pressing tasks is to analytically prove the universal form of the planar free energy (\ref{F0:universal}) that we have confirmed in various $\mN=2$ holographic SCFTs arising from the $N$ stack of M2-branes. One may attempt to improve the current saddle point ansatz (\ref{ansatz1}) and (\ref{ansatz2}) which are ignorant of the shift of the 't~Hooft coupling observed in the universal form (\ref{F0:universal}). Another promising approach is to employ the resolvent, whose associated spectral curve in the planar limit can be employed to derive the planar free energy of the corresponding matrix model via appropriate period integrals \cite{Ambjorn:1992gw,DiFrancesco:1993cyw,Akemann:1996zr,Dijkgraaf:2002fc,Aganagic:2002wv,Ashok:2002bi,Halmagyi:2003ze}, as well-reviewed in \cite{Marino:2011nm,Marino:2016new} for our context. Even though the resolvent method was applied only to the superconformal configuration of the ABJM theory among the examples of our interest, considering that the universal form (\ref{F0:universal}) has been observed beyond such a special case, we expect that there must be a systematic approach --- the resolvent method or another --- for the calculation of the planar free energy of generic $\mN=2$ holographic SCFTs with M-theory duals. It would be very interesting to figure out such an analytic method.

Another interesting direction is to extend our analytic/numerical saddle point analysis beyond the planar level by incorporating loop corrections. In principle, one may expect that the $N^0$ order corrections in the 't~Hooft limit can be tracked easily by including the 1-loop determinant in a saddle point approximation. There is a subtle issue in the saddle point approximation for the Hermitian matrix model of rank $N$ in the large $N$ expansion \cite{Liu:2019tuk,Liu:2020bih,Hong:2021bsb}, however, which yields a non-vanishing artifact at the linear $N$ order even after incorporating the 1-loop determinant. Exploring higher loop corrections in a saddle point approximation and thereby determining the genus-1 contribution correctly would therefore be a first step in this research direction. In due process, one can also investigate the $\log N$ order and thereby support the characteristic logarithmic behaviors with pure number coefficients employed in the recent work \cite{Bobev:2023dwx} for a larger class of holographic SCFTs. The results will offer useful guidance to understand stringy loop corrections in a dual IIA string theory path integral in the context of holography, which can be explored alternatively through M-theory, as well as a non-trivial test for the known Airy conjectures for the $S^3$ partition functions beyond the planar level.

It would also be intriguing to explore the $S^3$ partition functions of more generic $\mN=2$ holographic SCFTs dual to the $N$ stack of M2-branes and conjecture new Airy formulae for them. For instance, it is still non-trivial to employ the known methods such as the saddle point approximation (see Appendix \ref{app:detail:N010} for discussion on this point) and the free Fermi-gas formalism to analyze the $S^3$ partition function of the $N^{0,1,0}$ theory with different $r_{1,2}$ values beyond the $\lambda^{-1/2}N^2$ leading order in the double scaling limit. Figuring out how to investigate its $S^3$ partition function beyond the leading order and then exploring if the subleading structure allows for a new Airy conjecture is therefore an interesting open problem. Extending a similar analysis to generic $\mN=2$ holographic SCFTs with M-theory duals would then be a natural follow-up task.

The ultimate question when it comes to the Airy tale is how to prove the conjectured Airy formula for the $S^3$ partition functions of various $\mN=2$ holographic SCFTs. There are two well-established methods that have been employed to derive the Airy formulae: the free Fermi-gas formalism \cite{Marino:2011eh} and the chain of dualities relating the matrix model of our interest and the topological string theories \cite{Witten:1992fb,Aganagic:2002wv,Gopakumar:1998ki,Ooguri:2002gx,Bershadsky:1993cx,Dijkgraaf:2002fc,Eynard:2007hf,Marino:2009jd}, where the latter is outlined well in \cite{Fuji:2011km} for the ABJM theory. Their application to the $S^3$ partition functions of generic $\mN=2$ holographic SCFTs does not seem plausible at this point, however, since both approaches require certain necessary conditions that are not apparently satisfied in various examples of our interest. It would be very interesting if one figures out how to apply these methods in general setting or finds a brand new method, and thereby proves that this class of $\mN=2$ holographic SCFTs with M-theory duals always has the 3-sphere partition function captured by an Airy function. In the context of holographic duality, such a field theoretic derivation of the Airy formula will also enhance our understanding of the underlying Airy structure in quantum gravity path integrals involving AdS$_4$ backgrounds, as elucidated through supergravity localization \cite{Dabholkar:2014wpa,Caputa:2018asc}.

While this manuscript and the aforementioned open questions focus on perturbative contributions to the $S^3$ partition function, understanding the non-perturbative behaviors is another important direction in this research program. For example, non-perturbative corrections in the $S^3$ partition functions of the ABJM/ADHM theory and their stringy dual descriptions have been analyzed extensively in numerous references \cite{Hatsuda:2014vsa,Cagnazzo:2009zh,Drukker:2010nc,Hatsuda:2012dt,Nosaka:2015iiw,Gautason:2023igo,Beccaria:2023ujc,Nosaka:2024gle}. Such a non-perturbative analysis on the field theory side has been restricted to the cases where the perturbatively exact Airy formula is proven via the free Fermi-gas formalism, however, and therefore our numerical saddle point analysis could be a first step toward the non-perturbative analysis for generic $\mN=2$ holographic SCFTS without known Airy formulae. To achieve this, improving the precision in the numerics to capture the leading WS instanton corrections of order $\sim e^{-\#\sqrt{\lambda}}$ would be a primary task. Another related question on non-perturbative contributions concerns different saddles in the matrix model and their roles in phase transitions \cite{Nosaka:2015bhf,Nosaka:2016vqf}, which would be interesting to explore via the same numerical approach employed in this manuscript.

Motivated by a successful numerical analysis for the $S^3$ partition functions of $\mN=2$ holographic SCFTs arising from the $N$ stack of M2-branes, it is natural to ask if a similar all order exact (with respect to a large 't~Hooft coupling expansion) result for the planar free energy is available in different classes of holographic SCFTs. One of the promising targets is the 3d SCFTs associated with a non-vanishing Romans mass \cite{Gaiotto:2009mv,Guarino:2015jca}, whose $S^3$ partition functions have been analyzed in various ways \cite{Jafferis:2011zi,Marino:2011eh,Suyama:2010hr,Suyama:2013fua,Liu:2019tuk,Hong:2021bsb,Liu:2021njm}. The compact all order expression like (\ref{F0:universal}) has not yet been determined in this class of holographic SCFTs, however, and it would be very interesting if the numerical analysis can enlighten the hidden structure beyond the most recent observation in \cite{Liu:2021njm}.

Another interesting direction of generalization is to incorporate a non-trivial squashing parameter $b$ in the game. The matrix models capturing the squashed 3-sphere partition functions are given explicitly in the literature \cite{Hama:2011ea,Imamura:2011wg} and one can therefore analyze the $S^3_b$ planar free energy in the same way as we did for the round 3-sphere. One technical obstacle is that the supersymmetric localization results in a matrix model written in terms of a double sine function for the $S^3_b$ partition function \cite{Hama:2011ea,Imamura:2011wg}, not the special $\ell$-function exploited extensively in this manuscript, whose asymptotic expansion is less explicit beyond the leading order; see \cite{Hatsuda:2016uqa} mentioned in footnote~\ref{squashed:sqrt3} for the special case with $b=(\sqrt{3})^{\pm1}$ where this technical issue was overcome by employing dual topological strings. It would be very interesting to explore squashed 3-sphere partition functions beyond the leading order in the large 't~Hooft coupling limit by addressing this technical issue since the result can be employed to support the Airy conjecture involving the squashing parameter \cite{Bobev:2022jte,Bobev:2022eus,Bobev:2023lkx,Hristov:2022lcw}. The first step towards this research direction will be reported in \cite{Hong:2024}.

Lastly, to understand the $S^3$ planar free energies valid to all orders in the large 't~Hooft coupling expansion in the context of holographic duality, the primary task is to reproduce the subleading terms from higher derivative corrections on the gravity side. Recall that the leading $\lambda^{-1/2}$ behavior in the planar free energy has been successfully derived from the 2-derivative on-shell action in a dual supergravity side \cite{Drukker:2010nc,Herzog:2010hf,Cheon:2011vi,Martelli:2011qj}. The starting point for the subleading analysis is therefore to analyze the 8-derivative corrections in Type IIA supergravity or in 11-dimensional supergravity --- see \cite{Grimm:2017okk,Ozkan:2024euj} for example and references therein --- and evaluate their Euclidean on-shell actions on the AdS$_4\times X$ background where the internal manifold $X$ is specified by a dual 3d SCFT of interest, modulo a subtle issue in supergravity on-shell actions upon the Kaluza-Klein compactification \cite{Kurlyand:2022vzv,Beccaria:2023hhi}. In due process, it would be very interesting to figure out the origin of the simple universal structure of the planar free energy (\ref{F0:universal}) from the holographic viewpoint. For example, understanding a constant shift of the 't~Hooft coupling as depicted in the universal form (\ref{F0:universal}) could potentially be achieved through an examination of shifted D-brane charges (or M2 brane charges for a finite $k$) as investigated in \cite{Bergman:2009zh} for the ABJM theory\footnote{The first term in the M2-brane charge shift $-\fft{1}{24}(k-\fft1k)$ due to a topological 8-derivative correction to 11d supergravity \cite{Bergman:2009zh} seems identical to the $-\fft{k}{24}$ shift observed in (\ref{F:ABJM:Airy:B}). One has to fully investigate all the 8-derivative corrections to 11d supergravity in this regard, however, to establish such a matching between the shifts rigorously. We leave this for a future work.} and generalized for less supersymmetric 3d gauge theories in \cite{Aharony:2009fc}. Another avenue worth exploring is the investigation of the same shift in relation to the all-order renormalization of Newton constant as proposed in \cite{Bobev:2021oku,Bobev:2022eus}. The remaining $\lambda^{-2}$ behavior in (\ref{F0:universal}) might signal the on-shell action of the quartic $R^4$ term, considering that its coefficient involves the transcendental number $\zeta(3)$ \cite{Gross:1986iv,Grimm:2017okk,Ozkan:2024euj} and we have encountered the same $\zeta(3)$ factor in the ABJM/ADHM planar free energies. This line of exploration will offer a very concrete research direction towards comprehending string/M-theory beyond the 2-derivative low energy description.

\section*{Acknowledgments}

We are grateful to Nikolay Bobev, Pieter-Jan De Smet, Valentin Reys, Christoph F. Uhlemann, and Xuao Zhang for valuable discussions. SG is supported by the VUB Research Council through the Strategic Research Program High-Energy Physics. JH is supported by the Sogang University Research Grant of 202410008.01, the Basic Science Research Program of the National Research Foundation of Korea (NRF) funded by the Ministry of Education through the Center for Quantum Spacetime (CQUeST) with grant number NRF-2020R1A6A1A03047877, and the Fonds Wetenschappelijk Onderzoek--Vlaanderen (FWO) Junior Postdoctoral Fellowship with grant number 1203024N.

\appendix

\section{Useful functions}\label{app:fcts}
The $\ell$-function is defined as
\begin{equation}
	\ell(z)\equiv-z\log(1-e^{2\pi\ri z})+\fft{\ri}{2}\bigg(\pi z^2+\fft{1}{\pi}\text{Li}_2(e^{2\pi\ri z})\bigg)-\fft{\ri\pi}{12}\,,\label{ell}
\end{equation}
see \cite{Jafferis:2010un,Jafferis:2011zi} for example. The $\ell$-function has the following inversion properties
\begin{equation}
	\begin{split}
		\exp[\ell(u)]\exp[\ell(-u)]&=1\,,\\
		\exp[\ell(\fft12+\fft{\ri}{2\pi}u)]\exp[\ell(\fft12-\fft{\ri}{2\pi}u)]&=\fft{1}{2\cosh(u/2)}\,,\label{ell:inversion}
	\end{split}
\end{equation}
and its first derivative can be written in terms of a simple trigonometric function as
\begin{equation}
	\ell'(z)=-\pi z\cot(\pi z)\,.\label{ell:deriv}
\end{equation}
The $\ell$-function has the following expansion \cite{Martelli:2011qj}
\begin{equation}
	\ell(z) = \begin{cases}
	\displaystyle \frac{\ri \pi}{2} \left(z^2 - \frac{1}{6} \right) + \sum_{n=1}^\infty \left(\frac{z}{n} + \frac{\ri}{2 \pi n^2} \right) e^{2 \pi \ri nz} & (\Im[z]>0)\\
	\displaystyle -\frac{\ri \pi}{2} \left(z^2 - \frac{1}{6} \right) + \sum_{n=1}^\infty \left(\frac{z}{n} - \frac{\ri}{2 \pi n^2} \right) e^{-2 \pi \ri nz} & (\Im(z)<0)
    \end{cases}\,,\label{expl}
\end{equation}
which is useful to study the contribution from $\mN=2$ chiral multiplets to the matrix model for the $S^3$ partition function. For the contribution from $\mN=2$ vector multiplets the following expansion is useful:
\begin{align}
    \log \left(2 \sinh \frac{z}{2} \right) = \frac{z}{2} - \sum^\infty_{m=1} \frac{1}{m} e^{-mz} \quad \left(\Re[z] > 0 \right) \, . \label{expu}
\end{align}

The polylogarithm is defined within the unit disk $|z|<1$ as
\begin{equation}
	\text{Li}_s(z)=\sum_{k=1}^\infty\fft{z^k}{k^s}\qquad(|z|<1)\,,
\end{equation}
and extended to the entire complex plane via analytic continuation. The polylogarithm satisfies the following inversion formula
\begin{equation}
	\text{Li}_n(e^{2\pi\ri z})+(-1)^n\text{Li}_n(e^{-2\pi\ri z})=-\fft{(2\pi\ri)^n}{n!}\text{B}_n(z)~~~\begin{cases}
		0\leq\Re[z]<1~\&~\Im[z]\geq0 \\
		0<\Re[z]\leq1~\&~\Im[z]<0
	\end{cases}\,,\label{polylog:inversion}
\end{equation}
where $\text{B}_n(z)$ are the Bernoulli polynomials of degree $n$. Using the inversion formula (\ref{polylog:inversion}), it is straightforward to show ($x\in\mathbb{R}$)
\begin{equation}
    \begin{aligned}
        \sum_{n=1}^\infty \frac{1}{n^2} \cos 2 \pi n x &= \pi^2 \text{B}_2(\{x\}) \, , \\
        \sum_{n=1}^\infty \frac{1}{n^3} \sin 2 \pi n x &= \frac{2\pi^3}{3}  \text{B}_3(\{x\}) \,, \\
        \sum_{n=1}^\infty \frac{1}{n^4} \cos 2 \pi n x &= -\frac{\pi^4}{3} \text{B}_4(\{x\}) \,, \\
        \sum_{n=1}^\infty \frac{1}{n^5} \sin 2 \pi n x &= -\frac{2\pi^5}{15} \text{B}_5(\{x\}) \, , \label{eq:sos}
    \end{aligned}
\end{equation}
where $\{x\}$ is defined as the modded value $\{x\}\equiv x-\lfloor x\rfloor$.

\section{Details in the saddle point approximation}\label{app:detail}
In this Appendix we present technical details involved in the analytic calculation of the planar free energy via a saddle point approximation in section \ref{sec:ana}.

\subsection{The 1-node planar effective action}\label{app:detail:isol1}
Here we show that the contributions from $\mN=2$ chiral multiplets to the planar effective action, namely (\ref{free1:conti:split:adj}) and (\ref{free1:conti:split:fun}) for adjoint representations (in conjunction with the vector multiplet contribution) and the pairs of fundamental \& anti-fundamental representations, are expanded in the large $\lambda$ limit as (\ref{Free1adj}) and (\ref{Free1fun}) respectively. 

\subsubsection{Fundamental \& (anti) fundamental representations}\label{app:detail:isol1:fun}
To investigate the contribution from the pairs of fundamental \& anti-fundamental $\mN=2$ chiral multiplets, (\ref{free1:conti:split:fun}), we first derive
\begin{equation}
	\begin{split}
		&-\lambda^{-1}\int^{q_r}_{q_l}dq\, \rho(q) \ell\left(\frac{2-\Delta - \tDelta}{2} \pm \ri \frac{\lambda^{1/2}q}{2 \pi}  \right)\\
		&=\frac{2- \Delta- \tDelta}{4}\lambda^{-1/2}\int^{q_r}_{q_l} dq\, \rho(q) |q| \pm \ri (\cdots) \\
		&\quad-\lambda^{-1} \sum_{n=1}^\infty \int^{0}_{q_l} dq\, \rho(q)  \left(\frac{\frac{2-\Delta - \tDelta}{2} \pm \ri \frac{\lambda^{1/2}q}{2 \pi}}{n} \mp \frac{\ri}{2 \pi n^2} \right)e^{\pm\pi \ri n (\Delta + \tDelta)}e^{n\lambda^{1/2}q}  \\
		&\quad-\lambda^{-1} \sum_{n=1}^\infty \int^{q_r}_{0} dq\,\rho(q) \left(\frac{\frac{2-\Delta - \tDelta}{2} \pm \ri \frac{\lambda^{1/2}q}{2 \pi}}{n} \pm \frac{\ri}{2 \pi n^2} \right)e^{ \mp\pi \ri n (\Delta + \tDelta )}e^{-n\lambda^{1/2}q}
	\end{split}\label{fun:exp}
\end{equation}
based on the series expansion of the $\ell$-function (\ref{expl}). Note that we have not specified the imaginary part in the expansion since they will cancel out after all in the contribution from the pairs fundamental \& anti-fundamental representations. Substituting the expansion (\ref{fun:exp}) into (\ref{free1:conti:split:fun}) then yields
\begin{equation}
\begin{split}
	S_0^\text{1-fun}[\rho;\lambda]  & = \frac{2 - \Delta - \tDelta}{2}\lambda^{-1/2}\int^{q_r}_{q_l}dq\,\rho(q)  |q|  \\ 
	&\quad- \lambda^{-1} \sum_{n=1}^\infty \int_{q_l}^0 dq\,\rho(q) \left[ \left(\frac{\frac{2-\Delta - \tDelta}{2} + \ri\lambda^{1/2} \frac{q}{2 \pi}}{n}- \frac{\ri}{2 \pi n^2} \right)e^{ \pi \ri n \left(\Delta + \tDelta \right)} \right. \\ &\kern4em + \left. \left(\frac{\frac{2-\Delta - \tDelta}{2} - \ri \lambda^{1/2} \frac{q}{2 \pi}}{n}+ \frac{\ri}{2 \pi n^2} \right)e^{-\pi \ri n \left(\Delta + \tDelta \right)} \right]e^{n \lambda^{1/2}q} \\  
	&\quad- \lambda^{-1} \sum_{n=1}^\infty \int^{q_r}_0 dq\,\rho(q) \left[ \left(\frac{\frac{2-\Delta - \tDelta}{2} + \ri\lambda^{1/2} \frac{q}{2 \pi}}{n}+ \frac{\ri}{2 \pi n^2} \right)e^{- \pi \ri n \left(\Delta + \tDelta \right)} \right. \\ &\kern4em + \left. \left(\frac{\frac{2-\Delta - \tDelta}{2} - \ri \lambda^{1/2} \frac{q}{2 \pi}}{n}- \frac{\ri}{2 \pi n^2} \right)e^{\pi \ri n \left(\Delta + \tDelta \right)} \right]e^{-n \lambda^{1/2}q} \, . 
\end{split}\label{fun1:1}
\end{equation}
To proceed, we make the substitution of $t=-\lambda^{1/2}q$ in the before-last integral and $t=\lambda^{1/2}q$ in the last, so that we can better envision the saddle point approximation we are about to do. In particular, we will use
\begin{align}
	\int_0^{\lambda^{1/2} Q} d t \, e^{-nt} t^a = \int_0^{\infty} dt\, e^{-nt} t^a + \mO\left(e^{-n\lambda^{1/2}Q} \right) \label{saddlepoint}
\end{align}
upon the change of an integration variable, where $Q$ is either $-q_l$ or $q_r$. This allows us to take the boundaries of the last two integrals in (\ref{fun1:1}) to infinity when we ignore exponentially suppressed terms in the large $\lambda$ limit. Then we take the Taylor expansion of $\rho(q)$ within the last two integrals of (\ref{fun1:1}) around $q=0$ in the large $\lambda$ limit (recall $q=\mp\lambda^{-1/2}t$), which simplifies (\ref{fun1:1}) to
\begin{equation}
	\begin{split}
		S_0^\text{1-fun}[\rho;\lambda]
		&= \frac{2 - \Delta - \tDelta}{2}\lambda^{-1/2}\int^{q_r}_{q_l}dq\,\rho(q)  |q|  \\ 
		&\quad- 2\lambda^{-3/2} \rho\left(0 \right)\sum_{n=1}^\infty \int_{0}^\infty dt\,  \left[ \left(\frac{2-\Delta - \tDelta}{2n} - \ri \frac{t}{2 \pi n}- \frac{\ri}{2 \pi n^2} \right)e^{ \pi \ri n \left(\Delta + \tDelta \right)} \right. \\ 
		&\quad + \left. \left(\frac{2-\Delta - \tDelta}{2n} + \ri  \frac{t}{2 \pi n}+ \frac{\ri}{2 \pi n^2} \right)e^{-\pi \ri n \left(\Delta + \tDelta \right)} \right]e^{-n t}  + \mathcal{O}\left(\lambda^{-2} \right) \, . 
	\end{split}\label{fun1:2}
\end{equation}
Evaluating the $t$-integral in (\ref{fun1:2}) using the formula
\begin{align}
	\int_0^{\infty} dt \, e^{-nt} t^a =  \frac{a!}{n^{a+1}} \,, \label{eq:id}
\end{align}
we obtain
\begin{equation}
\begin{split}
	S_0^\text{1-fun}[\rho;\lambda]
	&= \frac{2 - \Delta - \tDelta}{2}\lambda^{-1/2}\int^{q_r}_{q_l}dq\,\rho(q)  |q|   \\
	&\quad-2\lambda^{-3/2} \rho\left(0\right)\sum_{n=1}^\infty \frac{1}{n^2}  \left[ (2-\Delta- \tDelta) \cos \pi n (\Delta + \tDelta )+\frac{2}{\pi n} \sin \pi n (\Delta + \tDelta )  \right]\\
	&\quad+ \mathcal{O} \left(\lambda^{-2} \right) \,.
\end{split}\label{fun1:3}
\end{equation}
Implementing the sums over $n$ in (\ref{fun1:3}) using the identities (\ref{eq:sos}), we finally arrive at the expression (\ref{Free1fun}) that gives the contribution from the pairs of fundamental \& anti-fundamental $\mN=2$ chiral multiplets to the planar effective action, keeping the first two leading terms in the large $\lambda$ expansion. Note that we have implicitly assumed $\Delta + \tDelta \leq 2$ to employ the identities (\ref{eq:sos}) without involving modded values, but this inequality does not pose any extra constraint; both the constraints for the ADHM theory (\ref{ADHM:constraints}) and the ones for the $V^{5,2}$ theory (\ref{V52:constraints}) already impose $\Delta + \tDelta \leq 2$. 

\subsubsection{Adjoint representations}\label{app:detail:isol1:adj}
To investigate the contribution from adjoint $\mN=2$ chiral multiplets (\ref{free1:conti:split:adj}), we make the substitution of $t=\lambda^{1/2}(q-q')$ for the first integration variable $q'$ as 
\begin{align}
	&S_0^\text{1-adj}[\rho;\lambda]\nn\\
	&= \lambda^{-1/2}\int_{q_l}^{q_r}dq\,\rho(q)^2\sum_{n=1}^\infty \int_0^{\lambda^{1/2}Q} dt\, \frac{2}{n} e^{-nt} \left[1-\sum_{I=1}^3 \bigg\{ (1-\Delta_I) \cos 2 \pi n \Delta_I + \frac{\frac{1}{n} + t}{2\pi} \sin 2 \pi n \Delta_I \bigg\} \right] \nn  \\
	&\quad-\lambda^{-1}\int_{q_l}^{q_r}dq\,\rho(q)\rho'(q) \sum_{n=1}^\infty \int_0^{\lambda^{1/2}Q} dt\,  \frac{2t}{n} e^{-nt} \left[1-\sum_{I=1}^3 \bigg\{ (1-\Delta_I) \cos 2 \pi n \Delta_I + \frac{\frac{1}{n} + t}{2\pi} \sin 2 \pi n \Delta_I \bigg\} \right] \nn  \\
	&\quad+\frac{1}{2}\lambda^{-3/2}\int_{q_l}^{q_r}dq\,\rho(q)\rho''(q) \sum_{n=1}^\infty \int_0^{\lambda^{1/2}Q} dt\,   \frac{2t^2}{n} e^{-nt} \left[1-\sum_{I=1}^3 \bigg\{ (1-\Delta_I) \cos 2 \pi n \Delta_I + \frac{\frac{1}{n} + t}{2\pi} \sin 2 \pi n \Delta_I \bigg\} \right] \nn \\
	&\quad+\mO(\lambda^{-2})\, , \label{adj1:1}
\end{align}
where we have implemented the Taylor expansion of $\rho(q')$ around $q'=q$ in the large $\lambda$ limit and also defined $Q=q-q_l$.

\medskip

We will again extend the boundary of $t$-integrals in (\ref{adj1:1}) from $\lambda^{1/2}Q$ to $\infty$ and use the identity (\ref{eq:id}) as we did for (anti-)fundamental representations in the previous Appendix \ref{app:detail:isol1:fun}. But this time we need to be a bit more careful since the $t$-integrals have to be integrated again over $q$ in (\ref{adj1:1}). Hence we should in principle keep track of the exponentially suppressed terms of order $\mO(e^{-n\lambda^{1/2}Q})$ upon the use of the identity (\ref{eq:id}), and check if they could contribute to the power law behaviors up to $\lambda^{-3/2}$ orders after the integration over $q$. In Appendix \ref{app:detail:exp}, we investigate this subtlety and argue that the exponentially suppressed terms do not contribute to planar free energy at the orders of our interest, and therefore we can follow the same procedure described in \ref{app:detail:isol1:fun}.

\medskip

Taking the analysis in Appendix \ref{app:detail:exp} into account, we can extend the integration ranges in (\ref{adj1:1}) from $\lambda^{1/2}Q$ to $\infty$ and thereby simplify it as
\begin{align}
		&S_0^\text{1-adj}[\rho;\lambda] \nn\\
		&= \lambda^{-1/2}\int_{q_l}^{q_r}dq\,\rho(q)^2 \sum_{n=1}^\infty   \frac{2}{n^2}  \left[1-\sum_{I=1}^3\bigg\{ (1-\Delta_I) \cos 2 \pi n \Delta_I + \frac{1}{\pi n } \sin 2 \pi n \Delta_I \bigg\} \right] \nn\\
		&\quad-\lambda^{-1}\int_{q_l}^{q_r}dq\,\rho(q)\rho'(q) \sum_{n=1}^\infty      \frac{2}{n^3} \left[1-\sum_{I=1}^3\bigg\{ (1-\Delta_I) \cos 2 \pi n \Delta_I + \frac{3}{2\pi n} \sin 2 \pi n \Delta_I \bigg\} \right] \nn\\
		&\quad+\frac{1}{2}\lambda^{-3/2}\int_{q_l}^{q_r}dq\,\rho(q)\rho''(q) \sum_{n=1}^\infty      \frac{4}{n^4} \left[1-\sum_{I=1}^3\bigg\{ (1-\Delta_I) \cos 2 \pi n \Delta_I + \frac{2}{\pi n} \sin 2 \pi n \Delta_I \bigg\} \right] \nn\\
		&\quad + \mathcal{O}(\lambda^{-2})\,.\label{adj1:2}
\end{align}
Implementing the sums over $n$ in (\ref{adj1:2}) using the identities (\ref{eq:sos}), we finally arrive at the expression (\ref{Free1adj}) that gives the contribution from adjoint $\mN=2$ chiral multiplets to the planar effective action, keeping the first two leading terms in the large $\lambda$ expansion. Note that we have implicitly assumed $\Delta_I\leq 2$ to employ the identities (\ref{eq:sos}) without involving modded values, but this inequality does not pose any extra constraint since the examples of our interest --- ADHM and $V^{5,2}$ theories --- satisfy a stronger constraint $\sum \Delta_I = 2$.

\subsection{The 2-node planar effective action}\label{app:detail:isol2}
Here we show that the contributions from $\mN=2$ chiral multiplets to the planar effective action, namely (\ref{free2:conti:split:bif}) and (\ref{free2:conti:split:fun}) for bi-fundamental representations (in conjunction with the classical CS contribution and the vector multiplet contribution) and the pairs of fundamental \& anti-fundamental representations, are expanded in the large $\lambda$ limit as (\ref{Free2bif}) and (\ref{Free2fun}) respectively. 

\subsubsection{Fundamental \& (anti) fundamental representations}\label{app:detail:isol2:fun}
The contribution from fundamental \& anti fundamental $\mN=2$ chiral multiplets, (\ref{free2:conti:split:fun}), can be investigated by following the same procedure described in Appendix \ref{app:detail:isol1:fun} for the 1-node case, which involves the series expansion of the $\ell$-function (\ref{expl}) and the change of an integration variable $t=\lambda^{1/2}|q|$. Once the dust settles, one can rewrite (\ref{free2:conti:split:fun}) as
\begin{align}
	S_0^\text{2-fun}[\rho,p;\lambda] &=\lambda^{1/2} \int_{q_a}^{q_d} dq\, \rho(q)  |q| \left[ \lambda^{-1}_\mu\left(1-\Delta - \frac{p(q)}{2 \pi} \right) +  \lambda^{-1}_\nu\left(1-\Delta + \frac{p(q)}{2 \pi} \right) \right] \nn\\ 
	&\quad - 4\lambda^{-1/2} \sum_{n=1}^\infty \frac{1}{n^2}\rho(0)  \Bigg[ \lambda^{-1}_\mu\left(1- \Delta - \frac{p(0)}{2 \pi}\right) \cos  n \left(2 \pi \Delta + p(0) \right) \nn\\
	&\quad +  \frac{\lambda^{-1}_\mu}{\pi n } \sin  n \left(2 \pi \Delta + p(0) \right) 
	+\lambda^{-1}_\nu\left(1- \Delta + \frac{p(0)}{2 \pi}\right) \cos  n \left(2 \pi \Delta - p(0) \right) \nn\\
	&\quad + \frac{\lambda^{-1}_\nu}{\pi n} \sin  n \left(2 \pi \Delta - p(0) \right)\Bigg]   + \mO(\lambda^{-2})\,.\label{fun2}
\end{align}
Implementing the sums over $n$ in (\ref{fun2}) using the identities (\ref{eq:sos}), we obtain the expression (\ref{Free2fun}) that gives the contribution from the pairs of fundamental \& anti-fundamental $\mN=2$ chiral multiplets to the planar effective action, keeping the first two leading terms in the large $\lambda$ expansion.

\subsubsection{Bi-fundamental representations}\label{app:detail:isol2:bif}
To begin with, we write the contribution from bi-fundamental $\mN=2$ chiral multiplets (\ref{free2:conti:split:bif}) as
\begin{equation}
	S_0^\text{2-bif}[\rho,p;\lambda]=\lambda^{-1/2}S_{0,1/2}^\text{2-bif}[\rho,p]+\lambda^{-3/2}S_{0,3/2}^\text{2-bif}[\rho,p]+\mO(\lambda^{-2})\,.\label{free2:conti:split:bif:exp}
\end{equation}
By substituting $t=\lambda^{1/2}(q-q')$ for the first integration variable $q'$ in (\ref{free2:conti:split:bif}) and then following the calculations of Appendix \ref{app:detail:isol1:adj} for adjoint chiral multiplets, one can expand (\ref{free2:conti:split:bif}) in the large $\lambda$ limit explicitly and thereby determine the expansion coefficients in (\ref{free2:conti:split:bif:exp}). In due process, we extend the $t$-integral boundary to infinity as in Appendix \ref{app:detail:isol1:adj}, which is justified in Appendix \ref{app:detail:exp}. We omit these intermediate steps parallel to the 1-node case and present the final results only.

\medskip

The $\lambda^{-1/2}$ order coefficient in (\ref{free2:conti:split:bif:exp}) is given by
\begin{equation}
\begin{split}
	S_{0,1/2}^\text{2-bif}[\rho,p] & = \int_{q_a}^{q_d} dq\, \rho(q)^2 \biggr[\frac{1}{\pi}  q p(q)+\frac{2}{3}\pi^2 \\
	&\quad-  2 \pi^2 \sum_{a=1}^{2}\left(1- \Delta_{a} -\frac{p(q)}{\pi}\right) \text{B}_2\left(\Delta_{a} +\frac{p(q)}{\pi} \right) \\
	&\quad - 2 \pi^2 \sum_{a=3}^{4}\left(1- \Delta_{a} +\frac{p(q)}{\pi}\right) \text{B}_2\left(\Delta_{a} -\frac{p(q)}{\pi} \right)  \\
	&\quad - \frac{4}{3}\pi^2 \sum_{a=1}^{2}\text{B}_3\left(\Delta_{a} +\frac{p(q)}{\pi}\right) - \frac{4}{3}\pi^2 \sum_{a=3}^{4} \text{B}_3\left(\Delta_{a} - \frac{p(q)}{\pi}\right) \biggl] \,.
\end{split}\label{free2bil}
\end{equation}
To arrive at (\ref{free2bil}), we have assumed
\begin{align}
	- \Delta_{2} \pi \leq p \leq \Delta_{4} \pi  \label{regime}
\end{align}
that allow us to ignore the modulo function in the formula (\ref{eq:sos}), where we have also assumed $\Delta_1 \geq \Delta_2$ and $\Delta_3 \geq \Delta_4$ without loss of generality. In the first glance, the stronger constraints
\begin{align}
	- \Delta_{2} \pi \leq p < (1-\Delta_1)\pi\qquad\&\qquad -(1-\Delta_3)\pi<p\leq \Delta_{4} \pi 
\end{align}
seem to be required to derive (\ref{free2bil}) without the modulo function, but thanks to the combination of Bernoulli polynomials in (\ref{free2bil}) one can show that the weaker constraint (\ref{free2bil}) is in fact enough. See \cite{Geukens2023} for more detailed discussion on this matter. Since the derivation of (\ref{free2bil}) is relying on the inequality (\ref{regime}), one has to check if the saddle point configuration obtained based on the expression (\ref{free2bil}) indeed satisfies (\ref{regime}) a posteriori. For the saddle point configuration presented in subsection \ref{sec:ana:2-node:lead}, it is straightforward to confirm this. Finally, using the constraint $\sum_{a=1}^4 \Delta_a = 2$, (\ref{free2bil}) can be simplified further as obtained in \cite{Jafferis:2011zi} for the ABJM theory, which we present in (\ref{Free2bif}).

\medskip

The $\lambda^{-3/2}$ order coefficient in (\ref{free2:conti:split:bif:exp}) is given by \cite{Geukens2023}
\begin{align}
		&S_{0,3/2}^\text{2-bif}[\rho,p] =  -2\int dq \, \rho(q) \Bigg[ \frac{\pi^2}{3} \rho(q)\left(p'(q)\right)^2 \nn\\
		&+ \sum_{a=1}^2 \Bigg\{ \frac{\pi^4}{3} \left(2\rho'(q)\frac{p'(q)}{2\pi} +\rho(q) \frac{p''(q)}{2\pi}\right) \text{B}_4 \left(\Delta_a + \frac{p(q)}{\pi} \right)  \nonumber\\ 
		& - \pi^2\rho(q)\left(p'(q) \right)^2 \left(1 - \Delta_a - \frac{p(q)}{\pi} \right) \text{B}_2 \left(\Delta_a + \frac{p(q)}{\pi} \right) \nonumber\\ 
		&  +\frac{2 \pi^3}{3} \left[-\left(2 \rho'(q)p'(q)+\rho(q) p''(q) \right)\left( 1 -\Delta_a - \frac{p(q)}{\pi} \right) + 2\rho(q) \frac{\left(p'(q) \right)^2}{2 \pi}  \right]\text{B}_3 \left(\Delta_a + \frac{p(q)}{\pi} \right)  \nonumber\\
		&- \frac{2 \pi^3}{3}\bigg[2 \rho'(q)p'(q) +  \rho(q)p''(q)\bigg] \text{B}_4 \left(\Delta_a + \frac{p(q)}{\pi} \right) -\frac{4 \pi^2}{3}\ \rho(q)\left(p'(q)\right)^2 \text{B}_3 \left(\Delta_a + \frac{p(q)}{\pi} \right) \Bigg\}\nonumber\\ 
		&+ \sum_{a=3}^4 \Bigg\{ - \frac{\pi^4}{3} \left(2\rho'(q)\frac{p'(q)}{2\pi} +\rho(q) \frac{p''(q)}{2\pi}\right) \text{B}_4 \left(\Delta_a - \frac{p(q)}{\pi} \right) \nonumber \\ 
		& - \pi^2\rho(q)\left(p'(q) \right)^2 \left(1 - \Delta_a + \frac{p(q)}{\pi} \right) \text{B}_2 \left(\Delta_a - \frac{p(q)}{\pi} \right) \nonumber \\ 
		&  +\frac{2 \pi^3}{3} \left[\left(2 \rho'(q)p'(q)+\rho(q) p''(q) \right)\left( 1 -\Delta_a + \frac{p(q)}{\pi} \right) + 2\rho(q) \frac{\left(p'(q) \right)^2}{2 \pi}  \right]\text{B}_3 \left(\Delta_a - \frac{p(q)}{\pi} \right)  \nonumber \\
		&+ \frac{2 \pi^3}{3}\bigg[2 \rho'(q)p'(q) +  \rho(q)p''(q)\bigg] \text{B}_4 \left(\Delta_a - \frac{p(q)}{\pi} \right) -\frac{4 \pi^2}{3}\ \rho(q)\left(p'(q)\right)^2 \text{B}_3 \left(\Delta_a - \frac{p(q)}{\pi} \right) \Bigg\} \Bigg] \nn\\
		&+\frac{2 \pi^4}{3}\int dq \, \rho(q)\rho''(q) \left[\frac{1}{15} + \sum_{a=1}^2 \Bigg\{ \left(1- \Delta_{a} -\frac{p(q)}{\pi}\right)\text{B}_4 \left(\Delta_{a} +\frac{p(q)}{\pi}  \right) + \frac{4}{5} \text{B}_5 \left(\Delta_{a} +\frac{p(q)}{\pi}  \right) \Bigg\}\right. \nonumber \\ 
		& +   \left. \sum_{a=3}^4 \Bigg\{ \left(1- \Delta_{a} +\frac{p(q)}{\pi}\right)\text{B}_4 \left(\Delta_{a} -\frac{p(q)}{\pi}  \right) +  \frac{4}{5}\text{B}_5 \left(\Delta_{a} -\frac{p(q)}{\pi}  \right) \Bigg\} \right] \,,\label{free2bis}
\end{align} 
where we have again imposed the condition (\ref{regime}), which has to be confirmed for a saddle point configuration afterward, under the choices $\Delta_1 \geq \Delta_2$ and $\Delta_3 \geq \Delta_4$ without loss of generality. The constraint $\sum_{a=1}^4 \Delta_a = 2$ does not simplify the expression (\ref{free2bis}) significantly, so we keep the expression in terms of Bernoulli polynomials.

\medskip

Note that we skipped the $\lambda^{-1}$ order contribution to the planar effective action from bi-fundamental $\mN=2$ chiral multiplets, which is excluded in the expansion (\ref{free2:conti:split:bif:exp}) from the beginning. In fact there are non-vanishing $\lambda^{-1}$ order contributions arising from (\ref{free2:conti:split:bif}) as in the 1-node case, which read
\begin{equation}
\begin{split}
	S_{0,1}^\text{2-bif}[\rho,p]&=\sum_{n=1}^\infty \int_{q_a}^{q_d} dq \,  \Bigg[\frac{1}{n^3}\frac{d }{d q} \bigg[\sum_{a=1}^4\rho(q)^2 \left(\frac{1}{2}-\left(1-\Delta_a - \sigma_a \frac{p(q)}{\pi} \right) \cos  2 n (\pi \Delta_a + \sigma_a p(q) ) \right) \bigg] \nn  \\
	&\kern6em~-\frac{1}{2 \pi}\frac{3}{n^4} \frac{d }{d q}\bigg[\sum_{a=1}^4\rho(q)^2 \sin 2n (\pi \Delta_a + \sigma_a p(q))  \bigg]\Bigg]\,.
\end{split}\label{free2bi12}
\end{equation}
The functional derivatives of (\ref{free2bi12}) with respect to $\rho(q)$ and $p(q)$ vanish, however, since (\ref{free2bi12}) is given by total derivative terms. Hence the $\lambda^{-1}$ order contribution to the planar effective action (\ref{free2bi12}) does not affect the saddle point equation as in the 1-node case discussed around (\ref{rho1:exp:12}), and its on-shell value also vanishes. This observation effectively justifies the expansion (\ref{free2:conti:split:bif:exp}) excluding the $\lambda^{-1}$ order contribution.

\subsection{Contribution from exponentially suppressed terms} \label{app:detail:exp}
Here we investigate the $t$-integrals in (\ref{adj1:1}) and see if the extension of the integration boundary from $\lambda^{1/2}Q$ (recall $Q=q-q_l$) to $\infty$ is indeed safe in the calculation of (\ref{adj1:1}) up to $\mO(\lambda^{-2})$, and therefore (\ref{adj1:2}) is truly valid. To be more explicit, we would like to prove  
\begin{equation}
	\begin{split}
		S_0^\text{1-adj,\,(exp)}[\rho;\lambda]&\equiv  \sum_{n=1}^\infty \Bigg[\frac{2}{n}J_1\bigg(1-\sum_{I=1}^3 \bigg\{ (1-\Delta_I) \cos 2 \pi n \Delta_I + \frac{1}{2n\pi} \sin 2 \pi n \Delta_I \bigg\} \bigg)  \\ 
		&\kern3em~~-  \frac{1}{\pi n} J_2 \sum_{I=1}^3 \sin 2 \pi n \Delta_I \Bigg]\\
		&\overset{!}{=}\mO(\lambda^{-2}) \,,
	\end{split}\label{claim}
\end{equation}
where (\ref{claim}) is introduced by subtracting the right hand side of (\ref{adj1:2}) from that of (\ref{adj1:1}). In (\ref{claim}) we have used
\begin{equation}
	J_a  =  \int_{q_l}^{q_r} dq \, \rho(q) \bigg(\lambda^{-1/2} \rho(q) I_{a-1}^\text{(exp)}(q) - \lambda^{-1} \rho'(q) I_a^\text{(exp)}(q) +\fft12\lambda^{-3/2} \rho''(q) I_{a+1}^\text{(exp)}(q) \bigg)\label{def:J}
\end{equation}
for a compact expression, based on the definitions
\begin{equation}
	I_a(q)\equiv \int_0^{\lambda^{1/2}Q} dt \, e^{-nt}t^a\,,\qquad I_a^\text{(exp)}(q)\equiv I_a(q) -\fft{a!}{n^{a+1}}\,.
\end{equation}
The recursion formula for $I_a(q)$ and the first few values of $I_a^\text{(exp)}(q)$ are given by
\begin{equation}
    \begin{split} 
    	I_a(q) &= \frac{a}{n}I_{a-1} - \frac{1}{n}\lambda^{a/2}Q^a e^{-n\lambda^{1/2}Q}\qquad(a\geq1)\,,\\
        I_0^\text{(exp)}(q) &= -\frac{1}{n}e^{-n \lambda^{1/2} Q}\, ,  \\
        I_1^\text{(exp)}(q) &= -\left(\frac{1}{n^2} + \frac{1}{n} \lambda^{1/2}Q\right)e^{-n \lambda^{1/2} Q}\, ,  \\
        I_2^\text{(exp)}(q) &= -\left(\frac{2}{n^3} + \frac{2}{n^2} \lambda^{1/2}Q + \frac{1}{n} \lambda Q^2\right)e^{-n \lambda^{1/2} Q}\,,\\
        I_3^\text{(exp)}(q) &= -\left(\frac{6}{n^4} + \frac{6}{n^3} \lambda^{1/2}Q + \frac{3}{n^2} \lambda Q^2+\fft1n\lambda^{3/2}Q^3\right)e^{-n \lambda^{1/2} Q}\, .
    \end{split}\label{integrals}
\end{equation}

To prove the claim (\ref{claim}), it suffices to show
\begin{equation}
	J_1\overset{!}{=}\mO(\lambda^{-2})\qquad\&\qquad J_2\overset{!}{=}\mO(\lambda^{-2})\,.\label{claim:simple}
\end{equation}
We first write out $J_1$ and $J_2$ explicitly by substituting (\ref{integrals}) into (\ref{def:J}) as 
\begin{subequations}
\begin{align}
	J_1&=\lambda^{-1/2}\int_{q_l}^{q_r} dq\, \rho(q) \bigg[- \rho(q) \frac{1}{n} + \lambda^{-1/2} \rho'(q) \left(\frac{1}{n^2} + \frac{1}{n} \lambda^{1/2}Q\right) \nn  \\
	&\quad-\fft12\lambda^{-1}\rho''(q)\left(\frac{2}{n^3} + \frac{2}{n^2} \lambda^{1/2}Q + \frac{1}{n} \lambda Q^2\right)\bigg]e^{-n \lambda^{1/2} Q} \,,\\
	J_2&=\lambda^{-1/2}\int_{q_l}^{q_r} dq\, \rho(q) \bigg[- \rho(q)\left(\frac{1}{n^2} + \frac{1}{n} \lambda^{1/2}Q\right) + \lambda^{-1/2} \rho'(q)\left(\frac{2}{n^3} + \frac{2}{n^2} \lambda^{1/2}Q + \frac{1}{n} \lambda Q^2\right) \nn \\
	&\quad-\fft12\lambda^{-1}\rho''(q)\left(\frac{6}{n^4} + \frac{6}{n^3} \lambda^{1/2}Q + \frac{3}{n^2} \lambda Q^2+\fft1n\lambda^{3/2}Q^3\right)\bigg]e^{-n \lambda^{1/2} Q}\,.
\end{align}\label{J}%
\end{subequations}
Then we take the change of variable $q=q_l+\lambda^{-1/2}s$ and implement the Taylor expansion for $\rho(q)$ and its derivatives around $q=q_l$. Focusing on $J_1$ ($J_2$ can be studied in the same manner), we obtain
\begin{equation}
\begin{split}
	J_1&=-\lambda^{-1}\sum_{k=0}^\infty\sum_{\ell=0}^\infty\lambda^{-(k+\ell)/2}\fft{\rho^{(k)}(q_l)\rho^{(\ell)}(q_l)}{k!\ell!}\int_{0}^{\lambda^{1/2}S} ds\, s^{k+\ell} \frac{1}{n}e^{-n s}\\
	&\quad+\lambda^{-3/2}\sum_{k=0}^\infty\sum_{\ell=0}^\infty\lambda^{-(k+\ell)/2}\fft{\rho^{(k)}(q_l)\rho^{(\ell+1)}(q_l)}{k!\ell!}\int_{0}^{\lambda^{1/2}S} ds\, s^{k+\ell} \left(\frac{1}{n^2} + \frac{s}{n}\right)e^{-n s}\\
	&\quad-\fft12\lambda^{-2}\sum_{k=0}^\infty\sum_{\ell=0}^\infty\lambda^{-(k+\ell)/2}\fft{\rho^{(k)}(q_l)\rho^{(\ell+2)}(q_l)}{k!\ell!}\int_{0}^{\lambda^{1/2}S} ds\, s^{k+\ell} \left(\frac{2}{n^3} + \frac{2s}{n^2} + \frac{s^2}{n}\right)e^{-n s}\,,
\end{split}\label{J1:1}
\end{equation}
where we have defined $S\equiv q_r-q_l$. Based on the integrals (\ref{integrals}), one can replace the upper boundary of the $s$-integrals with $\infty$ up to exponentially suppressed terms as
\begin{equation}
	\begin{split}
		J_1&=-\lambda^{-1}\sum_{k=0}^\infty\sum_{\ell=0}^\infty\Bigg[\lambda^{-(k+\ell)/2}\fft{\rho^{(k)}(q_l)\rho^{(\ell)}(q_l)}{k!\ell!}\int_{0}^{\infty} ds\, s^{k+\ell} \frac{1}{n}e^{-n s}+\mO(e^{-n\lambda^{1/2}S})\Bigg]\\
		&\quad+\lambda^{-3/2}\sum_{k=0}^\infty\sum_{\ell=0}^\infty\Bigg[\lambda^{-(k+\ell)/2}\fft{\rho^{(k)}(q_l)\rho^{(\ell+1)}(q_l)}{k!\ell!}\int_{0}^{\infty} ds\, s^{k+\ell} \left(\frac{1}{n^2} + \frac{s}{n}\right)e^{-n s}+\mO(\lambda^{1/2}e^{-n\lambda^{1/2}S})\Bigg]\\
		&\quad-\fft12\lambda^{-2}\sum_{k=0}^\infty\Bigg[\sum_{\ell=0}^\infty\lambda^{-(k+\ell)/2}\fft{\rho^{(k)}(q_l)\rho^{(\ell+2)}(q_l)}{k!\ell!}\int_{0}^{\infty} ds\, s^{k+\ell} \left(\frac{2}{n^3} + \frac{2s}{n^2} + \frac{s^2}{n}\right)e^{-n s}+\mO(\lambda e^{-n\lambda^{1/2}S})\Bigg]\,.
	\end{split}\label{J1:2}
\end{equation}
Note that now the $s$-integrals are independent of the expansion parameter $\lambda$. Ignoring the terms of order $\mO(\lambda^{-2})$ in (\ref{J1:2}), we arrive at
\begin{equation}
\begin{split}
	J_1&=-\lambda^{-1}\sum_{k,\ell=0}^{k+\ell\leq1}\Bigg[\lambda^{-(k+\ell)/2}\fft{\rho^{(k)}(q_l)\rho^{(\ell)}(q_l)}{k!\ell!}\int_{0}^{\infty} ds\, s^{k+\ell} \frac{1}{n}e^{-n s}\Bigg]\\
	&\quad+\lambda^{-3/2}\rho(q_l)\rho'(q_l)\int_{0}^{\infty} ds\, \left(\frac{1}{n^2} + \frac{s}{n}\right)e^{-n s}+\mO(\lambda^{-2})\,.
\end{split}\label{J1:3}
\end{equation}
For a generic eigenvalue density $\rho(q)$, this final expression (\ref{J1:3}) might not satisfy the claim (\ref{claim:simple}). This means that the contribution from adjoint $\mN=2$ chiral multiplets and the $\mN=2$ vector multiplet to the planar effective action, (\ref{adj1:2}), should be modified by taking the corrections (\ref{claim}) into account to be rigorous.

\medskip

But we can circumvent this issue by first pretending that the claim (\ref{claim:simple}) and therefore the expression (\ref{adj1:2}) are valid, finding the saddle point configuration $\rho^\star(q)$ based on that working assumption, and then confirming (\ref{J1:3}) is suppressed as $\mO(\lambda^{-2})$ on-shell. We follow this strategy and confirm that (\ref{J1:3}) is indeed suppressed as $\mO(\lambda^{-2})$ on the leading saddle point configuration (\ref{1-sol}) accompanied with higher order corrections in (\ref{rho1:exp}), thanks to the fact that
\begin{equation}
	\rho_0(q_l)=0\qquad\&\qquad \rho_{1/2}(q)=0\label{1-node:rho:condition}
\end{equation}
and all the terms in (\ref{J1:3}) that are not suppressed as $\mO(\lambda^{-2})$ are proportional to $\rho(q_l)$ that vanishes on shell up to $\mO(\lambda^{-1})$. One can do exactly the same analysis for $J_2$ and arrive at the same conclusion. This justifies the expression (\ref{adj1:2}) for the contribution from adjoint $\mN=2$ chiral multiplets to the planar effective action. 

\medskip

The contribution from bi-fundamental $\mN=2$ chiral multiplets in the 2-node case, (\ref{free2:conti:split:bif}), also involves double integrals and therefore one must investigate the errors arising from extending the first integration range to infinity carefully in the same manner. The corresponding calculation is parallel to the above analysis for adjoint $\mN=2$ chiral multiplets so we skip details here. In particular, for the ABJM/$N^{0,1,0}$ theories, the errors are suppressed as $\mO(\lambda^{-2})$ thanks to the eigenvalue density (\ref{2-sol:ABJMN010}) vanishing on-shell at the end points of the domain up to $\mO(\lambda^{-1})$, which is precisely the same conclusion we obtained above for adjoint $\mN=2$ chiral multiplets. For the $Q^{1,1,1}$ theory, however, the situation is slightly different and the eigenvalue density (\ref{2-sol:Q111}) does not vanish on-shell at the end points anymore. Fortunately, one can still show that the conclusion remains the same by using the property $p_0(\pm q_0)=\pi/2$ instead. Based on this observation, we safely ignored the errors arising from extending the first integration range to infinity in Appendix \ref{app:detail:isol2:bif} analyzing the contribution from bi-fundamental $\mN=2$ chiral multiplets in the 2-node case.

\subsection{Dirac-Delta function from a piecewise function}\label{app:detail:delta}
Here we explain how Dirac-Delta functions arise in the 2nd derivative of the eigenvalue density as in subsections \ref{sec:ana:1-node:sub} and \ref{sec:ana:2-node:sub}. To begin with, consider the following piecewise function 
\begin{align}
     f(x) =   \begin{cases} f_1(x) & (x < x_0)\\
                    f_2(x) & (x \geq x_0)
        \end{cases}\quad\text{with}\quad f_1(x_0) = f_2(x_0)~~\text{and}~~f'_1(x_0) \neq f'_2(x_0) \,,
\end{align}
where $f_1(x)$ and $f_2(x)$ are smooth functions on the real line. This function $f(x)$ is not differentiable at $x=x_0$ to be rigorous, and we have
\begin{equation}
	f'(x) = \begin{cases} 
		f_1'(x) & (x < x_0)\\
		f_2'(x) & (x > x_0)
	\end{cases}\qquad\&\qquad f''(x) = \begin{cases} 
	f_1''(x) & (x < x_0)\\
	f_2''(x) & (x > x_0)
	\end{cases}\,.
\end{equation}
Then the `effective value' of $f''(x_0)$ can be determined by making the fundamental theorem of calculus consistent with the definition of the integral of a piecewise function over an interval $(x_l,x_r)$ including the breaking point $x=x_0$. To be specific, the fundamental theorem of calculus tells us
\begin{equation}
	\int_{x_l}^{x_r}dx\,f''(x)=f'(x_r)-f'(x_l)=f_2'(x_r)-f_1'(x_l)\,.\label{2der:f:1}
\end{equation}
On the other hand, the definition of the integral of a piecewise function yields
\begin{equation}
\begin{split}
	\int_{x_l}^{x_r}dx\,f''(x)&=\int_{x_l}^{x_0}dx\,f''(x)+\int_{x_0}^{x_r}dx\,f''(x)\\
	&=f_2'(x_r)-f_1'(x_l)-f_2'(x_0)+f_1'(x_0)
\end{split}\label{2der:f:2}
\end{equation}
The two results (\ref{2der:f:1}) and (\ref{2der:f:2}) are inconsistent with each other due to the last two terms in (\ref{2der:f:2}). To remedy this, we introduce an `effective value' of $f''(x_0)$ in terms of the Dirac-Delta function as
\begin{align}
	f''(x) =  \begin{cases} 
		f_1''(x) & (x < x_0) \\ 
		\left(f'_2(x_0)-f'_1(x_0)\right)\delta(x-x_0) & (x = x_0) \\
		f_2''(x)  & (x > x_0) 
	\end{cases} \, .\label{secder}
\end{align}
The prescription (\ref{secder}) does not affect the result of (\ref{2der:f:1}) but modifies that of (\ref{2der:f:2}) since the contribution from the Dirac-Delta function is not captured by the typical definition of a piecewise function; the last two terms in (\ref{2der:f:2}) are now canceled by the contribution from the Dirac-Delta function in (\ref{secder}) and therefore (\ref{2der:f:2}) is modified to be consistent with (\ref{2der:f:1}). In subsections \ref{sec:ana:1-node:sub} and \ref{sec:ana:2-node:sub} we use this prescription (\ref{secder}) to consider the integral of the 2nd derivative of the eigenvalue density (and the imaginary part of the eigenvalue distribution in the 2-node examples) properly.

\subsection{The less constrained \texorpdfstring{$N^{0,1,0}$}{N010} theory}\label{app:detail:N010}
In order to deal with all our 2-node theories of interest in a universal way, we have imposed the shared constraints \eqref{2-node:constraints} and then for the $N^{0,1,0}$ theory further assumed $\lambda_\mu=\lambda_\nu$ (or $r_1=r_2$). In this Appendix we revisit the $N^{0,1,0}$ theory but with generic $r_1=r_\mu=\tir_\mu$ and $r_2=r_\nu=\tir_\nu$ values and the weaker constraint on $R$-charges for fundamental and anti-fundamental $\mN=2$ chiral multiplets, namely
\begin{equation}
	\Delta_\mu+\tDelta_\mu=\Delta_\nu+\tDelta_\nu=1\,,\label{N010:weaker}
\end{equation}
where we have identified the $R$-charges for fundamental multiplets as $\Delta_\mu = \Delta_{\mu_q}$ \& $\Delta_\nu = \Delta_{\nu_q}$, and similarly for the $R$-charges for anti-fundamental multiplets as $\tDelta_\mu = \tDelta_{\mu_q}$ \& $\tDelta_\nu = \tDelta_{\nu_q}$. Note that $\Delta_\mu$ and $\Delta_\nu$ are not necessarily the same here, which is distinguished from the shared constraints \eqref{2-node:constraints}.

Looking at the effective action in \eqref{free2} with generic $\Delta_\mu$ and $\Delta_\nu$ values, we see the need to shift the imaginary part of our ansatz such that the contributions from fundamental \& anti-fundamental $\mN=2$ chiral multiplets are complex conjugates of each other. This shift is similar in purpose to the $\lambda^0$ term in \eqref{ansatz1} for the 1-node case. To be more specific, we shift our original ansatz in \eqref{ansatz2} as
\begin{equation}
    \begin{aligned}
        \mu(q) = \lambda^{1/2} q + \ri p(q) \quad \rightarrow \quad \mu(q) = \lambda^{1/2} q + \ri p(q) + \ri \pi (1- \Delta_{\mu} - \Delta_{\nu} ),    \\
        \nu(q) = \lambda^{1/2} q - \ri p(q) \quad \rightarrow \quad \nu(q) = \lambda^{1/2} q - \ri p(q) + \ri \pi (1- \Delta_{\mu} - \Delta_{\nu} ), \label{eq:cont5}
    \end{aligned}
\end{equation}
Consequently, the bi-fundamental planar effective action (\ref{Free2bif})\footnote{In fact the classical CS term is only affected by the constant shift (\ref{eq:cont5}) here, which we included in the bi-fundamental planar effective action for convenience.} and the fundamental \& anti-fundamental one \eqref{free2:conti:split:fun} are modified as
\begin{subequations}
\begin{align}
	S_0^\text{2-bif}[\rho,p;\lambda] &=  \lambda^{-1/2}\int_{q_a}^{q_d}  dq\, \rho(q) \bigg[-4 \rho(q) p(q)^2 - 4\pi (\Delta_{1} \Delta_{2} - \Delta_{3} \Delta_{4}) \rho(q) p(q)  \nn\\ 
	&\quad  +2 \pi^2 \big(\Delta_{1} \Delta_{2}(\Delta_{3} +\Delta_{4}) +\Delta_{3} \Delta_{4}(\Delta_{1} +\Delta_{2})\big) \rho(q)\label{N010:S0:bif}\\
	&\quad+ \frac{1}{\pi}  p(q) q+\ri\lambda^{-1/2}(1-\Delta_\mu-\Delta_\nu)p(q)  \bigg]+\lambda^{-3/2}S^\text{2-bif}_{0,3/2}[\rho,p]+\mO(\lambda^{-2})  \,,\nn\\
	S_0^\text{2-fun}[\rho,p;\lambda]&= - \int_{q_a}^{q_d}  dq\,\rho(q)  \left[\lambda^{-1}_\mu \ell\left(\frac{1}{2} \left(1 + \Delta_{\nu}  - \Delta_{\mu}\right) - \frac{p(q)}{2\pi} + \ri \lambda^{1/2}\frac{q}{2\pi} \right) \right. \nn\\ 
	& \kern7em~ \left. +\lambda^{-1}_\mu \ell\left(\frac{1}{2} \left(1 + \Delta_{\mu}  - \Delta_{\nu}\right) + \frac{p(q)}{2\pi} - \ri \lambda^{1/2}\frac{q}{2\pi}  \right) \right.  \nn\\ 
	& \kern7em~ \left. + \lambda^{-1}_\mu \ell\left(\frac{1}{2} \left(1 + \Delta_{\mu}  - \Delta_{\nu}\right) + \frac{p(q)}{2\pi} + \ri \lambda^{1/2} \frac{q}{2\pi} \right) \right. \nn\\ 
	& \kern7em~ \left. +\lambda^{-1}_\mu  \ell\left(\frac{1}{2} \left(1 + \Delta_{\nu}  - \Delta_{\mu}\right) - \frac{p(q)}{2\pi} - \ri \lambda^{1/2} \frac{q}{2\pi} \right) \right.  \nn\\ 
	& \kern7em~ \left. + \lambda^{-1} \delta \ell\left(\frac{1}{2} \left(1 + \Delta_{\mu}  - \Delta_{\nu}\right) + \frac{p(q)}{2\pi} + \ri \lambda^{1/2} \frac{q}{2\pi} \right) \right. \nn\\ 
	& \kern7em~ \left. +\lambda^{-1}\delta  \ell\left(\frac{1}{2} \left(1 + \Delta_{\nu}  - \Delta_{\mu}\right) - \frac{p(q)}{2\pi} - \ri \lambda^{1/2} \frac{q}{2\pi} \right) \right] \, ,\label{N010:S0-fun}
\end{align}\label{N010:S0}%
\end{subequations}
for the $N^{0,1,0}$ theory with the aforementioned weaker constraints. Here we have introduced a new parameter $\delta\equiv\frac{1}{k}(r_2 - r_1)=\lambda(\lambda_\nu^{-1}-\lambda_\mu^{-1})$ to capture the difference between $r_1$ and $r_2$; recall that $\delta$ is held fixed under the 't~Hooft limit as discussed around (\ref{2-node:fixed}). Unlike the 1-node case, however, both (\ref{N010:S0:bif}) and (\ref{N010:S0-fun}) now have non-trivial imaginary terms; in particular, the last two terms of (\ref{N010:S0-fun}) cannot be made complex conjugates of each other in general. This implies that for the $N^{0,1,0}$ theory with the weaker constraints one needs more than just shifting the eigenvalue ansatz to get the real planar effective action (or possibly up to a constant imaginary term as in the 1-node case). 

\medskip

One can still investigate the $\lambda^{-1/2}$ leading order planar effective actions (\ref{N010:S0}) under the aforementioned weaker constraints, however, since the above issue involving the non-trivial imaginary contributions are beyond the $\lambda^{-1/2}$ leading order. Indeed, the bi-fundamental planar effective action (\ref{N010:S0:bif}) remains the same as (\ref{Free2bif}) at the $\lambda^{-1/2}$ leading order. The large $\lambda$ limit of the fundamental/anti-fundamental planar effective action (\ref{N010:S0-fun}) reads (recall $\alpha = \frac{r_1 + r_2}{k}$)
\begin{align}
   S_0^\text{2-fun}[\rho,p;\lambda] = \frac{1}{2} \alpha \lambda^{-1/2} \int_{q_a}^{q_d} dq\, \rho(q) |q|+o(\lambda^{-1/2}) \,,
\end{align}
which also agrees with the $\lambda^{-1/2}$ leading order result read off from \eqref{Free2fun} restricted to the superconformal fixed point and $r_1=r_2$. This confirms that the $\lambda^{-1/2}$ leading order planar free energy (\ref{F:N010:lead}) is in fact valid under the weaker constraints (\ref{N010:weaker}) and generic $r_{1,2}$ values, beyond the superconformal configuration (\ref{N010:constraints:sc}) and the extra condition $r_1=r_2$. The planar effective actions (\ref{N010:S0}) start deviating from (\ref{Free2:bif+fun}) at the first subleading order involving non-trivial imaginary contributions, however, and therefore we restrict ourselves to the superconformal configuration and the extra condition $r_1=r_2$ for the subleading analysis in the main text.

\section{Numerical analysis for \texorpdfstring{$S^3$}{S3} partition functions}\label{app:num}
In this Appendix we provide numerical data that support the analytic expressions of the planar free energies in section \ref{sec:num}.
\subsection{ADHM theory}\label{app:num:ADHM}
The numerical analysis described in the beginning of subsection \ref{sec:num:1-node} has been implemented for the ADHM theory with the following $(\lambda,\Delta)$ configurations 
\begin{equation}
\begin{split}
	\lambda&\in\{30,35,40\}\,,\\
	\bDelta&\in\bigg\{(\fft12,\fft12,1,0,\fft12,\fft12),\,(\fft14,\fft34,1,0,\fft12,\fft12),\,(\fft25,\fft35,1,0,\fft12,\fft12),\,(\fft25,\fft35,1,0,\fft14,\fft34),\\
	&\qquad(\fft38,\fft58,1,0,\fft12,\fft12),\,(\fft14,\fft34,1,\fft14,\fft12,\fft12),\,(\fft13,\fft23,1,\fft16,\fft12,\fft12),\,(\fft25,\fft35,1,\fft{1}{10},\fft12,\fft12),\\
	&\qquad(\fft25,\fft35,1,\fft14,\fft12,\fft12),\,(\fft38,\fft58,1,\fft15,\fft12,\fft12),\,(\fft13,\fft23,1,\fft17,\fft12,\fft12),\\
	&\qquad(\fft25,\fft23,\fft{14}{15},\fft{2}{15},\fft{8}{15},\fft{8}{15}),\,(\fft25,\fft57,\fft{31}{35},\fft{11}{70},\fft{39}{70},\fft{39}{70}),\,(\fft35,\fft35,\fft45,0,\fft35,\fft35),\\
	&\qquad(\fft38,\fft23,\fft{23}{24},\fft{7}{48},\fft{25}{48},\fft{25}{48}),\,(\fft58,\fft35,\fft{31}{40},\fft{1}{10},\fft{33}{80},\fft{13}{16})\bigg\}\,,
\end{split}\label{ADHM:config}
\end{equation}
where $\bDelta$ collectively represents $\bDelta=(\Delta_I,\chi,\Delta,\tDelta)$, on top of the configuration (\ref{ADHM:example}) investigated already in subsection \ref{sec:num:1-node:ADHM}. 

\medskip

For all configurations in (\ref{ADHM:config}), we confirmed that the imaginary part of the on-shell effective action is given by (\ref{1-node:S-onshell:Im}). For the real part, we split the $\bDelta$ configurations into two groups and implement the numerical analysis in a slightly different way as follows.

\medskip

\textbf{Case I.} For the first eleven $\bDelta$ configurations in (\ref{ADHM:config}) satisfying the constraints $\Delta_3=1$, we confirmed that the $N^2$ leading order coefficient in the \texttt{LinearModelFit} (\ref{1-node:lmf:improve}) is given by
\begin{align}
	S^\text{ADHM,(lmf)}_0(\lambda,\bDelta)\Big|_{\Delta_3=1}\simeq S^\text{ADHM}_0(\lambda,\bDelta)\Big|_{\Delta_3=1}&= \fft{4\pi\sqrt{2\tDelta_1\tDelta_2\tDelta_3\tDelta_4}}{3}\fft{\Big(\lambda-\fft{1-2(\tDelta_1+\tDelta_2)+\tDelta_1\tDelta_2}{24\tDelta_1\tDelta_2}\Big)^\fft32}{\lambda^2}\nn\\
	&\quad+\fft{\fft{\mA(2\tDelta_1)+\mA(2\tDelta_2)}{4}-\fft{\zeta(3)}{8\pi^2}(\tDelta_3^2+\tDelta_4^2)}{\lambda^2}\,.\label{ADHM-case1}
\end{align}
For example, for $\bDelta=(\fft25,\fft35,1,0,\fft12,\fft12)$ and $\bDelta=(\fft38,\fft58,1,\fft15,\fft12,\fft12)$, we obtained
\begin{table}[H]
	\centering
	\footnotesize
	\renewcommand{\arraystretch}{1.2}
	\begin{tabular}{ |c||c|c| } 
		\hline
		$\lambda$ & $S^\text{ADHM,(lmf)}_0(\lambda,\bDelta)$ & $S^\text{ADHM,(lmf)}_0(\lambda,\bDelta)-S^\text{ADHM}_0(\lambda,\bDelta)$ \\
		\hline\hline
		$30$ & $0.26658716652592388750$ & $7.723\times 10^{-20}$  \\
		\hline
		$35$ & $0.24659634672731647796$ & $8.186\times 10^{-19}$  \\
		\hline
		$40$ & $0.23051788871299883213$ & $2.384\times 10^{-18}$  \\
		\hline
	\end{tabular}
\end{table}
\noindent and 
\begin{table}[H]
	\centering
	\footnotesize
	\renewcommand{\arraystretch}{1.2}
	\begin{tabular}{ |c||c|c| } 
		\hline
		$\lambda$ & $S^\text{ADHM,(lmf)}_0(\lambda,\bDelta)$ & $S^\text{ADHM,(lmf)}_0(\lambda,\bDelta)-S^\text{ADHM}_0(\lambda,\bDelta)$ \\
		\hline\hline
		$30$ & $0.24149108534575765293$ & $1.163\times 10^{-15}$  \\
		\hline
		$35$ & $0.22337746519092736028$ & $2.403\times 10^{-17}$  \\
		\hline
		$40$ & $0.20880952063442259230$ & $-2.046\times 10^{-16}$  \\
		\hline
	\end{tabular}
\end{table}
\noindent respectively. This strongly supports the estimation (\ref{ADHM-case1}) and thereby the ADHM planar free energy (\ref{F:ADHM:num}) with (\ref{ADHM:mfc:special}). The other nine $\bDelta$ configurations satisfying the same constraint $\Delta_3=1$ yield similar results.

\medskip

\textbf{Case II.} For the remaining five $\bDelta$ configurations in (\ref{ADHM:config}) that do not satisfy the constraint $\Delta_3=1$, we could not deduce the analytic expression of $\mfc^\text{ADHM}(\bDelta)$ in (\ref{F:ADHM:num}). Hence we evaluated 
\begin{equation}
	\mD^\text{ADHM}(\lambda,\bDelta)\equiv\lambda^2\Bigg[S^\text{ADHM,(lmf)}_0(\lambda,\bDelta)-\fft{4\pi\sqrt{2\tDelta_1\tDelta_2\tDelta_3\tDelta_4}}{3}\fft{\Big(\lambda-\fft{1-2(\tDelta_1+\tDelta_2)+\tDelta_1\tDelta_2}{24\tDelta_1\tDelta_2}\Big)^\fft32}{\lambda^2}\Bigg]
\end{equation}
instead, using the $N^2$ leading order coefficient in the \texttt{LinearModelFit} (\ref{1-node:lmf:improve}) for the ADHM theory. The results turned out to be independent of $\lambda$, which confirms the ADHM planar free energy (\ref{F:ADHM:num}). For example, for $\bDelta=(\fft25,\fft35,1,0,\fft12,\fft12)$ and $\bDelta=(\fft25,\fft57,\fft{31}{35},\fft{11}{70},\fft{39}{70},\fft{39}{70})$, we found
\begin{table}[H]
	\centering
	\footnotesize
	\renewcommand{\arraystretch}{1.2}
	\begin{tabular}{ |c||c| } 
		\hline
		$\lambda$ & $\mD^\text{ADHM}(\lambda,\bDelta)$ \\
		\hline\hline
		$30$ & $-0.063350893118026318880$   \\
		\hline
		$35$ & $-0.063350893118051109276$  \\
		\hline
		$40$ & $-0.063350893118059503189$  \\
		\hline
	\end{tabular}
	\quad
	\begin{tabular}{ |c||c| } 
		\hline
		$\lambda$ & $\mD^\text{ADHM}(\lambda,\bDelta)$ \\
		\hline\hline
		$30$ & $-0.078279028757827250214$   \\
		\hline
		$35$ & $-0.078279028758408653624$  \\
		\hline
		$40$ & $-0.078279028758601609116$  \\
		\hline
	\end{tabular}
\end{table}
\noindent respectively where the small deviations for different $\lambda$ values are expected from non-perturbative corrections and numerical errors. The other three $\bDelta$ configurations yield similar results. Deducing the analytic expression of $\mfc^\text{ADHM}(\bDelta)$ for these general $\bDelta$ configurations beyond the constraint $\Delta_3=1$ is left for future research. 

\subsection{\texorpdfstring{$V^{5,2}$}{V52} theory}\label{app:num:V52}
The numerical analysis described in the beginning of subsection \ref{sec:num:1-node} has been implemented for the $V^{5,2}$ theory with the following $(\lambda,\Delta)$ configurations 
\begin{equation}
	\begin{split}
		\lambda&\in\{30,35,40\}\,,\\
		\bDelta&\in\bigg\{(\fft12,\fft56,\fft23,0,\fft13,\fft13),\,(\fft12,\fft56,\fft23,\fft14,\fft13,\fft13),\,(\fft59,\fft79,\fft23,\fft25,\fft16,\fft12),\\
		&\qquad(\fft{5}{12},\fft{11}{12},\fft23,\fft27,\fft13,\fft13),\,(\fft{5}{12},\fft{11}{12},\fft23,\fft27,\fft59,\fft19)\bigg\}\,,
	\end{split}\label{V52:config}
\end{equation}
where $\bDelta$ collectively represents $\bDelta=(\Delta_I,\chi,\Delta,\tDelta)$, on top of the configuration (\ref{V52:example}) investigated already in subsection \ref{sec:num:1-node:V52}. 

\medskip

For all configurations in (\ref{V52:config}), we confirmed that the imaginary part of the on-shell effective action is given by (\ref{1-node:S-onshell:Im}). Moving on to the real part, we evaluated 
\begin{equation}
	\mD^{V^{5,2}}(\lambda,\bDelta)\equiv\lambda^2\Bigg[S^{V^{5,2},\text{(lmf)}}_0(\lambda,\bDelta)-\fft{4\pi\sqrt{\tDelta_1\tDelta_2\tDelta_3\tDelta_4}}{3}\fft{\Big(\lambda-\fft{1-(\tDelta_1+\tDelta_2)+\tDelta_1\tDelta_2}{12\tDelta_1\tDelta_2}\Big)^\fft32}{\lambda^2}\Bigg]
\end{equation}
using the $N^2$ leading order coefficient in the \texttt{LinearModelFit} (\ref{1-node:lmf:improve}) for the $V^{5,2}$ theory. The results were confirmed to be independent of $\lambda$, which confirms the $V^{5,2}$ planar free energy (\ref{F:V52:num}). For example, for $\bDelta=(\fft12,\fft56,\fft23,\fft14,\fft13,\fft13)$, we found
\begin{table}[H]
	\centering
	\footnotesize
	\renewcommand{\arraystretch}{1.2}
	\begin{tabular}{ |c||c| } 
		\hline
		$\lambda$ & $\mD^{V^{5,2}}(\lambda,\bDelta)$ \\
		\hline\hline
		$30$ & $0.066013543651827395889$   \\
		\hline
		$35$ & $0.066013545463461466436$  \\
		\hline
		$40$ & $0.066013545878557205314$  \\
		\hline
	\end{tabular}
\end{table}
\noindent where the small deviations are expected from non-perturbative corrections and numerical errors. The other four $\bDelta$ configurations yield similar results. Deducing the analytic expression of $\mfc^{V^{5,2}}(\bDelta)$ in (\ref{F:V52:num}) is left for future research.

\subsection{ABJM theory}\label{app:num:ABJM}
The numerical analysis described in the beginning of subsection \ref{sec:num:2-node} has been implemented for the ABJM theory with the following $(\lambda,\Delta)$ configurations\footnote{For Data II and the first four $\Delta_a$ configurations in DATA III, we generated numerical data for $N=100\sim300$ instead of $N=100\sim350$ (both in step of $10$) but the numerical analysis proceeds exactly the same way.}

\medskip

\begin{subequations}
\noindent \text{\textbf{Data I.}}
\begin{equation}
	\lambda\in\{30,35,40\}\,,\qquad \Delta_a=(\fft12,\fft12,\fft12,\fft12)\,,
\end{equation}
\noindent \text{\textbf{Data II.}}
\begin{equation}
	\begin{split}
		\lambda&=35\,,\\
		\Delta_a&\in\bigg\{(\fft35,\fft15,\fft45,\fft25),\,(\fft13,\fft13,\fft13,1),\,(\fft25,\fft25,\fft45,\fft25),\,(\fft35,\fft35,\fft35,\fft15),\\
		&\qquad(\fft38,\fft38,\fft78,\fft38),\,(\fft78,\fft14,\fft58,\fft14),\,(\fft34,\fft18,\fft58,\fft12),\,(\fft78,\fft14,\fft12,\fft38)\bigg\}\,,
	\end{split}
\end{equation}
\noindent \text{\textbf{Data III.}}
\begin{equation}
	\begin{split}
		\lambda&=30\,,\\
		\Delta_a&\in\bigg\{(\fft58,\fft38,\fft34,\fft14),\,(\fft35,\fft25,\fft12,\fft12),\,(\fft58,\fft38,\fft12,\fft12),\,(\fft23,\fft13,\fft12,\fft12),\,(\fft35,\fft25,\fft{7}{10},\fft{3}{10}),\\
		&\qquad(\fft59,\fft49,\fft23,\fft13),\,(\fft47,\fft37,\fft57,\fft27),\,(\fft47,\fft47,\fft47,\fft27),\,(\fft{7}{12},\fft{5}{12},\fft56,\fft16),\,(\fft35,\fft25,\fft23,\fft13),\\
		&\qquad(\fft58,\fft14,\fft58,\fft12),\,(\fft35,\fft15,\fft{7}{10},\fft12),\,(\fft{7}{12},\fft13,\fft56,\fft14),\,(\fft47,\fft{5}{14},\fft{11}{14},\fft27),\,(\fft{9}{16},\fft38,\fft34,\fft{5}{16})\\
		&\qquad(\fft{5}{12},\fft{5}{12},\fft34,\fft{5}{12}),\,(\fft37,\fft37,\fft57,\fft37),\,(\fft{7}{16},\fft{7}{16},\fft{11}{16},\fft{7}{16}),\,(\fft{9}{16},\fft{9}{16},\fft{9}{16},\fft{5}{16}),\,(\fft78,\fft14,\fft58,\fft14)\\
		&\qquad(\fft12,\fft{5}{12},\fft34,\fft13),\,(\fft34,\fft38,\fft34,\fft{5}{12}),\,(\fft35,\fft{3}{10},\fft35,\fft12),\,(\fft{7}{12},\fft13,\fft{7}{12},\fft12),\,(\fft47,\fft{5}{14},\fft47,\fft12),\\
		&\qquad(\fft2\pi,\fft1\pi,2-\fft{9}{2\pi},\fft{3}{2\pi})\bigg\}\,,
	\end{split}
\end{equation}
\end{subequations}\label{ABJM:config}%
on top of the configuration (\ref{ABJM:example}) investigated already in subsection \ref{sec:num:2-node:ABJM}. For all configurations in (\ref{ABJM:config}), we confirmed that the $N^2$ leading order coefficient in the \texttt{LinearModelFit} (\ref{2-node:lmf:improve}) is given by
\begin{equation}
	S^\text{ABJM,(lmf)}_0(\lambda,\bDelta)\simeq S^\text{ABJM}_0(\lambda,\bDelta)= \fft{4\pi\sqrt{2\Delta_1\Delta_2\Delta_3\Delta_4}}{3}\fft{(\lambda-\fft{1}{24})^\fft32}{\lambda^2}+\fft{\mfc^\text{ABJM}(\bDelta)}{\lambda^2}\,.\label{ABJM-estimation}
\end{equation}
For example, we found
\begin{table}[H]
	\centering
	\footnotesize
	\renewcommand{\arraystretch}{1.2}
	\begin{tabular}{ |c||c|c| } 
		\hline
		$\bDelta$ & $S^\text{ABJM,(lmf)}_0(\lambda,\bDelta)$ & $S^\text{ABJM,(lmf)}_0(\lambda,\bDelta)-S^\text{ABJM}_0(\lambda,\bDelta)$ \\
		\hline\hline
		$(\fft25,\fft25,\fft45,\fft25)$ & $0.22618021668387649308$ & $-1.876\times 10^{-15}$  \\
		\hline
		$(\fft{7}{12},\fft{5}{12},\fft56,\fft16)$ & $0.21724767024706407391$ & $-2.956\times 10^{-14}$  \\
		\hline
		$(\fft2\pi,\fft1\pi,2-\fft{9}{2\pi},\fft{3}{2\pi})$ & $0.25294583781860191930$ & $6.359\times 10^{-18}$  \\
		\hline
	\end{tabular}
\end{table}
\noindent which strongly supports the estimation (\ref{ABJM-estimation}) and thereby the analytic expression for ABJM planar free energy (\ref{F:ABJM:num}) with (\ref{ABJM:mfc}). The other $\Delta_a$ configurations in (\ref{ABJM:config}) omitted in the above table yield similar results.

\subsection{\texorpdfstring{$N^{0,1,0}$}{N010} theory}\label{app:num:N010}
The numerical analysis described in the beginning of subsection \ref{sec:num:2-node} has been implemented for the $N^{0,1,0}$ theory at the superconformal configuration (\ref{N010:constraints:sc}) with the following $(\lambda,\alpha)$ configurations

\medskip

\begin{subequations}
	\noindent \text{\textbf{Data I.}}
	\begin{equation}
		\lambda\in\{30,32,34,36,38,40\}\,,\qquad \alpha=1\,,\label{N010:data1}
	\end{equation}
	\noindent \text{\textbf{Data II.}}
	\begin{equation}
		\lambda=\{30,35,40\}\,,\qquad \alpha\in\bigg\{\fft14,\fft13,\fft12,2,3\bigg\}\,.\label{N010:data2}
	\end{equation}\label{N010:config}
\end{subequations}%

\begin{figure}[t]
	\centering
	\includegraphics[width=0.5\textwidth]{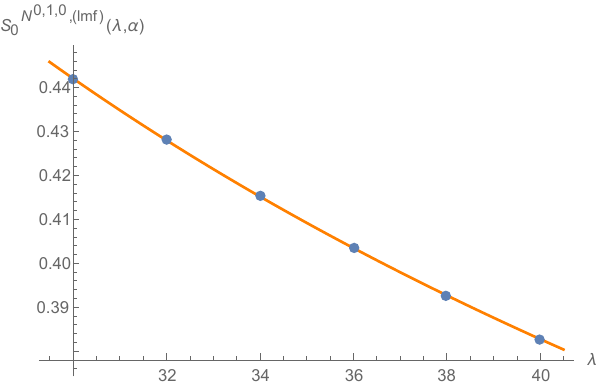}
	\caption{Blue dots represent the numerical values of $S_0^{N^{0,1,0},\text{(lmf)}}(\lambda,\alpha)$ and the orange line corresponds to the fitting curve in (\ref{N010:lmf:lambda}) for Data I (\ref{N010:data1}).}
	\label{fig:N010}
\end{figure}

\noindent The numerical analysis is exactly parallel to the procedure for the ABJM theory described in subsection \ref{sec:num:2-node:ABJM} and Appendix \ref{app:num:ABJM}, and all the configurations in (\ref{N010:config}) confirm the $N^{0,1,0}$ planar free energy (\ref{F:N010:num}). For example, the fitting data for the $N^{0,1,0}$ planar free energy is given for Data I as
\begin{table}[H]
	\centering
	\footnotesize
	\renewcommand{\arraystretch}{1.3}
	\begin{tabular}{ |c|c||c|c| } 
		\hline
		$R^{N^{0,1,0},\mfb}(\alpha)$ & $\sigma^{N^{0,1,0},\mfb}(\alpha)$ & $\mfc^{N^{0,1,0},\text{(lmf)}}(\alpha)$ & $\sigma^{N^{0,1,0},\mfc}(\alpha)$ \\
		\hline\hline
		$5.617\times10^{-13}$ & $3.016\times10^{-13}$ & $0.021381272141328572347$ & $5.988\times10^{-11}$ \\
		\hline
	\end{tabular}
\end{table}
\noindent where the \texttt{LinearModelFit} is implemented as
\begin{equation}
	S_0^{N^{0,1,0},\text{(lmf)}}(\lambda,\alpha)=\mfb^{N^{0,1,0},\text{(lmf)}}(\alpha)\fft{(\lambda-\fft{2-3\alpha}{48})^\fft32}{\lambda^2}+\mfc^{N^{0,1,0},\text{(lmf)}}(\alpha)\fft{1}{\lambda^2}\,,\label{N010:lmf:lambda}
\end{equation}
which is similar to the ABJM theory case (\ref{ABJM:lmf:lambda}). Note that the analytic expression for the first fitting coefficient can be read off from (\ref{F:N010:sub}) as
\begin{equation}
	\mfb^{N^{0,1,0}}(\alpha)\equiv\fft{2(1+\alpha)\pi}{3\sqrt{2+\alpha}}\,.
\end{equation}
Fig.~\ref{fig:N010} also shows that the \texttt{LinearModelFit} (\ref{N010:lmf:lambda}) captures the numerical values of the $N^2$ leading order term in the on-shell effective action $S_0^{N^{0,1,0},\text{(lmf)}}(\lambda,\alpha)$ precisely.
This concludes that the numerical Data I is consistent with the planar free energy (\ref{F:N010:num}). We have also investigated numerical Data II (\ref{N010:data2}) following the procedure for the $V^{5,2}$ theory described in Appendix \ref{app:num:V52}, and confirmed that the results again support the planar free energy (\ref{F:N010:num}).

\bibliography{S3PtnFctSubleading}
\bibliographystyle{JHEP}


\end{document}